\documentclass[prd,singlecolumn,amsmath,nobibnotes,nofootinbib,superscriptaddress,preprintnumbers]{revtex4-1}

\usepackage{bm} \usepackage{graphicx}
\usepackage{color} \usepackage{amssymb}
\usepackage{epstopdf}
\usepackage{rotating}
\usepackage{url}
\usepackage{subfigure}

\def\ba{\begin{eqnarray}}
\def\ea{\end{eqnarray}}
\def\be{\begin{equation}}
\def\ee{\end{equation}}
\def\({\left(}
\def\){\right)}
\def\[{\left[}
\def\]{\right]}
\def\<{\left<}
\def\>{\right>}

\newcommand{\prob}
  {{\rm{Pr}}}
\newcommand{\ns}
  {{N_{\rm{s}}}}
\newcommand{\nblob}
  {{N_{\rm{b}}}}
\newcommand{\nsavge}
  {{\bar{N}_{\rm{s}}}}
  \newcommand{\aobs}
  {\Omega_{\rm{obs}}}
  \newcommand{\src}
  {s}

\newcommand{\npix}
  {{N_{\rm{pix}}}}
\newcommand{\npixsq}
  {{N_{\rm{pix}}^2}}
\newcommand{\nside}
  {{N_{\rm{side}}}}
\newcommand{\nsidesq}
  {{N_{\rm{side}}^2}}
\newcommand{\nsidefull}
  {{N_{\rm{side, full}}}}
\newcommand{\nsidedeg}
  {{N_{\rm{side, deg}}}}
  \newcommand{\ndeg}
  {{n_{\rm deg}}}
  \newcommand{\degfwhm}
  {{f_{\rm deg}}}
\newcommand{\lmax}
  {{\ell_{\rm{max}}}}
\newcommand{\lmaxcov}
  {{\ell_{\rm{max, cov}}}}
\newcommand{\cl}
  {{C_\ell}}
\newcommand{\nl}
  {{N_\ell}}
\newcommand{\muk}
  {{\mu\mathrm{K}}}

\newcommand{\template}
  {\mathbf{t}}
\newcommand{\model}
  {\mathbf{m}}
  \newcommand{\fsky}
  {f_{\rm{sky}}}
  \newcommand{\data}
  {\mathbf{d}}
  \newcommand{\lcdm}
  {{\Lambda {\rm CDM}}}
  \newcommand{\eg}
	{{e.g.}}
	\newcommand{\diff}
  {{\rm{d}}}
  \newcommand{\matr}[1]
        {\mbox{\bf \sf{#1}}}
  
  \newcommand{\blob}
  {b}
  \newcommand{\zc}
  {z_{\rm crit}}
  \newcommand{\thetac}
  {\theta_{\rm crit}}
  
  \newcommand{\legl}
  {P_{\ell}}
  \newcommand{\beamlwmap}
  {B_{\rm \ell, WMAP}}
  \newcommand{\beamlsmooth}
  {B_{\rm \ell, deg}}
  \newcommand{\winldeg}
  {W_{\rm \ell, deg}}
  \newcommand{\sigmawmap}
  {\sigma_{\rm WMAP}}

\newcommand{\scale}{\ensuremath{R}}
\newcommand{\thetacrit}{\ensuremath{\theta_{\rm crit}}}
\newcommand{\stwo}{{\tt S2}}
\newcommand{\stwofil}{{\tt S2FIL}}
\newcommand{\comb}{{\tt COMB}}
\newcommand{\degrees}{\ensuremath{{^\circ}}}

\newcommand{\healpix}{{\tt HEALPix}}
\newcommand{\multinest}{{\tt MultiNest}}
\newcommand{\lapack}{{\tt LAPACK}}

\begin{document}

\title{Hierarchical Bayesian Detection Algorithm for Early-Universe Relics in the \mbox{Cosmic Microwave Background}}
\date{\today}

\author{Stephen M. Feeney}
\email{stephen.feeney.09@ucl.ac.uk}
\affiliation{Department of Physics and Astronomy, University College London, London WC1E 6BT, U.K.}
\author{Matthew C. Johnson}
\email{mjohnson@perimeterinstitute.ca}
\affiliation{Department of Physics and Astronomy, York University, Toronto, Ontario, Canada M3J 1P3} 
\affiliation{Perimeter Institute for Theoretical Physics, Waterloo, Ontario, Canada N2L 2Y5} 
\author{\mbox{Jason D. McEwen}}
\email{jason.mcewen@ucl.ac.uk}
\affiliation{Department of Physics and Astronomy, University College London, London WC1E 6BT, U.K.}
\author{Daniel J. Mortlock}
\email{mortlock@ic.ac.uk}
\affiliation{Astrophysics Group, Imperial College London, Blackett Laboratory, Prince Consort Road, London SW7 2AZ, U.K.}
\affiliation{Department of Mathematics, Imperial College London, London SW7 2AZ, U.K.}
\author{Hiranya V. Peiris}
\email{h.peiris@ucl.ac.uk}
\affiliation{Department of Physics and Astronomy, University College London, London WC1E 6BT, U.K.}

\begin{abstract}
A number of theoretically well-motivated additions to the standard cosmological model predict weak signatures in the form of spatially localized sources embedded in the cosmic microwave background (CMB) fluctuations. We present a hierarchical Bayesian statistical formalism and a complete data analysis pipeline for testing such scenarios. We derive an accurate approximation to the full posterior probability distribution over the parameters defining any theory that predicts sources embedded in the CMB, and perform an extensive set of tests in order to establish its validity. The approximation is implemented using a modular algorithm, designed to avoid \emph{a posteriori} selection effects, which combines a candidate-detection stage with a full Bayesian model-selection and parameter-estimation analysis. We apply this pipeline to theories that predict cosmic textures and bubble collisions, extending previous analyses by using: (1) adaptive-resolution techniques, allowing us to probe features of arbitrary size, and (2) optimal filters, which provide the best possible sensitivity for detecting candidate signatures. We conclude that the WMAP 7-year data do not favor the addition of  either cosmic textures or bubble collisions to $\lcdm$, and place robust constraints on the predicted number of such sources. The expected numbers of bubble collisions and cosmic textures on the CMB sky within our detection thresholds are constrained to be fewer than $4.0$ and $5.2$ at 95\% confidence, respectively.
\end{abstract}

\preprint{}

\maketitle

\section{Introduction}

The cosmic microwave background (CMB) radiation provides our best picture of the primordial universe, and therefore the best set of observations available to confront theories of the early universe with data. The angular power spectrum of the CMB, together with complementary datasets (e.g., from large-scale structure and supernova surveys), has established the standard model of cosmology, a spatially flat universe dominated by a cosmological constant and cold dark matter ($\Lambda$CDM). However, many theories of high-energy physics predict that there should be deviations from the isotropic and purely Gaussian density fluctuations predicted by $\Lambda$CDM. In this paper, we are concerned with the question of how to optimally test theories that predict spatially-localized sources embedded in the CMB. We present a statistical formalism and a set of approximations that are implemented in a full analysis pipeline to construct the posterior probability distribution over the parameters describing a class of theories. We implement a two-step algorithm in which we first locate the most promising candidate signatures, and then use the information about the number, location, and properties of the candidate sources to construct an approximation to the full posterior probability distribution.

To illustrate the application of this pipeline, we focus on two signatures that are predicted by theories with spontaneous symmetry breaking giving rise to phase transitions in the early universe: cosmic textures and cosmic bubble collisions. Cosmic textures are a type of topological defect produced when a non-Abelian global symmetry is broken~\cite{Turok:1989ai}. Textures are not stable, but instead undergo collapse as they come within the expanding cosmological horizon, eventually unwinding into scalar radiation~\cite{Turok:1989ai,Turok:1991qq,Spergel:1990ee,Pen:1993nx,Turok:1990gw}. CMB photons passing through a collapsing texture will be red-shifted, while those passing through an exploding texture will be blue-shifted, giving rise to a set of features in the CMB~\cite{Turok:1990gw}. Cosmic bubble collisions are predicted by theories of eternal inflation, where our observable universe is postulated to be embedded inside one bubble among many, formed during a first-order phase transition out of an inflating false vacuum (for a review of eternal inflation see, e.g., Refs.~\cite{Aguirre:2007gy,Guth:2007ng}). The density perturbations induced by collisions between our bubble and others can lead to localized features in the CMB, providing an observable signature of the dynamics of eternal inflation~\cite{Aguirre:2007an}. 

In previous work, we presented the first constraints on theories giving rise to cosmic bubble collisions~\cite{Feeney_etal:2010dd,Feeney_etal:2010jj} and the first full-sky constraints on cosmic textures~\cite{Feeney_etal:2012jf}. The present paper focuses on:
\begin{itemize}
\item Generalizing the statistical formalism and approximation scheme used in Refs.~\cite{Feeney_etal:2010dd,Feeney_etal:2010jj,Feeney_etal:2012jf};
\item Implementing an adaptive-resolution analysis, allowing us to overcome the limitations in Refs.~\cite{Feeney_etal:2010dd,Feeney_etal:2010jj} on the size of candidate bubble collisions;
\item Including and refining the candidate detection scheme using optimal filters presented in Ref.~\cite{McEwen:2012uk};
\item Performing a complete suite of tests of the formalism, approximations, and analysis pipeline;
\item Performing a new analysis of the posterior probability distribution for bubble collisions and cosmic textures that includes new candidates from the optimal filtering step in combination with the upgraded adaptive-resolution analysis pipeline.
\end{itemize}

The paper is structured as follows. In Sec.~\ref{sec:formalism}, we outline the formalism and approximations we use. In Sec.~\ref{sec:sources}, we describe the theoretical predictions for cosmic textures and bubble collisions. The algorithm used to calculate the approximated posterior is described in Secs.~\ref{sec:source_detection}-\ref{sec:patch_evidence} and tested in Secs.~\ref{sec:adaptive_res_tests} and~\ref{sec:approximation_tests}. A null test of the pipeline is carried out in Sec.~\ref{sec:null_test} before the pipeline is applied to CMB data from the Wilkinson Microwave Anisotropy Probe (WMAP)~\cite{Bennett:2003ba} in Sec.~\ref{sec:wmap_analysis}. The results of this analysis are compared with previous analyses in Sec.~\ref{sec:new_vs_old}, and our conclusions are summarised in Sec.~\ref{sec:conclusions}.


\section{Hierarchical Bayesian source detection formalism}\label{sec:formalism}

\subsection{The theory}

The observed fluctuations in the CMB can be modeled as the realization of a random field on the sphere, which, under the assumption of isotropy and Gaussianity, is completely characterized by its angular power spectrum. A number of extensions of this model predict various populations of distinct sources embedded in the background random field. This includes astrophysical sources such as clusters of galaxies (which affect the CMB through the Sunyaev-Zel'dovich effect~\cite{Sunyaev_Zeldovich:1972}), and primordial sources such as cosmic textures and cosmic bubble collisions. We restrict our attention to cases where the temperature anisotropies can be described as 
\begin{equation}\label{eq:sumofsources}
\frac{\Delta T}{T}(\theta, \phi) = \delta(\theta, \phi) + n(\theta, \phi) + \sum_{i = 1}^{\ns} t_i(\theta, \phi).
\end{equation}
Here, $\delta(\theta, \phi)$ is a realization of the background random field, $n(\theta, \phi)$ describes the instrumental noise as well as residual foregrounds, and $t_i(\theta, \phi)$ are templates for the temperature anisotropies laid down by each of $\ns$ distinct sources. We assume that the sources under consideration are non-overlapping. All terms other than the instrumental noise are assumed to include the effects of a finite instrumental beam. Such a theory can be described by
\begin{itemize}
\item {\bf Model parameters:} This includes the parameters describing the background random field and the source templates.\footnote{The formalism could also be extended to allow for marginalization over any imperfectly known experimental parameters. For simplicity, we assume that the parameters of the WMAP experiment are perfectly known.} The background random field is described by the parameters of $\Lambda$CDM, which we denote by the vector $\model_{\Lambda {\rm CDM}}$. These parameters include: the fraction of energy density in baryons, $\Omega_b h^2$; cold dark matter, $\Omega_{\rm CDM} h^2$; and dark energy, $\Omega_{\Lambda}$; the scalar spectral index, $n_s$; the primordial scalar amplitude, $A_s$; and the optical depth to reionization, $\tau$. Modeling the instrument gives a characterization of the expected noise properties. No model of the Galactic foreground residuals is available for the dataset considered in this analysis, and we therefore resort to null tests of simulations including foreground residuals in order to determine their effects (although a model of the foreground residuals could, in principle, be included in the formalism). 

It is convenient to treat the extension hypothesis as a hierarchical Bayesian model (e.g. Ref.~\cite{Loredo:2012}) in which the population-level parameters are considered separately from the lower-level parameters describing the individual sources. The parameters describing the templates are hence divided into two categories: global parameters, $\model_0$, which describe the source population as a whole; and local parameters, $\model_i$, characterizing individual sources. Any model will possess at least one global parameter -- $\nsavge$, the expected total number of detectable sources -- in addition to any properties common to all templates. Further, any model will possess at least one set of local parameters: $\{ \theta_i, \phi_i \}$, the central position of the $i^{\rm th}$ template. Other properties that can differ from template to template (e.g., size) are also classified as local parameters. Global template parameters, in addition to the parameters of $\Lambda$CDM, can be thought of as labeling different theories, characterizing the background cosmology and the type of source. Local parameters characterize the properties of sources in the context of a specific theory.

\item {\bf Theoretical priors:} An important component of the theory is the prior probability distribution over the model parameters. In principle, a complete theory of cosmology would provide an explanation for the observed properties of the population of sources {\em and} the background random field. In general such a full theory is not available. To make progress, we will assume that there are no correlations between the properties of the background field and the properties of the sources, rendering the prior separable. In the context of a specific theory of sources, the prior over local parameters can be fully determined. The priors over the parameters of $\Lambda$CDM and the global template parameters are somewhat less certain in the absence of a theoretical construction in which different models can be compared.\footnote{Of course, the best example is the eternally inflating multiverse, in which regions with diverse physical properties are sampled. Defining the theoretical prior in this case is difficult due to the infinite number of regions that must be compared; this is the ``Measure Problem" of eternal inflation (see Refs.~\cite{Freivogel:2011eg,Salem:2012} for recent reviews).} A reasonable assumption is therefore to use an uninformative prior, which assigns equal probability to all possibilities. The use of uninformative priors requires care to be taken when defining the prior range, and will be discussed in later sections.

\item {\bf Model statistics:} Given a set of model parameters, it is necessary to understand how particular realizations of the temperature anisotropies are determined. For the sources, this is most efficiently encoded in the theoretical prior over local model parameters of the templates. For the background random field and instrumental noise, this is most efficiently encoded in the two-point correlation function. Under $\Lambda$CDM, for perfect data, the correlation in the temperature between two positions on the sphere is given by
\begin{equation}
C_{ij} \equiv C (\theta_{ij}) = \sum_{\ell} \frac{2 \ell + 1}{4 \pi} C_{\ell} (\model_{\Lambda {\rm CDM}}) P_{\ell} (\cos(\theta_{ij})),
\end{equation}
where $\theta_{ij}$ is the angular distance between two points on the sphere labeled by $i$ and $j$, and $C_{\ell} (\model_{\Lambda {\rm CDM}})$ is the angular power spectrum, which is dependent on the choice of parameters $\model_{\Lambda {\rm CDM}}$. The characterization of the instrumental noise and beam depends on the experiment in question, and will be described in greater detail below for the WMAP experiment.
\end{itemize}

\subsection{The full posterior}

Having fully specified the theory, we can now ask how to test it. Our goal is to construct the posterior probability distribution over the global source parameters, given a dataset $\data$ consisting of pixelized temperature measurements covering a solid angle $\aobs = 4 \pi \fsky$ of the sky (and, optionally, any statistics derived from them). Bayes' theorem gives the posterior as 
\begin{equation}\label{eq:bayes}
\prob(\model_0 | \data, \fsky) 
  = \frac{\prob(\model_0)
    \, \prob(\data | \model_0, \fsky)}{\prob(\data | \fsky)},
\end{equation}
where $\prob(\model_0)$ is the prior distribution on the global parameters, $\model_0$, $\prob(\data | \model_0, \fsky)$ is the likelihood of getting the observed data, and $\prob(\data | \fsky)$ ensures that the posterior is normalized. The posterior can also be used to derive summary statistics, such as confidence intervals on the global model parameters. The quantity $\nsavge$ is always included in the set of global model parameters $\model_0$, and $\lcdm$ is specified by $\nsavge = 0$. Therefore, we can also perform model selection by comparing the posterior probability of a model for which $\nsavge = 0$ and one which admits $\nsavge > 0$.

The model likelihood $\prob(\data | \model_0, \fsky)$ is obtained by marginalizing over: 
\begin{enumerate}
\item The parameters of $\lcdm$;
\item The actual number of sources present on the observable sky (given the expected number of sources, $\nsavge$, the actual number is a realization of a Poisson-like process of mean $\fsky \nsavge$);
\item The local template parameters.
\end{enumerate}
Unpacking the model likelihood, we therefore have:
\begin{eqnarray}
\label{equation:modellikelihood}
\prob(\data | \model_0, \fsky)
  & = & \sum_{\ns = 0}^\infty 
    \frac{(\fsky \nsavge)^\ns e^{-\fsky \nsavge}}{\ns!}  \\
    && \times \int \diff \model_\lcdm \prob(\model_\lcdm)
    \int \diff \model_1 \ldots \diff \model_\ns \prob(\model_1, \ldots , \model_{\ns}) \prob (\data | \ns, \fsky, \model_\lcdm, \model_0, \model_1, \ldots , \model_\ns), \nonumber
\end{eqnarray}
where $\prob(\model_\lcdm)$ is the prior over the parameters of $\lcdm$, $\prob(\model_1, \ldots \model_{\ns})$ is the prior over the local model parameters for $\ns$ sources, and $\prob (\data | \ns, \fsky, \model_\lcdm, \model_0, \model_1, \ldots \model_\ns)$ is the likelihood. For measurements of the CMB temperature anisotropies under the assumption of $\lcdm$ as the background random field, the likelihood is written as
\begin{equation}\label{eq:likelihood}
\prob (\data | \ns, \fsky, \model_\lcdm, \model_0, \model_1, \ldots , \model_\ns) = \frac{1}{(2 \pi)^{\npix / 2} |\matr{C}|}
    e^{- (\data - \sum_{i=1}^{\ns} \template(\model_i) ) 
      \matr{C}^{-1} 
    (\data - \sum_{i=1}^{\ns} \template(\model_i) )^{\rm{T}} / 2},
\end{equation}
where $\matr{C}$ is the pixel-pixel covariance matrix including a $\lcdm$ CMB signal and instrumental noise. 

One must evaluate Eq.~\ref{eq:likelihood} in order to construct the full posterior, Eq.~\ref{eq:bayes}. Evaluating the full posterior exactly is impossible, as it requires marginalizing over a formally infinite dimensional parameter space (since one must consider realizations with an arbitrary number of templates). Even if the parameter space were finite, the enormous size of modern CMB datasets, such as those produced by the WMAP or Planck~\cite{Planck_Ade:2011} experiments, makes inverting the (non-diagonal) covariance matrix prohibitively expensive. Nevertheless, it is possible to apply a controlled and testable sequence of approximations to estimate the full posterior, as we now describe.

\subsection{Approximation to the posterior}

In order to proceed, we must construct a suitable approximation to the model likelihood. Let us first deal with the cosmological parameters $\model_\lcdm$. We assume that our inferences on $\model_i$ do not vary significantly for the range of $\lcdm$ parameters allowed by current cosmological data. Under this assumption, we can proceed as if $\prob(\model_\lcdm) = \delta(\model_\lcdm - \bar{\model}_\lcdm)$, where $\bar{\model}_\lcdm$ is the best-fit concordance cosmological model. Performing the integral over $\model_\lcdm$ we obtain the approximation to the model likelihood
\begin{eqnarray}
\label{equation:modellikelihood_approx}
\prob(\data | \model_0, \fsky)
  & \simeq & \sum_{\ns = 0}^\infty 
    \frac{(\fsky \nsavge)^\ns e^{-\fsky \nsavge}}{\ns!}  \prob (\data | \ns),
    \end{eqnarray}
where we define the helpful short-hand
\begin{equation}\label{eq:single_evidence}
\prob (\data | \ns) \equiv \int \diff \model_1 \ldots \diff \model_\ns \prob(\model_1, \ldots , \model_{\ns}) \prob (\data | \ns, \fsky, \bar{\model}_\lcdm, \model_0, \model_1, \ldots , \model_\ns).
\end{equation}

\subsubsection{Likelihood Derivation for One Candidate Source}

\begin{figure}
\includegraphics[width=10.0cm]{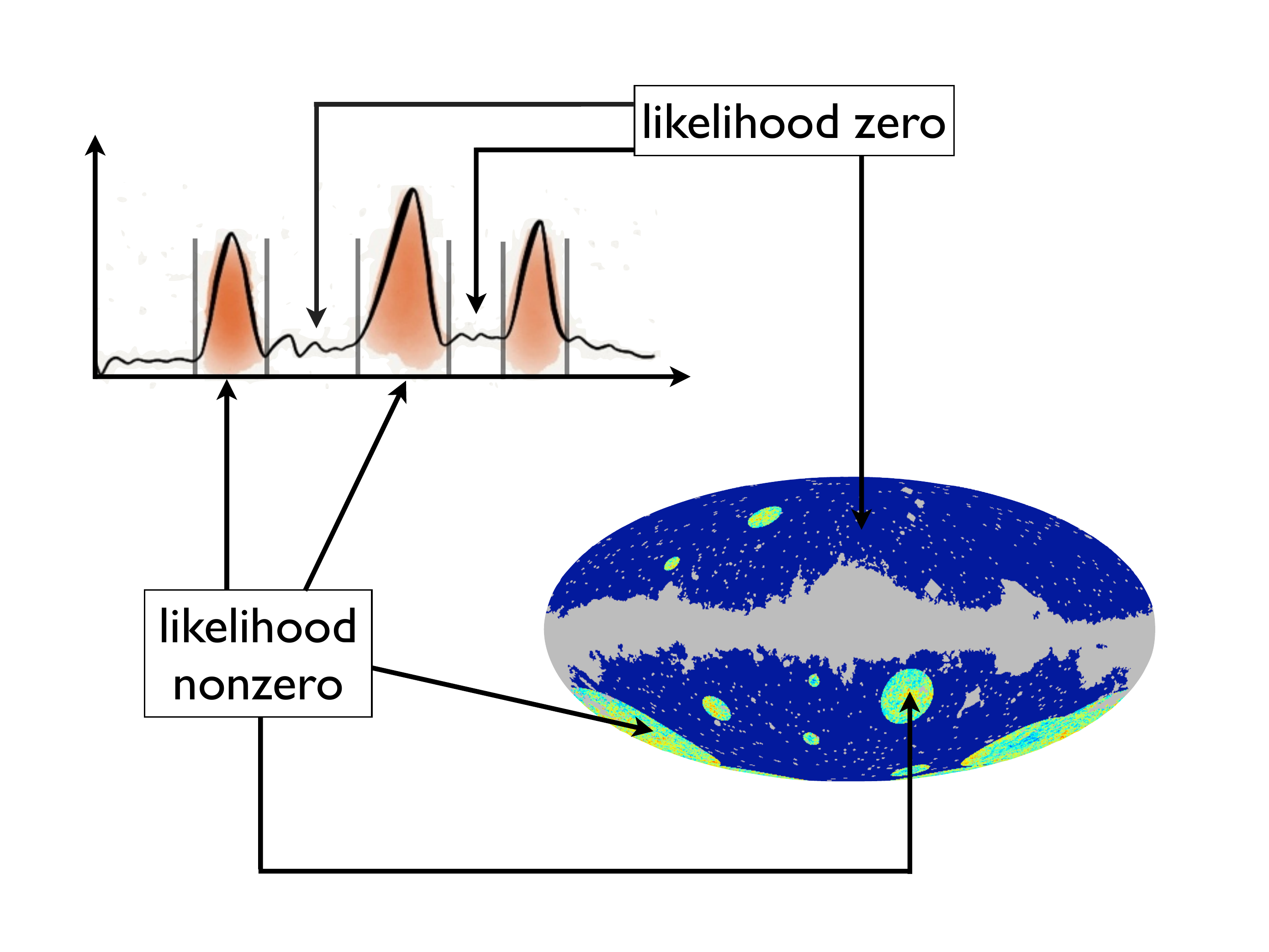}
\caption{A schematic depicting the approximation scheme we employ. By locating a set of well-separated candidate sources, it is possible to determine in which regions of parameter space the likelihood function is appreciably different from zero. Eq.~\ref{eq:single_evidence} can be approximated by integrating over only those regions where the likelihood is large. This collapses the sum in Eq.~\ref{equation:modellikelihood_approx} to a finite number of terms. Finally, we neglect correlations of the random Gaussian background CMB between pixels inside and outside each blob.
  \label{fig-likelihoodsummary}
}
\end{figure}

In a similar spirit, if we knew something about the dependence of the likelihood on the local model parameters, it would be possible to approximate the remaining integrals. This is depicted schematically in Fig.~\ref{fig-likelihoodsummary}. To see how this works in detail, let us begin with a particular example. Imagine that there is a region of the sky which has been judged by some independent method to be a good candidate source. We can segment the data into a ``blob", containing the candidate source, and the rest of the sky. The details of the size and shape of the blob are treated in abstract here, and will depend on the particular theory of the sources being tested. We can now evaluate the sum over $\ns$ in Eq.~\ref{equation:modellikelihood_approx} term by term. The likelihood in the first term, for $\ns =0$, is simply given by
\begin{equation}
 \prob (\data | \ns=0) = \frac{1}{(2 \pi)^{\npix / 2} |\matr{C}|} e^{- \data \matr{C}^{-1}  \data^{\rm{T}} /2},
\end{equation}
which is the likelihood for the null hypothesis, i.e., no sources. Here, and in what follows, $\matr{C}$ is evaluated at the best fit values $\bar{\model}_\lcdm$. Moving on to the $\ns = 1$ term, we must evaluate the integral over $\model_1$. Recall that the local model parameters always include the location at which the template is centered and, if relevant, its size. We can therefore separate the integral over $\model_1$ into the region inside the blob containing our candidate source, which we will refer to as region $b$, and the rest of the sky, which we will refer to as region $\bar{b}$: 
\begin{eqnarray}
\prob (\data | \ns=1)&=&  \int_b \diff \model_1  \prob(\model_1)  \frac{1}{(2 \pi)^{\npix / 2} |\matr{C}|}
    e^{- (\data - \template)
      \matr{C}^{-1}
    (\data - \template)^{\rm{T}} /2} \nonumber \\
   &+&   \int_{\bar{b}} \diff \model_1  \prob(\model_1)  \frac{1}{(2 \pi)^{\npix / 2} |\matr{C}|}
    e^{- (\data - \template)
      \matr{C}^{-1}
    (\data - \template)^{\rm{T}} /2}.
\end{eqnarray}

If there are no sources in the region $\bar{b}$, then we can approximate the likelihood by integrating over region $b$ alone. The accuracy of this approximation depends on our ability to locate the candidate source; however, it will always provide a {\em lower} (i.e., conservative) bound on the likelihood (since we are integrating a positive-definite function). We therefore have
\begin{eqnarray}\label{eq:margm1}
\prob (\data | \ns=1)  &\simeq&  \int_b \diff \model_1  \prob(\model_1)  \frac{1}{(2 \pi)^{\npix / 2} |\matr{C}|}
    e^{- (\data - \template)
      \matr{C}^{-1}
    (\data - \template)^{\rm{T}} /2}.
 \end{eqnarray}
While we have reduced the parameter space over which we must integrate, this expression is still numerically intractable for large datasets, since we must invert the large, generally non-diagonal covariance matrix. To make progress, we must make a few further approximations. Expanding the exponential in the likelihood, we have
 \begin{eqnarray}
  \frac{1}{(2 \pi)^{\npix / 2} |\matr{C}|}
    e^{- (\data - \template)
      \matr{C}^{-1}
    (\data - \template)^{\rm{T}} /2}   &=& \frac{1}{(2 \pi)^{\npix / 2} |\matr{C}|} 
    e^{- (d_i - t_i)
      {C}_{ij}^{-1}
    (d_j - t_j) /2} 
    \times
    e^{- d_{\alpha} 
      {C}_{\alpha \beta}^{-1}
    d_{\beta} /2}
    \times
    e^{- (d_i - t_i)
      {C}_{i \alpha}^{-1}
    d_{\alpha}}	,
\end{eqnarray}
where the indices $i$ and $j$ refer to pixels in region $b$ while the indices $\alpha$ and $\beta$ correspond to pixels in region $\bar{b}$. We have used the fact that the template vanishes in region $\bar{b}$. Re-arranging, we obtain
 \begin{eqnarray}
    \frac{1}{(2 \pi)^{\npix / 2} |\matr{C}|}  e^{- (\data - \template)
      \matr{C}^{-1}
    (\data - \template)^{\rm{T}} /2} &=& 
    \frac{e^{- (d_i - t_i)
      {C}_{ij}^{-1}
    (d_j - t_j) /2} }{ e^{- d_i
      {C}_{ij}^{-1}
    d_j/2}  }
    \times
    \left( \frac{1}{(2 \pi)^{\npix / 2} |\matr{C}|}
    e^{- \data
      \matr{C}^{-1}
    \data^{\rm{T}} /2} \right)
    \times
    \left( e^{t_i
      {C}_{i \alpha}^{-1}
    d_{\alpha}} \right)	\\
    &\simeq& 
         \frac{   e^{- (\data - \template)
      {\matr{C}^{(bb)}}^{-1}
    (\data - \template)^{\rm{T}} /2} }{  e^{- \data
      {\matr{C}^{(bb)}}^{-1}
    \data^{\rm{T}} /2}  }
    \times
    \left( \frac{1}{(2 \pi)^{\npix / 2} |\matr{C}|}
    e^{- \data
      \matr{C}^{-1}
    \data^{\rm{T}} /2} \right)
    \times
    \left( e^{t_i
      {C}_{i \alpha}^{-1}
    d_{\alpha}} \right) \\
    & \simeq & 
         \frac{ e^{- (\data - \template)
      {\matr{C}^{(bb)}}^{-1}
    (\data - \template)^{\rm{T}} /2} }{  e^{- \data
      {\matr{C}^{(bb)}}^{-1}
    \data^{\rm{T}} /2}  }
    \times
 \prob (\data | \ns=0).
\end{eqnarray}
In these expressions, $\matr{C}^{(bb)}$ is the covariance matrix constructed using only the data in region $b$. We have made two approximations in deriving this final expression. First, we have neglected correlations between the template and the data in region $\bar{b}$, which is equivalent to assuming
\begin{equation}\label{eq:bbbarneglected}
e^{2 t_i
      {C}_{i \alpha}^{-1}
    d_{\alpha}^{\rm{T}}}  \simeq 1.
\end{equation}
This is justified in the limit where the inverse covariance falls off sufficiently fast with angular distance. Our second approximation was to assume that we can make the replacement
\begin{equation}\label{eq:correlationneglect2}
(d_i - t_i) {C}_{ij}^{-1} (d_j - t_j) \rightarrow  (\data - \template) {\matr{C}^{(bb)}}^{-1} (\data - \template)^{\rm{T}} /2,
\end{equation}
which is justified to the extent that the subgroup of the inverse of the full covariance matrix corresponding to pixels in region $b$ can be approximated as the inverse of a covariance matrix defined only using pixels in region $b$. For a diagonal covariance, this is exact. For the non-diagonal covariance matrix of $\lcdm$ it is only approximate. We comment on the validity of these approximations under $\lcdm$ in Section~\ref{sec:approximation_tests}. 

Finally, performing the integral over $\model_1$ in Eq.~\ref{eq:margm1}, we obtain
\begin{equation}
\prob (\data | \ns=1) \simeq  \prob (\data | \ns=0 ) \, \rho_b (\model_0),
\end{equation}
where the patch-based evidence ratio $\rho_b$ is given by
\begin{equation}
\label{equation:rho_int_over_local_params}
\rho_b (\model_0) \equiv \frac{\int_b \diff \model_1  \prob(\model_1) e^{- (\data - \template)
      {\matr{C}^{(bb)}}^{-1}
    (\data - \template)^{\rm{T}} /2} }{  e^{- \data
      {\matr{C}^{(bb)}}^{-1}
    \data^{\rm{T}} /2} }.
\end{equation}
This is a measure of how much better the theory with a template at fixed $\model_0$ fits the patch of data than the theory with only the random field.

Now, we evaluate the $\ns = 2$ term. Again, we approximate the full integral as the integral over region $b$ alone. This yields
\begin{equation}
\prob (\data | \ns=2) =  \int_b \int_b \diff \model_1\diff \model_2  \prob(\model_1) \prob(\model_2) \frac{1}{(2 \pi)^{\npix / 2} |\matr{C}|}
    e^{- (\data - \template(\model_1) - \template(\model_2))
      \matr{C}^{-1}
    (\data - \template(\model_1) - \template(\model_2))^{\rm{T}} /2}.
\end{equation}
Making the same approximation about the covariance matrix as above, we have
\begin{equation}
\prob( \data | \ns = 2) \simeq \prob (\data | \ns=0 ) \frac{\int_b \int_b \diff \model_1 \diff \model_2  \prob(\model_1) \prob(\model_2) e^{- (\data - \template (\model_1) - \template(\model_2))
      {\matr{C}^{(bb)}}^{-1}
    (\data - \template(\model_1) - \template(\model_2) )^{\rm{T}} /2} }{  e^{- \data
      {\matr{C}^{(bb)}}^{-1}
    \data^{\rm{T}} /2} }.
\end{equation}
If there is in fact only a single source in region $b$ then the addition of another template will not increase the likelihood: in effect, we would be trying to fit a single feature with a template possessing {\em twice} the number of parameters ($\model_1$ and $\model_2$). As a concrete example, assume that the source location is the only local model parameter. If the source can be located anywhere on the sky with equal probability, the theory prior is simply
\begin{equation}
\prob(\model_i) = \frac{1}{4 \pi}.
\end{equation}
In the case where the likelihood function is roughly equal for all positions inside $b$, with the solid angle contained in $b$ given by $\Omega_b$, and there is no improvement from adding a second template, the relative size of the $\ns = 1$ and $\ns = 2$ terms can be estimated as
\begin{equation}
\frac{\prob( \data | \ns=1) }{\prob( \data | \ns=2) } \simeq \frac{4 \pi}{\Omega_b}.
\end{equation}
Assuming the blob does not cover the entire sky, this is always larger than one. Subsequent terms in the $\ns$ expansion will be penalized by higher powers of this ratio. While this is a toy model, this property is expected to hold generally.

More generally, if there are multiple sources in a single blob, increasing the number of templates will improve the fit. If the sources are well-separated, however, this will be a very unlikely occurrence, and terms with $\ns > 1$ can therefore be neglected. We operate under this assumption, providing justification in the context of the specific sources considered in Sec.~\ref{sec:sources}. Relaxation of this assumption would require the inclusion of all terms in the sum in Eq.~\ref{equation:modellikelihood_approx}. 

For a single blob, we can therefore approximate the full-sky posterior Eq.~\ref{eq:bayes} as
\begin{equation}
\prob(\model_0 | \data, \fsky) \simeq
 \frac{\prob(\model_0) \, \prob (\data | \ns=0 ) \, e^{- \fsky \nsavge} \, \left( 1 + (\fsky \nsavge) \, \rho_b (\model_0) \right) }{\prob (\data|\fsky)},
\end{equation}
where
\begin{equation}
\prob (\data|\fsky) \equiv \int \diff \model_0 \prob(\model_0) e^{- \fsky \nsavge}
\left( 1 + 
 (\fsky \nsavge) 
   \, \rho_b (\model_0) \right)
\end{equation}
is the evidence which ensures $\prob(\model_0 | \data, \fsky)$ is normalized to unity. Recall that $\nsavge$ is included in the vector of parameters $\model_0$.

\subsubsection{Likelihood Derivation for an Arbitrary Number of Candidate Sources}
 
We now move on to discuss the general case, where there are $\nblob$ blobs, labelled $\blob_1 \ldots \blob_{\ns}$, identified as containing a candidate source. We assume that the blobs in question do not overlap. For $N_{\rm r} = \nblob + 1$ regions on the sky, making the approximation that the template likelihood is small when evaluated outside a blob and neglecting correlations between blobs, we obtain for the general case
\begin{eqnarray}
\label{equation:likelihoodfinal}
\prob(\data | \ns) 
  = 
  \left\{
    \begin{array}{lll}
    0, & {\rm if} & \ns > \nblob , \\
    & & \\    \prob(\data | \ns = 0)
\sum_{\blob_1, \blob_2, \ldots, \blob_{\ns} = 1}^{\nblob}
    \Delta^{b_1 b_2 \ldots b_{\ns}}
    \prod_{\src = 1}^{\ns} 
      \rho_{b_s} (\model_0),
    & {\rm if}
        & \ns \leq \nblob.
        \end{array}
        \right.
\end{eqnarray}
The quantity $\Delta^{b_1 b_2 \ldots b_{\ns}}$ is one when all indices take distinct values and zero otherwise: the sum hence generates all permutations of $\ns$ sources located in $\nblob$ blobs, assuming no more than one source per blob. If there are fewer blobs on the sky than proposed sources, then the likelihood is very small: this would involve fitting more than one template within a single blob, and incurring the penalisation previously discussed. If there are at least as many blobs as proposed sources, then the likelihood takes the form of a sum that includes every possible association of the $\ns$ sources with the $\nblob$ blobs, provided that no two sources are matched to the same blob. For example, the approximation to the $\ns = 1$ term in the presence of multiple blobs is
\begin{equation}
\prob(\data | \ns = 1) = \prob(\data | \ns = 0) \left[ \rho_1 + \rho_2 \ldots + \rho_\nblob \right],
\end{equation}
which is simply an approximation to the integral over positions including only the contributions where the likelihood is large. The approximation to the $\ns = 2$ term is
\begin{equation}
\prob(\data | \ns = 2) = \prob(\data | \ns = 0) \left[ 2 \rho_{1} \rho_{2} + 2 \rho_{1} \rho_3 + 2 \rho_2 \rho_3 + \ldots \right],
\end{equation}
where the product results from the double integral over positions and again we only consider contributions from regions where the likelihood is large. The factor of two multiplying the blob evidences derives from the fact that there are two ways of placing two sources in two separate blobs. Note that terms for $\ns \geq 2$ would be altered if our assumption of well-separated blobs was violated. 

Substituting Eq.~\ref{equation:likelihoodfinal} into Eq.~\ref{equation:modellikelihood_approx}, the expression for the approximation to the full posterior is given by
\begin{equation}
\label{equation:posteriorfinalapp}
\prob(\model_0 | \data, \fsky) 
  \simeq \frac{\prob(\model_0)
    \, \prob(\data | \ns = 0) }{\prob(\data | \fsky)}   e^{-\fsky \nsavge} \sum_{\ns = 0}^\nblob 
    \frac{(\fsky \nsavge)^\ns }{\ns!}  
\sum_{\blob_1, \blob_2, \ldots, \blob_{\ns} = 1}^{\nblob}
    \Delta^{b_1 b_2 \ldots b_{\ns}}
    \prod_{\src = 1}^{\ns} 
      \rho_{b_s} (\model_0).
\end{equation}
Eq.~\ref{equation:posteriorfinalapp} is the main result of this calculation, from which all following results are derived.  In the limit of a single isolated observation, Eq.~\ref{equation:posteriorfinalapp} reproduces the Bayesian source detection formalism developed in Refs.~\cite{Hobson_McLachlan:2003,Hobson_etal:2010}.


\section{Sources}\label{sec:sources}

In this paper, we consider two theories that give rise to localized sources in the CMB: cosmic bubble collisions in the eternal inflation scenario and cosmic textures. For bubble collisions, the only global parameter is $\nsavge$, and the final result of the analysis is  a one-dimensional posterior probability distribution. The first analysis of cosmic bubble collisions using a variant of the approximation scheme outlined in the previous section was presented in Refs.~\cite{Feeney_etal:2010jj,Feeney_etal:2010dd}, where, in addition to the location, three local model parameters (size, edge discontinuity, and amplitude) were included. Cosmic textures have two global parameters: $\nsavge$ and a measure of the symmetry breaking scale $\epsilon$, and therefore the final product is a two-dimensional posterior probability distribution. An analysis of textures, which also used a variant of the approximation scheme outlined above, was presented in Ref.~\cite{Feeney_etal:2012jf}. In this study a model of textures with one local parameter (size) in addition to the position was considered. Previous work on testing for the signature of textures in the CMB was presented in Refs.~\cite{2008MNRAS.390..913C,2011MNRAS.410...33V,Cruz:2004ce,Cruz:2007pe,Cruz:2009nd}. Below, we outline our models of these two types of sources, including the prior probability distribution on the local model parameters. For bubble collisions, we update the model assumptions of Refs.~\cite{Feeney_etal:2010dd,Feeney_etal:2010jj} in light of improved theoretical understanding, and remove the edge discontinuity parameter from our analysis.

\subsection{Bubble collisions}
For an overview of the theory of eternal inflation and the observable effects of bubble collisions, we refer the reader to the reviews~\cite{Aguirre:2007gy,Aguirre:2009ug}. For a detailed discussion of the expected signature of bubble collisions in the CMB, we refer the reader to Refs.~\cite{Chang_Kleban_Levi:2009,Feeney_etal:2010dd,Czech:2010rg,Gobbetti_Kleban:2012,Kleban_Levi_Sigurdson:2011}; here, we provide only a brief overview.

Based on the symmetry of the bubble collision spacetime, the existence of a causal boundary splitting the bubble interior into regions affected and not affected by a collision event, and the fact that a bubble collision is a pre-inflationary relic, the most general template for the temperature fluctuation caused by a single bubble collision is given by~\cite{Chang_Kleban_Levi:2009,Feeney_etal:2010jj,Feeney_etal:2010dd}
\begin{equation}\label{eq-collision-template}
t(\theta,\phi) = \left ( \frac{\zc - z_0 \cos \thetac}{1 - \cos \thetac} + \frac{z_0 - \zc}{1 - \cos \thetac} \cos \theta \right ) \Theta(\thetac - \theta),
\end{equation}
where $\thetac$ is the angular scale of the source, corresponding to the causal boundary of the collision event, $z_0$ and $\zc$ are the amplitudes at the center and edge of the template, $\Theta$ is the Heaviside step function, and we have centered the template on the Galactic north pole. In the limit of small amplitude, this is an additive contribution to the CMB temperature anisotropies as in Eq.~\ref{eq:sumofsources}. Theoretical work~\cite{Gobbetti_Kleban:2012,Kleban_Levi_Sigurdson:2011} which appeared subsequent to the previous analysis suggests that there is no discontinuity in temperature at the causal boundary, and we therefore restrict our attention to $\zc = 0$.

The bubble collision model contains only one global parameter, the expected number of detectable sources $\nsavge$. This is partially a function of the properties of the potential sourcing inflation, and as such is impossible to predict without a model for the potential. In the context of an inflationary landscape, $\nsavge$ can be considered as a continuous parameter with some prior distribution reflecting the typical vacua produced, but without a measure for the landscape it is difficult to estimate even an order of magnitude for this number. It may even be quite likely that $\nsavge \gg 1$, in which case the approximation of looking for widely separated sources is not valid. Ref.~\cite{Kozaczuk:2012sx} has derived the observational signatures in this case (see their Fig. 7), and has used WMAP measurements of the low-$\ell$ CMB temperature power spectrum to constrain a combination of the number of observable collisions, their average amplitude and the current energy density in curvature. We shall concentrate on the small-$\nsavge$ limit and, in the absence of a detailed theory, assume that $\nsavge$ can take values of $\mathcal{O}(1)$, and set $\nsavge$ to be uniform in the range $0 \le \nsavge \le 10$. As we shall see, this parameter is constrained by data, and the precise choice of upper limit has no effect on the analysis. In this model, $\Lambda$CDM corresponds to $\nsavge = 0$.

The local parameters are the collision signature's central amplitude, $z_0$, size, $\thetac$, and location, $\{\theta_0, \phi_0\}$. The modulations are equally likely to be hot or cold and are isotropically distributed across the sky. Theory does not fix the expected amplitude of the collisions, so we assume that the amplitude is uniform in the range $-10^{-4} \le z_0 \le 10^{-4}$ (as stronger collisions would have been obvious in previous CMB data). Neglecting the back-reaction of the collision on the geometry of the bubble interior, the distribution of source sizes is proportional to $\sin \thetac$~\cite{Aguirre:2007an,Aguirre:2007wm,Freivogel_etal:2009it,Aguirre:2009ug}. Further assuming no correlation between the various local parameters, the final {\em normalized} prior on the local parameters is 
\begin{equation}
\prob(\model_1) = \prob(z_0) \, \prob(\theta_0, \phi_0) \, \prob(\thetac) = \frac{1}{2 \times 10^{-4}} \left ( \frac{\sin \theta_0}{4 \pi} \right) \left ( \frac{\sin \thetac}{\cos \thetac^{\rm min} - \cos \thetac^{\rm max}} \right ),
\end{equation}
where $0 \le \theta_0 \le \pi$, $0 \le\phi_0 < 2\pi$, and the extrema of the size distribution, $\{ \thetac^{\rm min}, \thetac^{\rm max} \}$ are chosen such that the bubble collisions are detectable.\footnote{This restriction to ``detectable'' signatures comes about as the overall number of collisions on the sky, $\nsavge$, is the primary parameter of interest, as opposed to, for example, the parameters describing any one putative signature. Once this approach is taken it is only possible to place data-driven constraints on the number of collisions with parameters that would make them plausibly detectable: it is impossible to conclude anything about the number of, say, small-$\thetac$ bubble collisions that would have been completely smeared out by the WMAP beam.} The lower limit, $\thetac^{\rm min} = 2^\circ$, stems from the fact that the CMB contains considerable power on the degree-scale, greatly increasing the difficulty of detection. The observable signature of a bubble collision with $\thetac$ larger than the upper limit, $\thetac^{\rm max} = 90^\circ$, would be indistinguishable from the signature of a collision of size $180^\circ - \thetac$.

\subsection{Cosmic textures}

For a detailed discussion of the production, evolution, and observational signature of cosmic textures we refer the reader to the original literature~\cite{Cruz:2007pe,Pen:1993nx,Spergel:1990ee,Turok:1989ai,Turok:1990gw,Turok:1991qq}. In brief, CMB photons passing through a collapsing or exploding texture will be red- or blue-shifted, producing an azimuthally symmetric feature on the CMB sky of angular size $\thetac$ with temperature profile of the form  
\begin{eqnarray}\label{eq-texture-template}
t(\theta, \phi) =  \frac{(-1)^{p} \, \epsilon}{\sqrt{1+4 \left( \frac{\theta}{\thetac} \right)^2 }}, \ \ \theta \leq \theta_*; \ \ \  t(\theta, \phi)=  \frac{(-1)^p \epsilon}{2} \exp \left( - \frac{(\theta^2 - \theta_*^2)}{2 \thetac^2} \right), \ \ \ \theta > \theta_*;
\end{eqnarray}
where the amplitude $\epsilon = 8 \pi^2 G \eta^2$ depends on the scale of symmetry breaking, $\eta$, $\thetac$ is the size of the texture, $\theta_* = \sqrt{3} \thetac /2$, $p=\{0,1\}$ and the template is centered at the Galactic north pole.

Assuming that all textures are produced in a single symmetry-breaking phase transition, the texture model has two global parameters: the dimensionless symmetry-breaking scale $\epsilon$ and the expected number of detectable sources $\nsavge$. Texture unwinding events produce features that all have the same amplitude on the CMB sky. The total expected number of unwinding events depends on the particular model giving rise to textures. Simulations~\cite{Cruz:2007pe} of SU(2) textures indicate that we can expect to have causal access to $\sim 7$ textures with $\thetac > 2^{\circ}$ in the CMB; the precise number of {\em detectable} unwinding events further depends on our particular realization of the background CMB, the dominant source of noise in the analysis. We adopt a uniform prior on $\nsavge$ between $0 \leq \bar{N}_s \leq 10$ to encode our ignorance of the precise theory and the effect of our CMB realization on detectability. Again, we shall see that this parameter is constrained by data, and thus the choice of upper limit has no effect on the analysis. Requiring that the symmetry-breaking scale for textures is below the scale of cosmological inflation (bounded to be lower than $\sim 10^{16} $ GeV by the absence of observed B-mode polarization), we place an upper bound on $\epsilon$ of $10^{-4}$. We assume a uniform prior on $\epsilon$ down to $2.5 \times10^{-5}$, which corresponds to the estimated detection limit of our pipeline (details of which can be found in Sec.~\ref{sec:source_detection}). Under the assumption of a uniform prior, the posterior is simply proportional to the likelihood; results for a different prior can be obtained easily by re-weighting the posterior.

The local parameters for textures are the size, $\thetac$, location, $\{ \theta_0, \phi_0 \}$ and $p$, which specifies whether the texture is hot or cold.  Textures are expected to be isotropically distributed over the sky, with a distribution of sizes proportional to $\thetac^{-3}$ (see Ref.~\cite{Cruz:2007pe} for a derivation), and so the normalized prior on the local texture parameters is
\begin{equation}
{\rm Pr}(\model_1) = \frac{\sin \theta_0}{4 \pi \thetac^3} \left( \frac{1}{{(2^\circ)}^2}  - \frac{1}{{(50^\circ)}^2} \right)^{-1} ,
\label{eq-local-priors}
\end{equation}
where $0 \le \theta_0 \le \pi$, $0 \le\phi_0 < 2\pi$, and we take $2^{\circ} \leq \thetac \leq 50^{\circ}$. The lower limit on $\thetac$ results from the large power on degree scales in the CMB; the upper limit stems from the fact that templates with $\thetac > 50^\circ$ are large enough to cover the whole sky and overlap themselves, rendering Eq.~\ref{eq-texture-template} invalid. To marginalize over the different signs, we can simply sum the likelihoods evaluated at the same magnitude of $\epsilon$, but where the template takes opposite sign.

We now describe our implementation of the formalism derived in Sec.~\ref{sec:formalism}. There are two main steps in the algorithm -- candidate-source detection and patch-based evidence calculation -- in which the non-zero regions of the likelihood function are first estimated and then evaluated. In the first step, optimal filters matched to the signals of interest are used to identity the positions and sizes of the most likely sources in the dataset. We describe this procedure in Sec.~\ref{sec:source_detection} below. In the second step, the patch-based evidence ratios, Eq.~\ref{equation:rho_int_over_local_params}, are calculated for each candidate using the nested sampler \multinest~\cite{Feroz_Hobson:2008,Feroz_Hobson_Bridges:2009}, before combining to form the posterior, Eq.~\ref{equation:posteriorfinalapp}, on the global model parameters. We describe this part of the calculation in Sec.~\ref{sec:patch_evidence}.


\section{Candidate detection with optimal filters}\label{sec:source_detection}

In order to effectively approximate the full posterior distribution describing the population of candidate sources, it is necessary to first locate the regions that provide significant contributions to the likelihood.  We follow the approach of Ref.~\cite{McEwen:2012uk}, using optimal filters defined on the sphere that are matched to the source profile of either bubble collision or texture signatures.  First, we construct optimal matched filters for the purpose of detecting candidate sources embedded in full-sky WMAP 7-year data and assess their performance.  Second, we briefly describe the optimal-filter-based algorithm for detecting sources of unknown and differing sizes, highlighting differences between the bubble collision and texture cases.  Third, we calibrate the algorithm on an end-to-end simulation of WMAP observations, before assessing its sensitivity.

\subsection{Optimal bubble collision and cosmic texture filters}

We construct two sets of matched filters: one set that enhances the contributions of bubble collision signatures and one set that enhances the contributions of texture signatures.  The matched filters are constructed to enhance the source profile in a specified stochastic background.  A stochastic background of CMB fluctuations is assumed, characterized by the lensed $\Lambda$CDM power spectrum that best fits the WMAP 7-year data, baryon acoustic oscillations and supernovae observations \cite{Larson_etal:2011} (hereafter referred to as the lensed WMAP7$+$BAO$+{\rm H_0}$ power spectrum).  The bubble collision and texture source profiles for which we search are relatively large-scale; thus we consider spherical harmonics up to the band-limit $\ell_{\rm max}=256$ only.  Since we eventually apply these filters to W-band WMAP observations, we assume observations are made in the presence of a Gaussian beam of full-width-half-maximum ${\rm FWHM}=13.2$ arcminutes and isotropic white noise of $N_\ell=0.02\ \mu{\rm K}^2$.  The optimal matched filters are constructed in harmonic space: thus the assumption of an azimuthally-symmetric beam profile and isotropic noise simplifies their construction and application considerably, while remaining highly accurate for the relatively low band-limit considered.  Once the source profile and stochastic background are defined, the filters are constructed on the sphere as outlined by Ref.~\cite{mcewen:2006:filters}.

\begin{figure*}
\centering
\subfigure[$\thetacrit=5\degrees$]{
  \includegraphics[height=45mm]
  {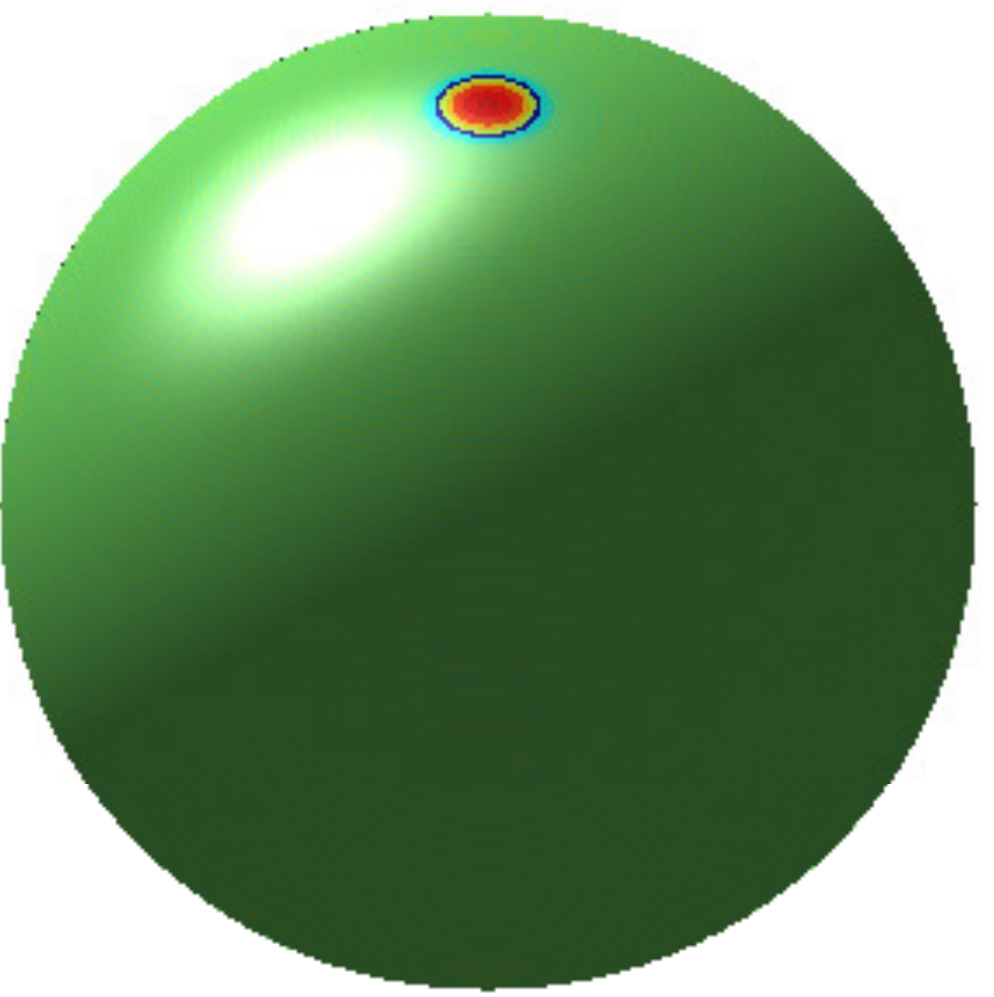}} \quad \quad
\subfigure[$\thetacrit=20\degrees$]{
  \includegraphics[height=45mm]
  {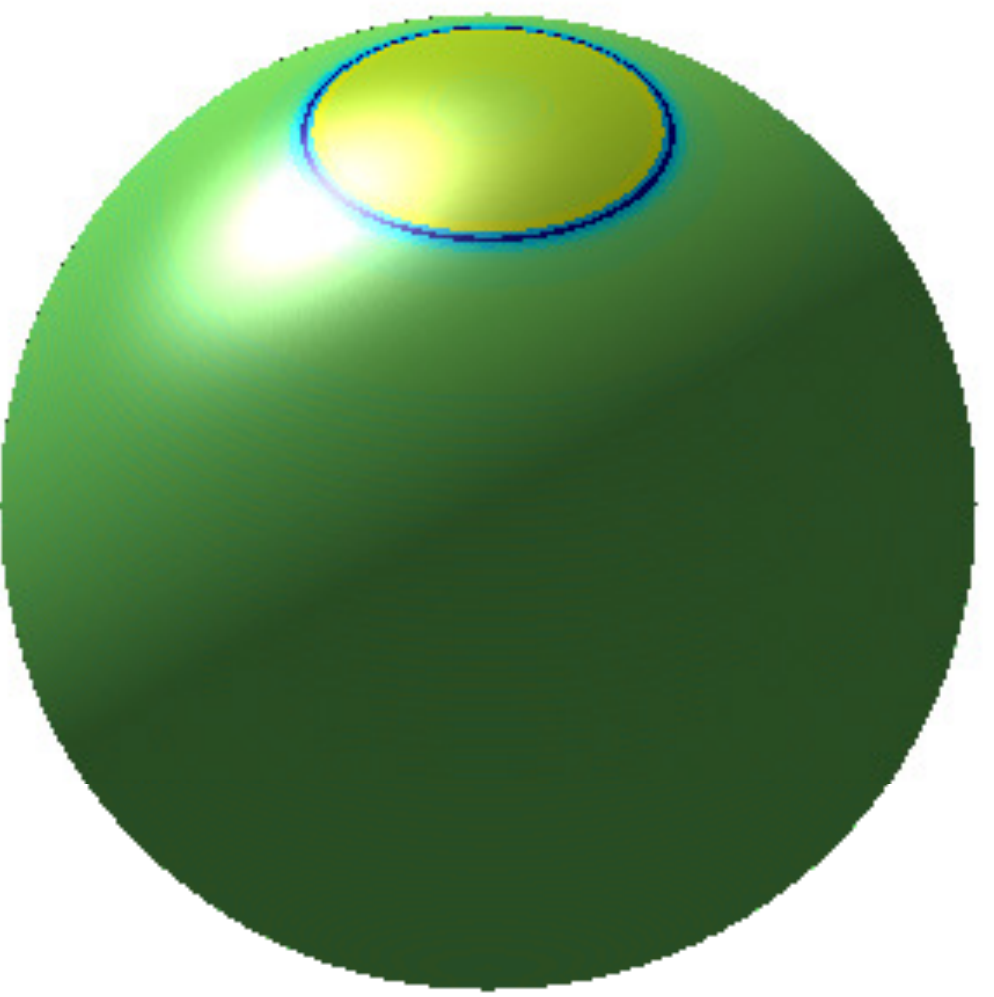}} \quad \quad
\subfigure[$\thetacrit=60\degrees$]{
  \includegraphics[height=45mm]
  {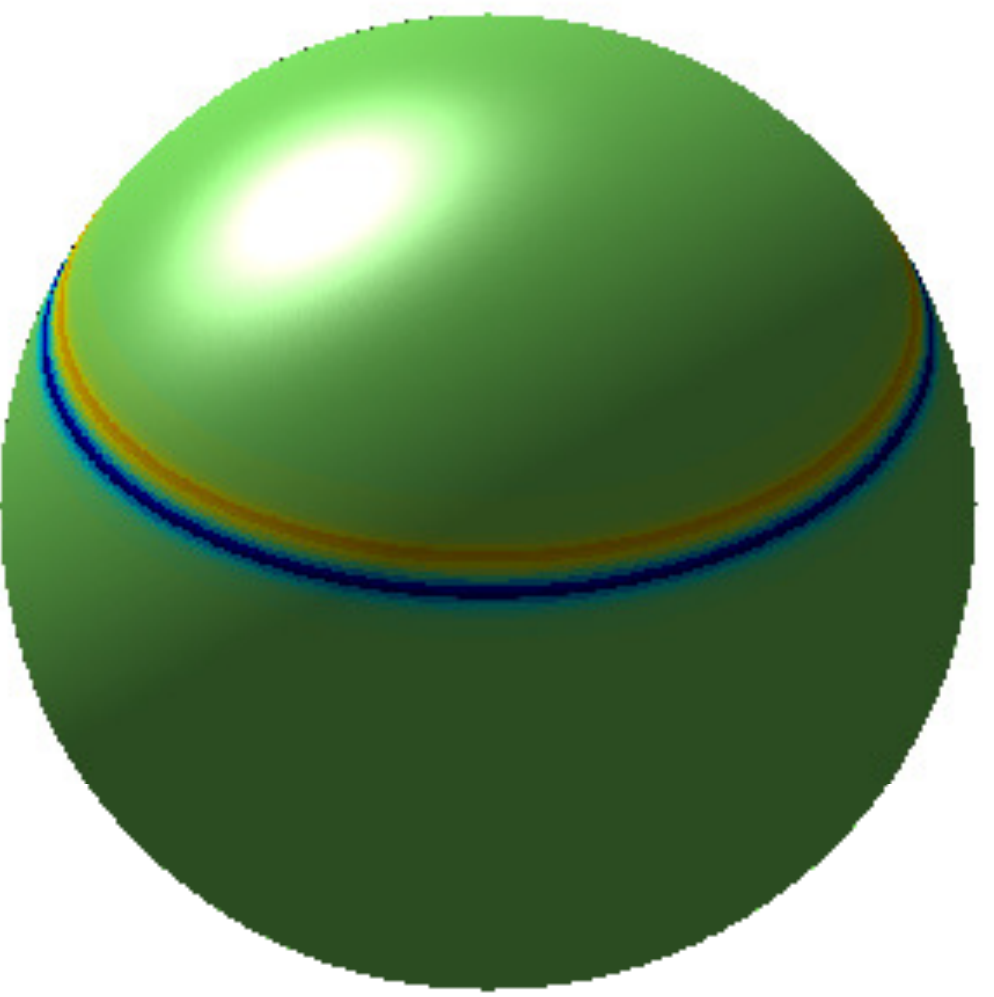}}
\caption{Matched filters optimized to bubble collision signatures of varying size embedded in a $\Lambda$CDM CMB background.}
\label{fig:mf_bubbles}
\end{figure*}

\begin{figure*}
\centering
\subfigure[$\thetacrit=5\degrees$]{
  \includegraphics[height=45mm]
  {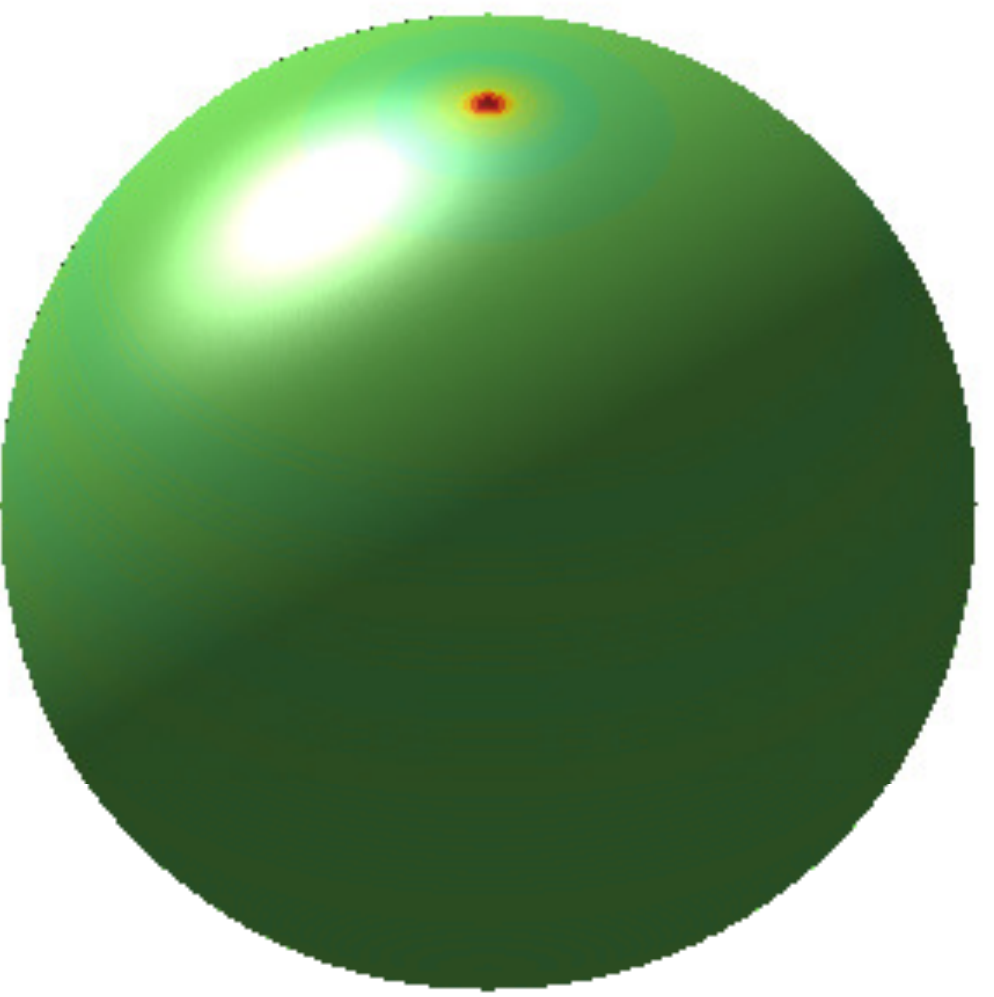}} \quad \quad
\subfigure[$\thetacrit=10\degrees$]{
  \includegraphics[height=45mm]
  {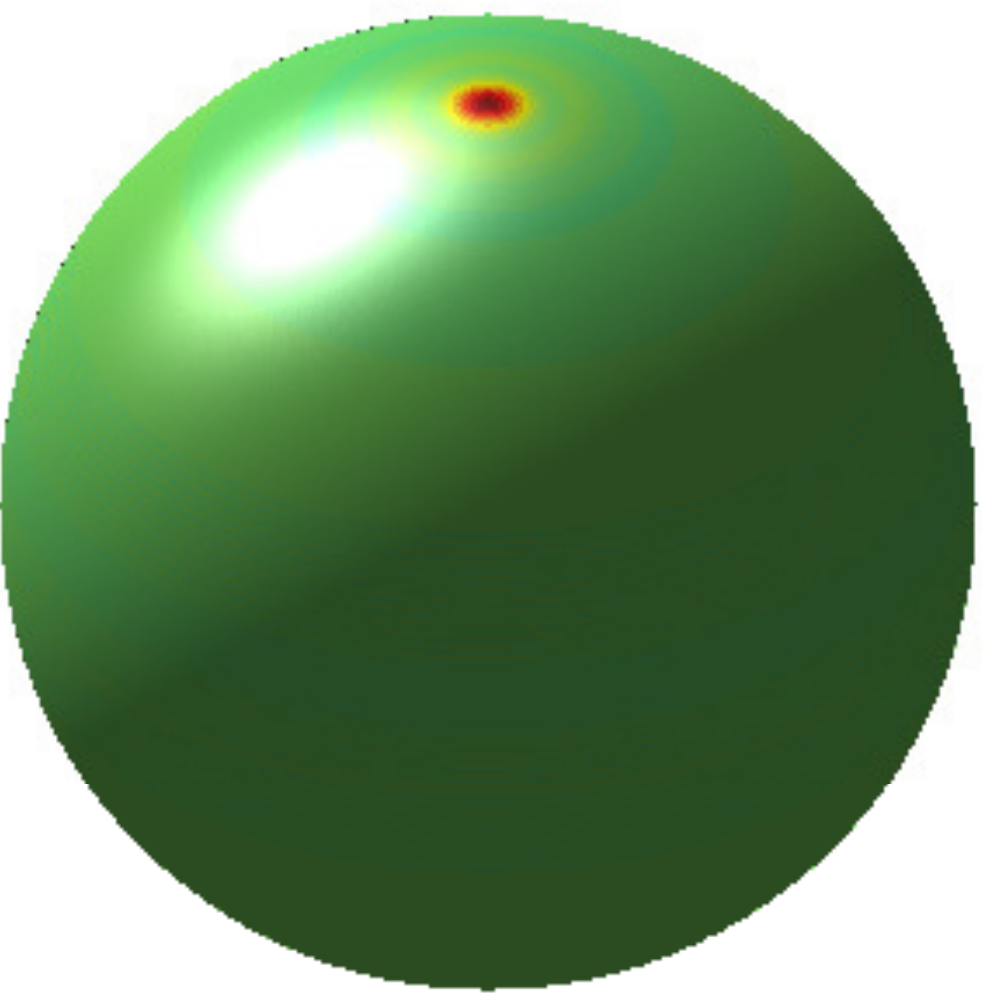}} \quad \quad
\subfigure[$\thetacrit=20\degrees$]{
  \includegraphics[height=45mm]
  {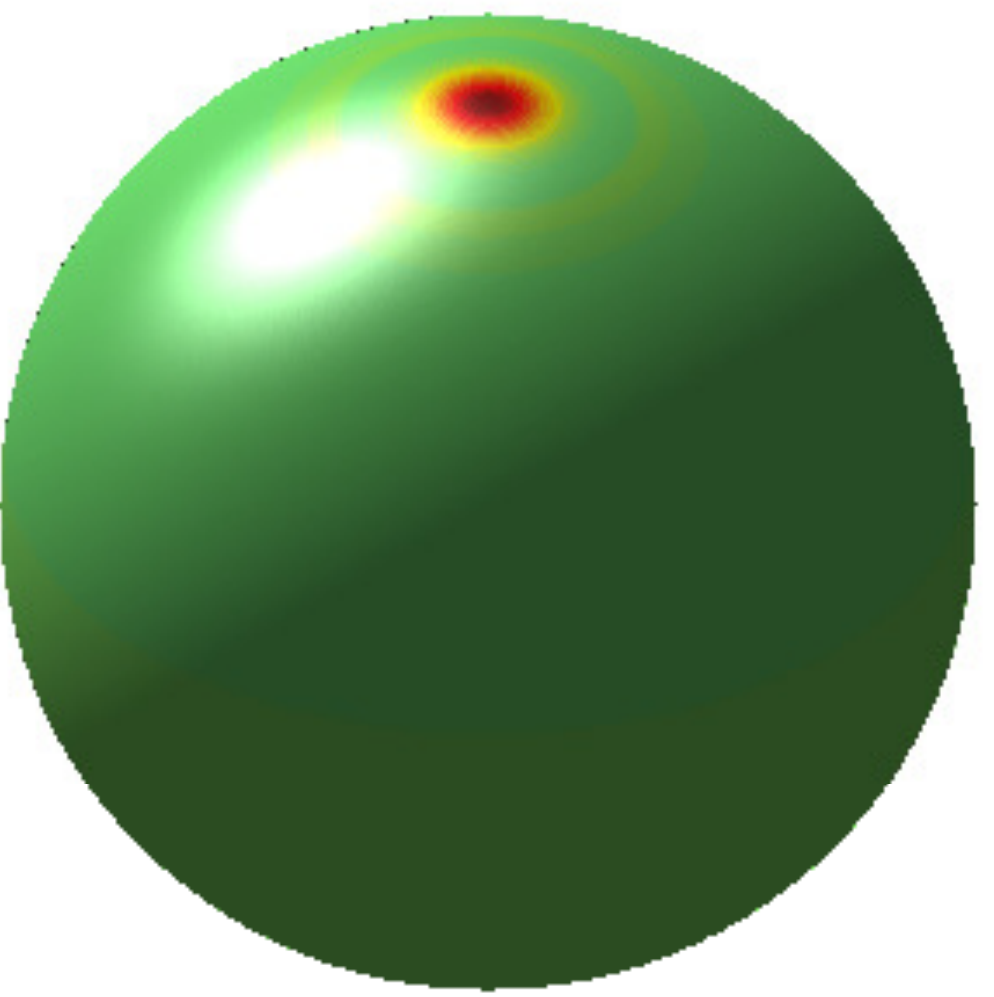}}
\caption{Matched filters optimized to cosmic texture signatures of varying size embedded in a $\Lambda$CDM CMB background.}
\label{fig:mf_textures}
\end{figure*}

In Fig.~\ref{fig:mf_bubbles} and Fig.~\ref{fig:mf_textures} we show the matched filters recovered for the bubble collision and texture profiles, respectively, for a range of source sizes.  Notice that the bubble collision filters on smaller scales contain a central broad hot region to enhance the main bubble collision contribution, surrounded by hot and cold rings to enhance the transition from the collision to the background. On larger scales, however, the matched filters contain only the hot and cold rings that enhance the transition. Since the CMB has more power on large scales, the matched filters on large scales do not respond to the large-scale features of the bubble collision signature but rather the transition region near the location where the template goes to zero. The texture source profile has a smooth, Gaussian transition to the background, and consequently the matched filters recovered for textures contain only a central broad region, without any strong contribution from the perimeter of the profile.

The matched filters constructed are optimal in the sense that no other filter can yield a greater enhancement of the signal-to-noise ratio (SNR) of the filtered field.  It is possible to calculate analytically the SNR of the filtered field for various filter types, as derived for example by Ref.~\cite{McEwen:2012uk}.  In Fig.~\ref{fig:snr} we plot the SNR computed for bubble collision and texture profiles for the unfiltered field, for the optimal filters constructed here, and for needlets~\cite{Marinucci:2007aj,Scodeller:2010mp}, which have been used previously to detect candidate sources \cite{Feeney_etal:2010jj,Feeney_etal:2010dd,Feeney_etal:2012jf}. Note that the lack of a sharp transition in the texture template means that the SNR for textures are lower than those of bubble collisions. Nevertheless, it is clear that matched filters yield the highest SNR, in accordance with expectations.

\subsection{Candidate object detection algorithm}

Although we have constructed optimal filters for a range of source sizes, we have not yet addressed the problem of detecting sources of unknown and differing sizes.  We adopt the algorithm described in detail by Ref.~\cite{McEwen:2012uk} for this purpose, which we review here briefly.  First, matched filters are constructed for a grid of source sizes \mbox{$\scale \in \{ \thetacrit^k \}_{k=1}^{N_{\thetacrit}}$}. All filters are then applied to the full-sky observed data by convolving the matched filter kernel with the observed data, which may be computed efficiently in harmonic space (see \eg\ Ref.~\cite{McEwen:2012uk}). Significance maps are then computed by normalizing the filtered field to the mean and standard deviation of filtered fields computed from realizations of the background process (i.e.\ CMB fluctuations and instrumental noise) in the absence of sources.  The significance maps are then thresholded (the calibration of threshold levels is discussed below), before potential candidate sources are found from the localized peaks of the thresholded significance maps.  Potential candidate sources are eliminated if a stronger source is found on adjacent scales, where the set of scales adjacent to scale \scale\ is defined by the set \mbox{$\{ \scale_{\rm adj } \in \{ \thetacrit^k \}_{k=1}^{N_{\thetacrit}}:$} \mbox{$| \scale_{\rm adj} - \scale | \leq \theta_{\rm adj} \}$}, i.e. where the distance between \scale\ and $\scale_{\rm adj}$ is less than the parameter $\theta_{\rm adj}$.  Once candidate sources are detected, the parameters of the source size, location and amplitude are estimated from the corresponding filter scale, peak position of the thresholded significance map and amplitude of the filtered field, respectively.

The construction of optimal filters is implemented in the \stwofil\ code \cite{mcewen:2006:filters} (which in turn relies on the codes \stwo\ \cite{mcewen:2006:fcswt} and \healpix\ \cite{Gorski:2004by}), while the \comb\ code  \cite{mcewen:2006:filters} has been used to simulate bubble collision signatures embedded in a CMB background.\footnote{\stwofil, \stwo\ and \comb\ are available from \url{http://www.jasonmcewen.org/}, while \healpix\ is available from \url{http://healpix.jpl.nasa.gov/}.}  The candidate object detection algorithm described here is implemented in a modified version of \stwofil\ that will soon be made publicly available.

\begin{figure*}
\centering
\subfigure[Bubble collision signatures of known source size]{
  \includegraphics[height=55mm]
  {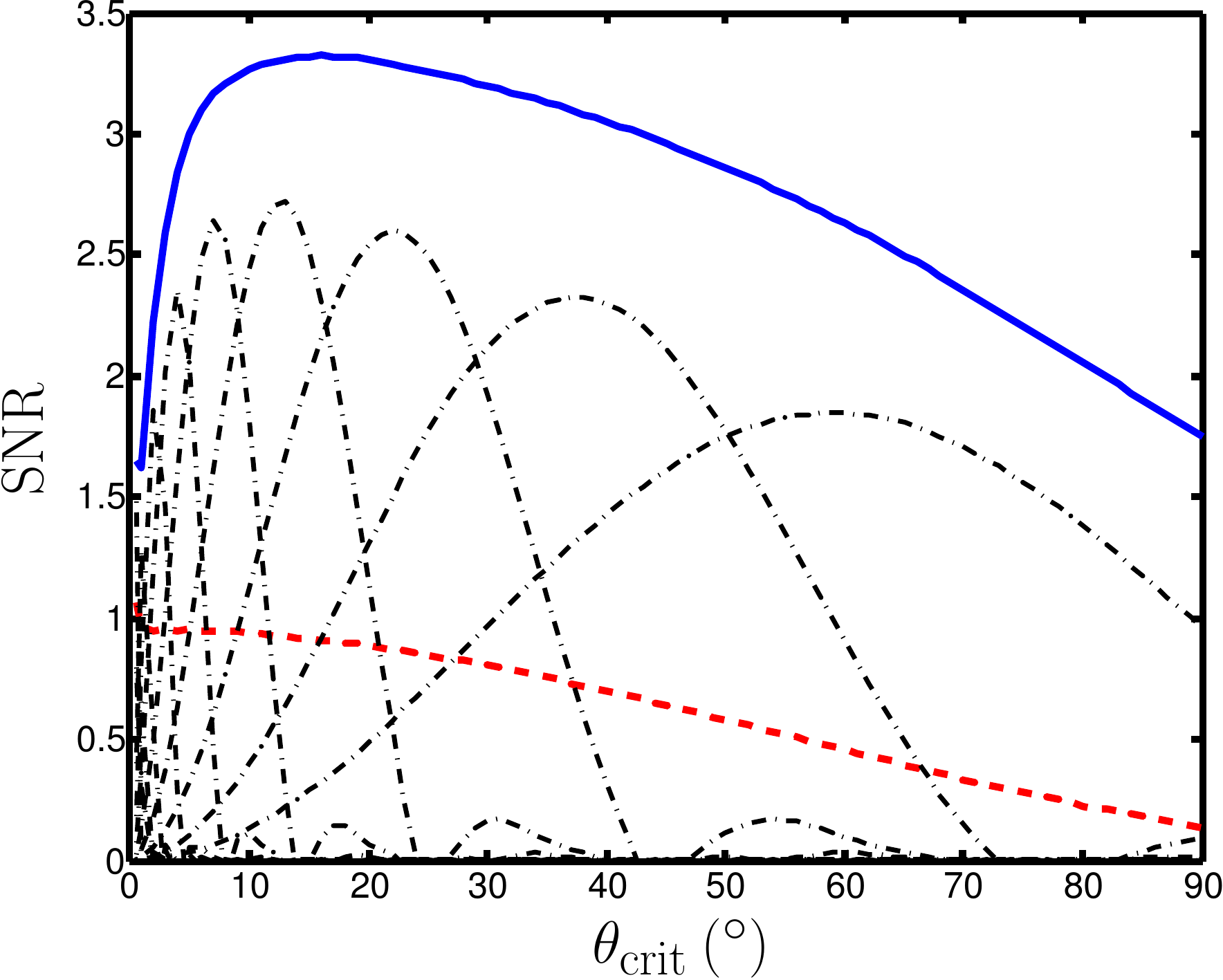}} \quad \quad
\subfigure[Bubble collision signatures of unknown source size]{
  \includegraphics[height=55mm]
  {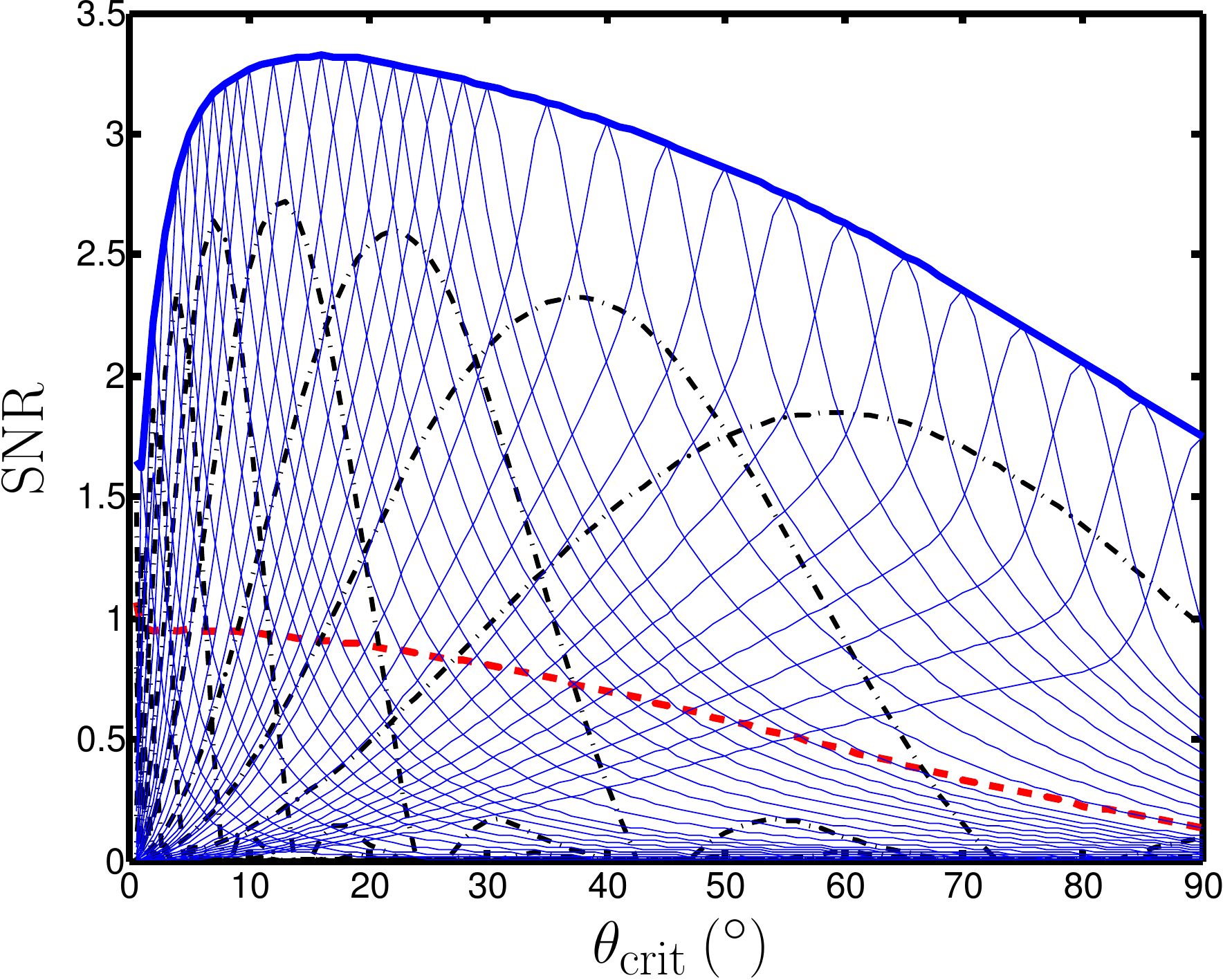}} \\
\subfigure[Cosmic texture signatures of known source size]{
  \includegraphics[height=55mm]
  {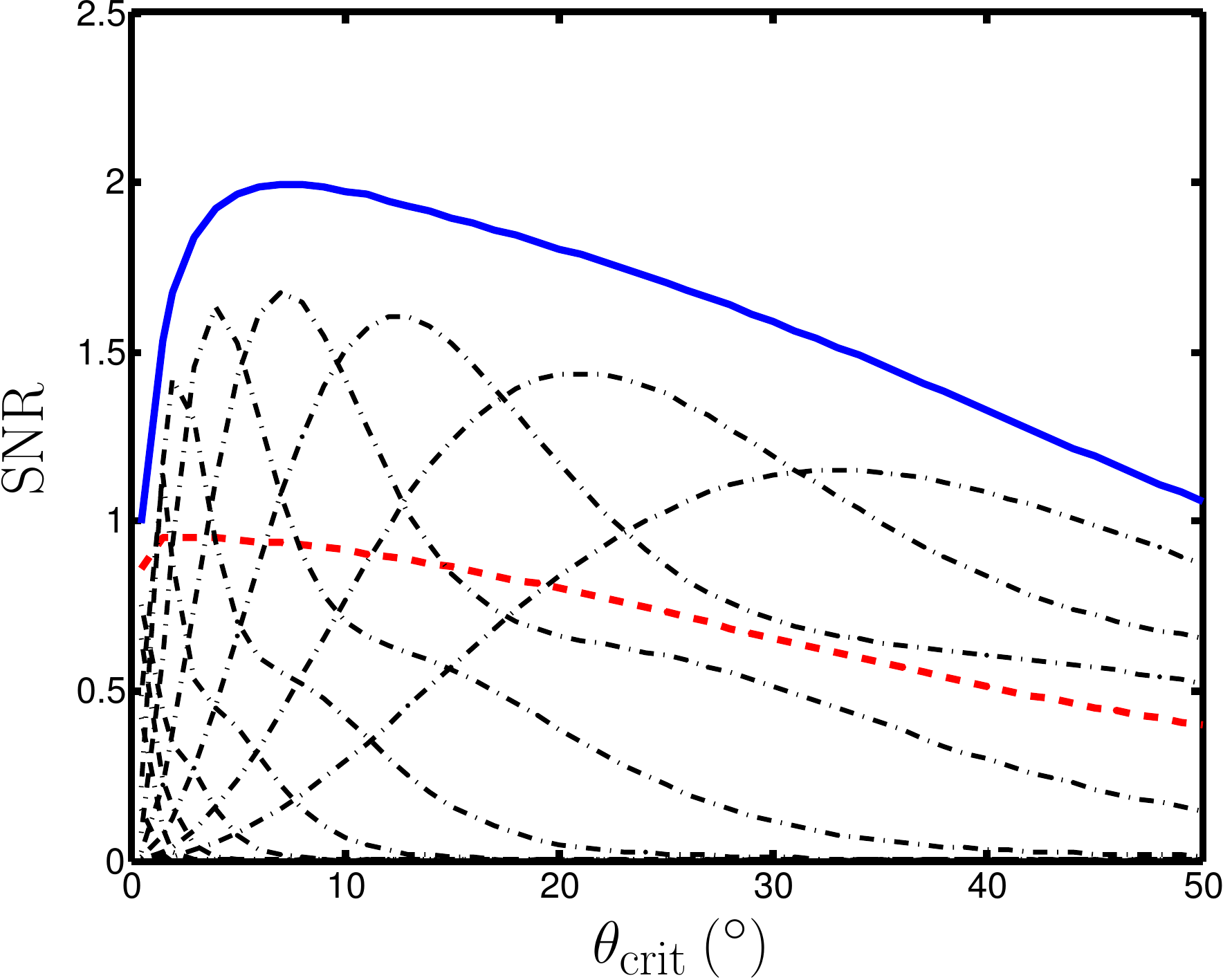}} \quad \quad
\subfigure[Cosmic textures of unknown source size]{
  \includegraphics[height=55mm]
  {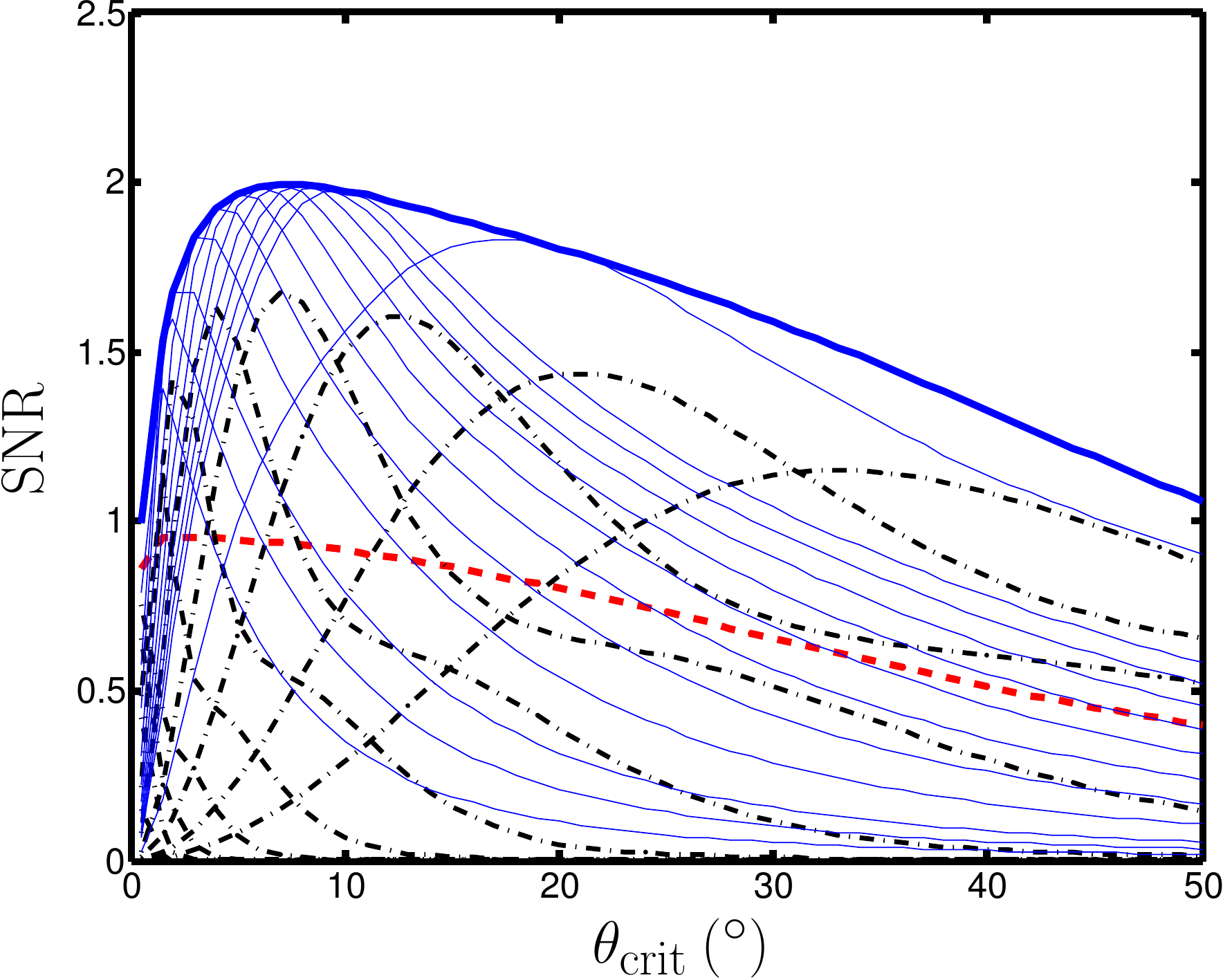}} 
\caption{SNRs of bubble collision (top row) and texture (bottom row) signatures of varying size with amplitude $100\ \mu$K embedded in a $\Lambda$CDM CMB background.  SNR curves are plotted for matched filters (solid blue curve), needlets~\citep{Marinucci:2007aj} with scaling parameter $B=1.8$ for a range of needlet scales $j$ (dot-dashed black curves) and for the unfiltered field (dashed red curve). In panels~(b) and (d) SNR curves for the matched filters constructed at a given scale and applied at all other scales are also shown (thin solid blue curves).  The scale for which the filters are constructed may be read off the plot from the intersection of the heavy and light solid blue curves.  Provided the \thetacrit\ grid is sampled sufficiently densely, the matched filters remain superior to needlets.}
\label{fig:snr}
\end{figure*}

There is no guarantee that the peak in the filtered field across scales will coincide with the size of the unknown source.  Nevertheless, for bubble collision signatures embedded in the CMB this has indeed been found to be the case \cite{McEwen:2012uk}.  For texture profiles, however, we have found this phenomenon to hold below scales $\thetacrit\sim10\degrees$ only.  Through numerical experiments we found that the difference in the behavior of the filtered field between bubble collisions and textures on large scales is due to the absence of a well-defined transition region from the source to the background in the texture profile.  For large texture sizes, not only is there no peak in the filtered field at the scale of the unknown source, but the SNR of the filtered field does not drop off rapidly when applied to nearby scales (Fig.~\ref{fig:snr}), as is the case for bubble collisions.  We trivially modify the candidate detection algorithm described above to account for this behavior.  For textures sizes below $\thetacrit=10\degrees$ we look across adjacent scales as usual to find the most significant potential candidate source, whereas for sizes above $\thetacrit=10\degrees$ we do not (by a judicious choice of the adjacency parameter $\theta_{\rm adj }$ there is in fact no need to modify the algorithm, as described below). 

Although the candidate detection algorithm considers a grid of candidate scales $\scale$, it is overwhelmingly probable that the signal for any given source peaks at scales between the samples of the grid.  It is thus important to examine how sensitive the matched filters are to small errors in the source size.  In Fig.~\ref{fig:snr}~(b) and (d) we plot SNR curves for matched filters constructed on the grid of candidate scales.  For bubble collision profiles, a sharp degradation in the SNR away from the scale used to construct each filter is clearly apparent.  For texture profiles the degradation is much less pronounced, especially at large scales (as discussed above).  Provided that the \thetacrit\ grid is sampled sufficiently densely, the matched filters for both bubble collision and texture profiles remain effective and are superior to needlets.

\subsection{Candidate object detection calibrated to WMAP}

We define the parameters of the optimal-filter-based candidate object detection algorithm here and calibrate the threshold levels for WMAP observations.  Throughout the calibration we apply the WMAP KQ85{\em yr7} mask \cite{Gold:2010fm}, since this is the mask adopted when analysing WMAP data subsequently.  We select the less-conservative KQ85 mask so as to reduce the variance in reconstructing large-scale information masked by the Galactic sky cut~\cite{Feeney_Pontzen_Peiris:2011}. This choice has the additional advantage of revealing more of the sky and hence increasing the number of candidates. The Bayesian stage of the pipeline will assess the overall evidence for the source model in each candidate region: if any candidate is found to contribute evidence in favor of the source model, then it will be examined closely for frequency-dependence signifying potential foreground contamination.

For bubble collisions we consider the grid of scales set out in Ref.~\cite{McEwen:2012uk}, as defined in Table~\ref{tab:mf_thresholds} (left).  The $\thetac$ prior range is smaller for textures than for bubble collisions -- the texture profile extends well beyond $\thetac$, covering the full sky for $\thetac \gtrsim 50^\circ$ -- and hence for textures we consider a smaller grid of scales, also defined in Table~\ref{tab:mf_thresholds} (right). Since we found the matched filters for textures to be sensitive to a large range of nearby scales, we nevertheless remain sensitive to the full prior range of sizes.  The SNR curves for the matched filters constructed for these scales for bubble collisions and textures are shown in Fig.~\ref{fig:snr}~(b) and (d).  These grids of scales are thus sufficiently sampled to ensure that the matched filters remain effective for scales between the samples of the grid.  We set the adjacency parameter to $\theta_{\rm adj} = 5\degrees$ for both bubble collisions and textures.  For textures this ensures that we look across scales for sizes below $\thetacrit=10\degrees$ but not above, whereas for bubble collisions we always look across adjacent scales.

\begin{table*}
\begin{tabular}[t]{c c}
\hline
\hline
\  $\thetac ({}\degrees)$ \ & \  $N_{\sigma_\scale}$ \ \\
\hline
1 & 4.25 \\
1.5 & 4.25 \\
2 & 4.25 \\
3 & 4.00 \\
4 & 4.00 \\
5 & 4.00 \\
6 & 4.00 \\
7 & 4.00 \\
8 & 4.00 \\
9 & 3.75 \\
10 & 3.75 \\
12 & 3.75 \\
14 & 3.75 \\
16 & 3.75 \\
18 & 3.50 \\
20 & 3.50 \\
22 & 3.50 \\
24 & 3.50 \\
26 & 3.50 \\
28 & 3.25 \\
30 & 3.25 \\
35 & 3.25 \\
40 & 3.25 \\
45 & 3.25 \\
50 & 3.25 \\
55 & 3.25 \\
60 & 3.25 \\
65 & 3.25 \\
70 & 3.25 \\
75 & 3.25 \\
80 & 3.00 \\
85 & 3.00 \\
90 & 3.00 \\
\hline
\hline
\end{tabular}
\
\
\
\
\
\begin{tabular}[t]{c c}
\hline
\hline
\  $\thetac ({}\degrees)$ \ & \  $N_{\sigma_\scale}$ \ \\
\hline
1 & 4.25 \\
1.5 & 4.00 \\
2 & 3.75 \\
3 & 3.50 \\
4 & 3.50 \\
5 & 3.25 \\
6 & 3.00 \\
7 & 3.00 \\
8 & 3.00 \\
9 & 2.75 \\
10 & 2.75 \\
20 & 2.50 \\
\hline
\hline
\end{tabular}
 \begin{center}
 \caption{The $\thetac$ grid and threshold levels $N_{\sigma_\scale}$ adopted for the optimal-filter-based candidate source detection algorithm for bubble collisions (left) and textures (right).  Threshold levels are calibrated to the WMAP end-to-end simulation to allow at most three false detections on each scale.
   \label{tab:mf_thresholds}}
 \end{center}
\end{table*}

We use 3,000 Gaussian CMB simulations to calculate the mean and standard deviation of the filtered field at each scale in the absence of sources, in order to compute significance maps. For these simulations, and for the WMAP data analysed subsequently, we perform Wiener filtering to recover spherical harmonic coefficients with $\ell \le 10$ from masked CMB maps \cite{Feeney_Pontzen_Peiris:2011}, where we adopt a Gaussian prior for the harmonic coefficients specified by the lensed WMAP7$+$BAO$+{\rm H_0}$ power spectrum.  Note that this differs from the maximum likelihood reconstruction~\cite{deOliveiraCosta:2006zj} of harmonic coefficients performed by Refs.~\cite{Feeney_etal:2010jj,Feeney_etal:2010dd,Feeney_etal:2012jf} and the cut-sky estimation performed by Ref.~\cite{McEwen:2012uk}. This can alter the spherical harmonic coefficients recovered on large scales, and thus the detected candidate sources, in a non-negligible manner.  However, Wiener filtering should give the most reliable reconstruction of the large-scale harmonic modes \cite{Feeney_Pontzen_Peiris:2011}.  

Finally, we calibrate the threshold levels $N_{\sigma_\scale}$ applied to the significance maps for each filter scale from a realistic WMAP simulation that does not contain embedded sources.  The thresholds are chosen to allow a manageable number of false detections while remaining sensitive to weak sources.  For this calibration we use a complete end-to-end simulation of the WMAP experiment provided by the WMAP Science Team~\cite{Gold:2010fm}.
The temperature maps in this simulation are produced from a simulated time-ordered data stream, which is processed using the same algorithm as the actual data. The data for each frequency band are obtained separately using simulated diffuse Galactic foregrounds and CMB fluctuations, and include realistic noise, smearing from finite integration time, finite beam size, and other instrumental effects. We use the foreground-reduced W-band simulation for calibration.  
The threshold levels $N_{\sigma_\scale}$ are selected to allow at most three false detections on each scale on this simulated map (recall that detections on one scale can be eliminated by stronger detections made on adjacent scales). These threshold levels are less conservative that those set by Refs.~\cite{Feeney_etal:2010jj,Feeney_etal:2010dd,Feeney_etal:2012jf,McEwen:2012uk}, in order to increase the number of false detections passed by the candidate source detection stage of the analysis pipeline and hence improve its sensitivity.  The calibrated threshold levels for both bubble collisions and textures are shown in Table~\ref{tab:mf_thresholds}. Once candidate sources are detected by the optimal-filter-based detection algorithm, we discard those objects that are significantly masked.  For the WMAP end-to-end simulation, we make 12 false detections of candidate bubble collisions and 4 false detections of candidate textures.

In the analysis of the WMAP end-to-end simulation (and in the analysis of the WMAP data considered subsequently), some of the bubble collision candidates that we detect differ from those found with optimal filters previously \citep{McEwen:2012uk}.  This is expected since we now use Wiener filtering to recover spherical harmonic coefficients, have included a more accurate model of the WMAP noise in the optimal filter construction (noise was neglected previously), and have reduced the threshold levels in order to increase the sensitivity of the entire pipeline.  Further, the thresholding-based nature of the candidate source detection algorithm means that small differences in the filtered field can have an impact of the final candidates detected if regions move below or above the threshold.  In the cases where candidates disappeared, we nevertheless found peaks in the filtered field; these were simply no longer above the threshold or were eliminated by stronger detections on nearby scales or positions.  Given these differences are due to improvements made to the pipeline, the results given here are to be preferred to those presented previously.

Following the candidate source-detection stage of the analysis pipeline, we pass to the Bayesian stage an estimate of the domain of parameter space over which the likelihood is expected to be non-negligible.  These regions of parameter space are estimated from each of the candidate sources detected. For the size of each source, the relevant region is determined first by finding the range of nearby filter scales for which significance maps exceed their threshold level at the source position. This range of scales is extended to the next smallest and largest filter scales (or the edge of the prior if encountered) to yield an estimate of the range of scales over which the likelihood is non-negligible. For example, for a bubble collision candidate found to be significant by the $40^\circ$ and $45^\circ$ optimal filters, we would estimate the likelihood to be non-negligible over the range $35$ -- $50^\circ$.
To estimate the integration limits of the central positions for each source, we first find all pixels within a radius $r$ of the source position estimated by the optimal filters, where $r$ is 25\% of the maximum source size estimated from the previous step.
The extrema $\{ \theta_{0}^{\rm min}, \theta_{0}^{\rm max}, \phi_{0}^{\rm min}, \phi_{0}^{\rm max} \}$ of these pixels are found, and the source positions are then sampled from the region defined by $\theta_{0}^{\rm min} \le \theta_{0} \le\theta_{0}^{\rm max}$, $\phi_{0}^{\rm min} \le \phi_{0} \le \phi_{0}^{\rm max}$.
Tests of these assumptions are included in the suite of pipeline tests detailed in later sections. 

\begin{figure*}
\centering
\subfigure[Bubble collisions]{
  \includegraphics[height=55mm]
  {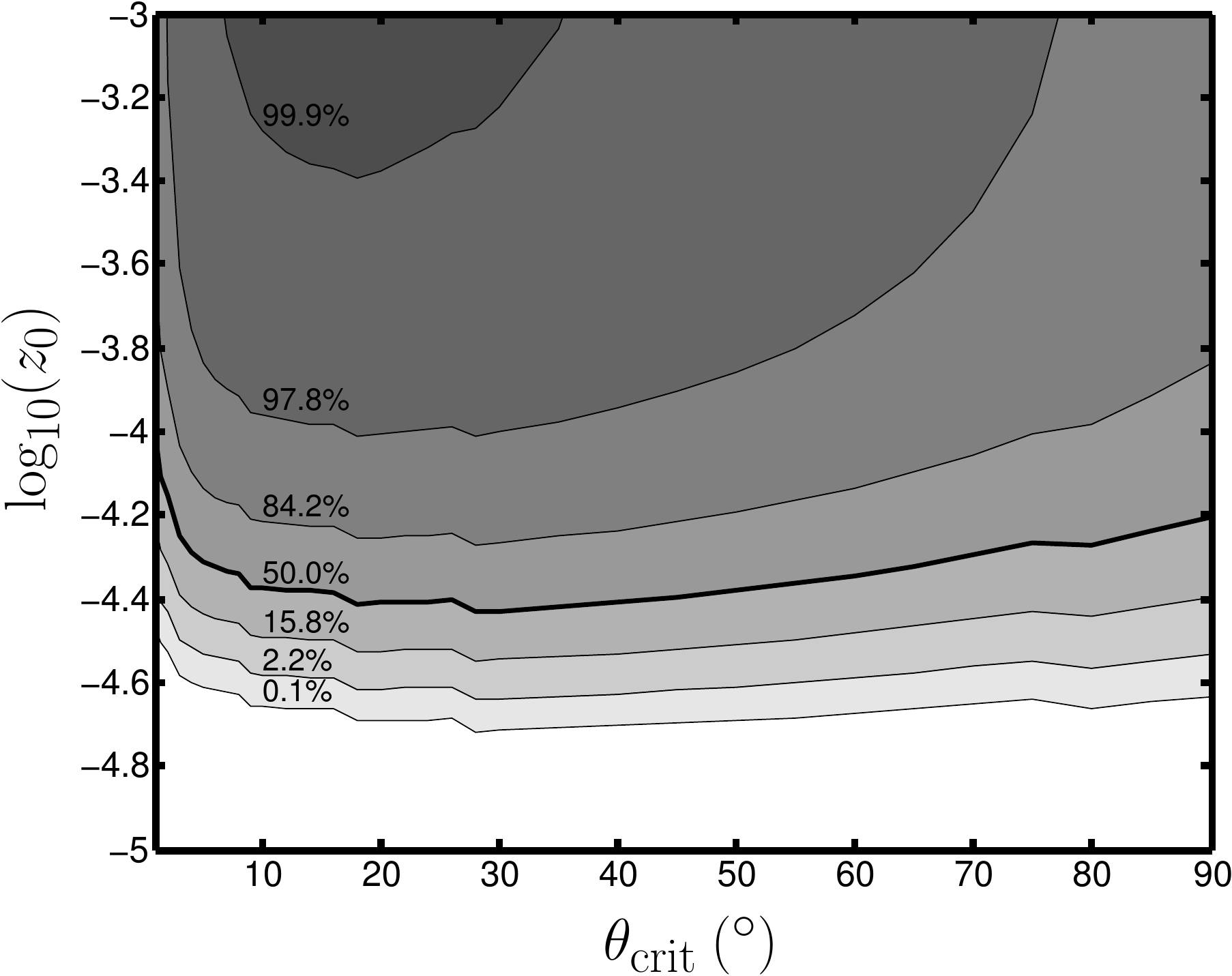}} \quad \quad
\subfigure[Cosmic textures]{
  \includegraphics[height=55mm]
  {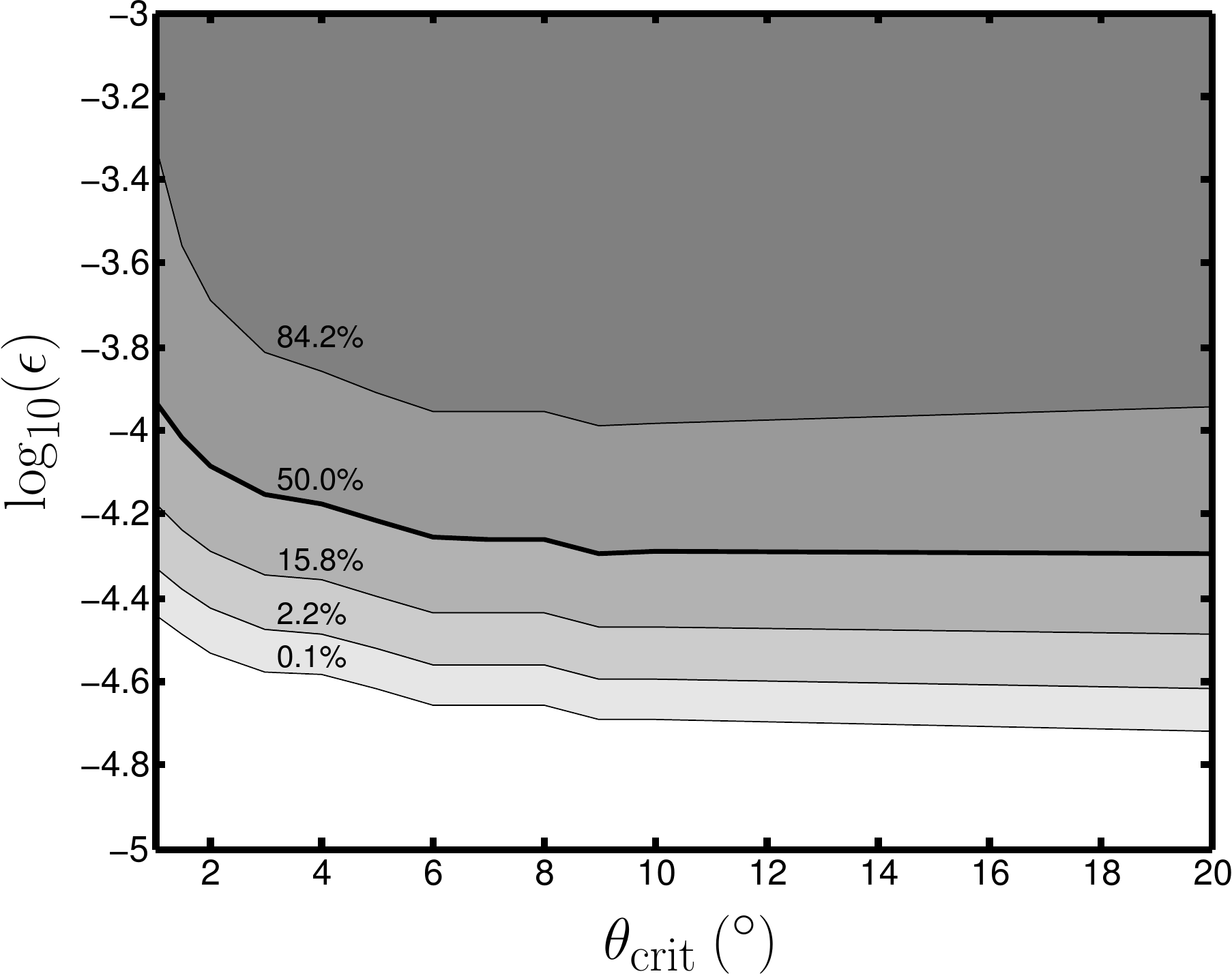}} 
\caption{Sensitivity of the optimal-filter-based candidate detection algorithm, with completeness curves plotted for one, two and three standard deviations from the 50\% completeness curve.  The completeness curves are computed in the following manner.  For each scale $\thetac$ we compute the source amplitude ($z_0$ for bubbles; $\epsilon$ for textures) that would be required to ensure that the SNR reaches the threshold specified in Table~\ref{tab:mf_thresholds}.  This level defines the 50\% completeness curve, since we expect half of the sources with this amplitude to fall below the curve and half to fall above.  Similarly, we compute approximate completeness curves for one-, two- and three-standard-deviation differences from the 50\% completeness curve.  The probabilities quoted on each completeness curve are computed assuming a Gaussian distribution of the filtered field at the source position.}
\label{fig:mf_sensitivity}
\end{figure*}

We conclude this section by assessing the level to which the optimal-filter-based candidate detection algorithm is sensitive for each source type.  In previous studies, simulations were performed for this purpose \citep{Feeney_etal:2010jj,Feeney_etal:2010dd,McEwen:2012uk}.  Here we instead take a probabilistic approach based on the analytic SNRs of the filters computed previously (see Fig.~\ref{fig:snr}).  This allows us to probe the source size-amplitude parameter space at higher resolution and accuracy than would be achievable with modest numbers of simulations (to reach an equivalent resolution and accuracy through simulations would be extremely computationally demanding).  In Fig.~\ref{fig:mf_sensitivity} we plot the sensitivity of the matched filters constructed for bubble collisions and cosmic textures.  These plots are produced as follows.  For each scale $\thetac$ we compute the source amplitude ($z_0$ for bubbles; $\epsilon$ for textures) that would be required to ensure that the SNR reaches the threshold specified in Table~\ref{tab:mf_thresholds}.  This level defines the 50\% completeness curves shown in Fig.~\ref{fig:mf_sensitivity} since, in the presence of noise, we expect half of the sources with this amplitude to be detected and half to be missed.  Similarly, we compute approximate completeness curves for one-, two- and three-standard-deviation differences from the 50\% completeness curve (note that the probabilities quoted on each curve are computed assuming a Gaussian distribution of the filtered field at the source position).  For the 50\% completeness curve, the bubble collision matched filters are sensitive to $z_0 \sim 10^{-4.4}$, while the cosmic texture matched filters are sensitive to $\epsilon \sim 10^{-4.2}$.  Note that the sensitivities computed in this manner are similar to those computed previously through simulations \citep{Feeney_etal:2010jj,Feeney_etal:2010dd,McEwen:2012uk}, both in terms of the sensitivity levels obtained and the shape of the sensitivity regions.  Further, we see that optimal matched filters are $\sim1.7$ times more sensitive than needlets for detecting bubble collision signatures, as found previously \citep{Feeney_etal:2010jj,Feeney_etal:2010dd,McEwen:2012uk}.

\section{Adaptive-resolution evidence calculation}\label{sec:patch_evidence}

\begin{figure}
\includegraphics[width=14.0cm]{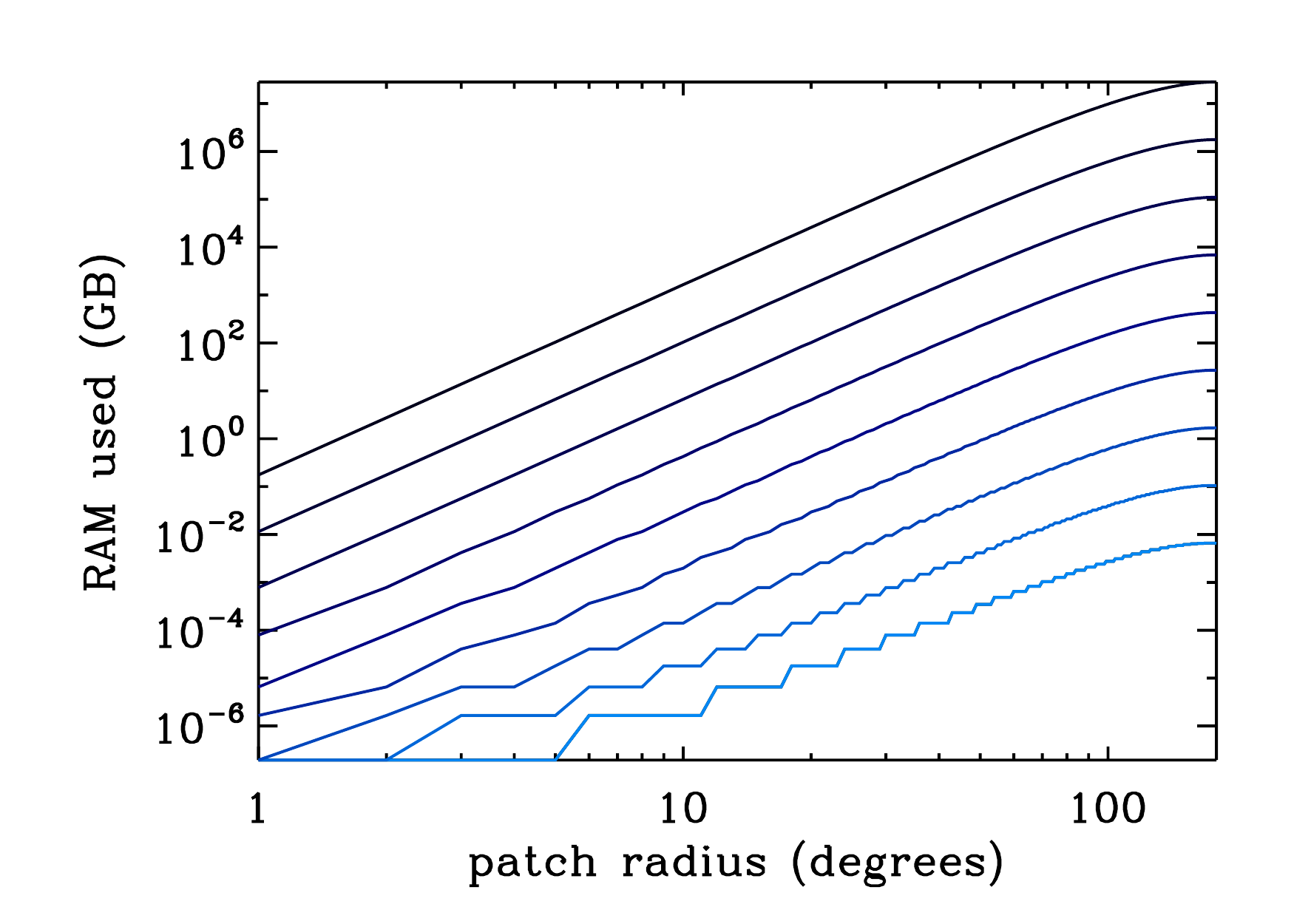}
\caption{The memory needed to calculate the covariance matrix for patches of a given size at \healpix\ resolutions ranging from $\nside = 8$ (lightest-blue, lowest curve) to $\nside = 2048$ (highest, darkest curve). The effects of pixel size are visible at small patch radii and low resolutions.
  \label{fig-patch_memory}
}
\end{figure}

Modern CMB experiments map the sky with extremely high resolution: the beam of the Planck experiment in the main CMB bands is expected to be $\sim5^\prime$~\cite{Planck_Ade:2011}, resulting in maps pixelated on the arcminute scale. While this is necessary for pinning down the secondary CMB anisotropies at small scales, it means that calculating pixel-space covariance matrices  becomes extremely memory-intensive. We illustrate this point in Fig.~\ref{fig-patch_memory}, which shows the memory needed to calculate covariance matrices from $1^\circ$ to $180^\circ$ in radius at \healpix\ resolutions ranging from $\nside = 8$ to $\nside = 2048$ (i.e., Planck resolution).\footnote{The quantity plotted corresponds to a total number of matrix elements equal to $\sim 1.5 \times \npixsq$. Our algorithm calculates the full $\npixsq$ covariance matrix in order to make use of the \lapack\ inversion routines~\cite{lapack}, then compresses the inverted matrix to a 1-D array containing its upper triangle to reduce memory costs while sampling.} It is clear that the memory costs, which to a good approximation rise as angular radius to the fourth power, make processing even relatively small patches prohibitive at full Planck resolution.

In previous work~\cite{Feeney_etal:2010jj,Feeney_etal:2010dd}, we chose to truncate both our patches and the integration limits of $\thetac$ to the maximum radius invertible with our memory constraints. While this allowed us to at least partially process almost all features at full WMAP resolution, it meant that we were unable to probe the large-$\thetac$ region of parameter space for which the prior for bubble collisions is highest. If we are to do so, it is clear that the input maps must be processed at degraded resolution: the larger the patch, the lower the resolution at which it can be processed. The maximum memory accessible per core in this analysis is $\sim 90$ GB, which means that the full-sky covariance matrix can be inverted at $\nside = 64$: this is therefore the minimum resolution at which any feature will be processed. The degradation process will now be described in detail, along with the suite of tests performed to assess its performance.

\subsection{Processing maps}

The candidate detection stage returns estimates for the size and position of features of interest within a map, defining the set of patches to be considered in the evidence calculation. The memory cost of computing each covariance matrix is derived from the number of unmasked pixels within the patch: if this is greater than the memory available, the patch must be processed at a degraded resolution.

In the \healpix\ pixelization scheme, each step down in resolution reduces the pixel count, $\npix$,  by a factor of four. As the covariance matrices are $\npixsq$ in size, an estimate of the number of steps down required is therefore given by
\begin{equation}
\ndeg = \mathrm{ceiling} \left (\frac{\log{(m_{\rm est} / m_{\rm max})}}{\log{16}} \right),
\end{equation}
where $m_{\rm est}$ and $m_{\rm max}$ are the estimate of the memory required and the memory available. The corresponding estimate of the finest resolution at which the patch can be processed is then
\begin{equation}
\nsidedeg = \frac{\nsidefull}{2 ^ \ndeg}.
\end{equation}

This estimate is tested by re-counting the number of unmasked pixels in the patch at the new resolution. An $\nsidedeg$ mask is created by averaging within degraded-resolution pixels: any $\nsidedeg$ pixel which is more than half masked at the input resolution is considered to be masked. A precise calculation of the memory cost of the degraded covariance matrix is then made: if this is below the memory threshold, as expected, the degraded resolution is accepted and the algorithm proceeds; if not, the resolution is decreased once more.

Once the required resolution has been determined, the input CMB temperature map can be degraded. As the CMB is a smooth field, it is not sufficient to simply average within $\nsidedeg$ pixels: doing so will introduce large pixelization effects, which will act as an extra noise term unaccounted for in the pixel-pixel covariance matrix. This can be avoided by smoothing such that the input map is smooth on the degraded-pixel scale prior to reducing the resolution. This is equivalent to introducing a band-limit, $\lmax$, in harmonic space. Choosing the band-limit is a balance. If the smoothing scale is set too large (i.e., $\lmax$ is too low), too much information will be discarded with each degrade and performance will suffer. If the smoothing scale is set too small (i.e., smaller than an $\nsidedeg$ pixel), the smoothed maps will contain pixelization artefacts.

The choice of $\lmax$ is somewhat arbitrary, but experimentation shows that the degradation is stable if the harmonic-space Gaussian smoothing kernel is 1\% of its maximum at $\lmax = 2 \nsidedeg$. This defines a smoothing scale at each resolution: the FWHM, $\degfwhm$, of the pixel-space kernel is given by
\begin{equation}
\degfwhm = \sqrt{ \frac{ 8 \, \log{2} \, \log{100} }{ \lmax (\lmax + 1) } }.
\end{equation}
Assuming the $12\nsidesq$ pixels in a \healpix\ map of given resolution are flat and square (a safe assumption at high-resolution), the smoothing scale is approximately 2.5 times the size of a pixel, and the maps are clearly smooth on the pixel-scale. The smoothing kernel sizes used for the three degraded resolutions considered in this work are listed in Table~\ref{tab:smoothing_scales}. Note that for speed and simplicity the smoothing is carried out on the full-sky, rather than within a patch: the time taken to smooth in pixel-space scales poorly with patch size.

\begin{table*}
\begin{tabular}{c c c c}
\hline
\hline
$\nsidedeg$ & \ $\npix$ \ & \ Pixel Scale (arcmin)\ & \ Smoothing FWHM \\
\hline
256 & 786432 & 13.7 & 33.9 \\
128 & 196608 & 27.5 & 67.7 \\
64 & 49152 & 55.0 & 135.2 \\
\hline
\hline
 \end{tabular}
 \begin{center}
 \caption{The full-widths at half-maximum of the Gaussian kernels used to smooth input maps prior to degradation. Also tabulated are the pixel count at each resolution, and an approximate pixel scale, derived assuming each pixel is square. 
   \label{tab:smoothing_scales}}
 \end{center}
\end{table*}

\subsection{Calculating the degraded evidence}

Care must be taken when calculating the covariance matrix for use in the likelihood: the covariance matrix must include as faithful as possible a representation of the components of the data. We must therefore capture every important feature of both the CMB ``signal'' (the CMB is, in fact, the dominant noise in the analysis) and instrumental noise measured in the input map, as well as the effects of the degradation process. It is helpful to break up the full covariance matrix, $\matr{C}$, into its CMB, $\matr{S}$, and noise, $\matr{N}$, constituents, as they are morphologically different.

\subsubsection{CMB covariance}

\begin{figure}
\includegraphics[width=8.5cm]{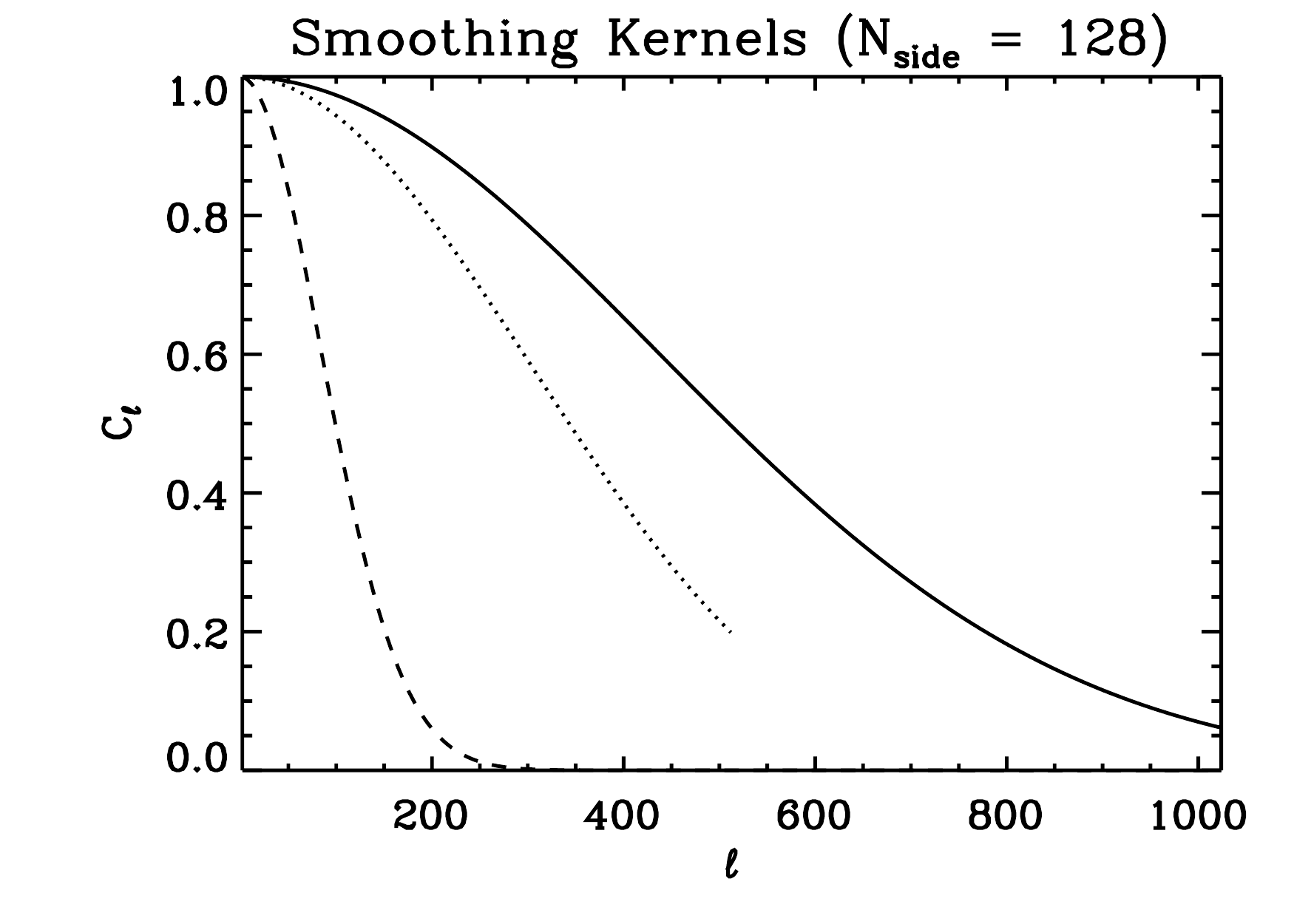}
\caption{Window functions of the various smoothing kernels appearing in the adaptive-resolution analysis. Shown are the WMAP7 W-band beam (approximated as a Gaussian, solid line), the Gaussian smoothing applied before degradation to $\nsidedeg = 128$ (dashed), and the pixel window function at this resolution (dotted). Note that the pixel window function is only defined up to $\ell = 4 \nsidedeg$, the maximum multipole allowed by the \healpix\ software.
\label{fig:smoothing_kernels}
}
\end{figure}

The CMB, as a correlated random field on the sphere, is most simply defined in harmonic space by its power spectrum, $\cl$. At full resolution, the CMB power is smoothed by the instrumental beam, which we approximate in this analysis with a Gaussian of FWHM $f_{\rm WMAP}$. In patches that are processed at reduced resolution, the CMB signal is also smoothed by the anti-aliasing beam (another Gaussian, of FWHM $\degfwhm$) and further by the pixel window function of the degraded resolution map. This final effect, shown for the relevant beams in Fig.~\ref{fig:smoothing_kernels}, is small but non-negligible. Ignoring the degraded pixel window function means that the covariance contains an overestimate of the CMB power -- in our analysis, a noise term -- and log-evidences can be underestimated by as much as 1 when degrading to $\nsidedeg = 64$.

Taking all of these effects into account, the CMB covariance between two pixels $i$ and $j$ is
\begin{equation}
S_{ij} = \sum_\ell \frac{2 \ell + 1}{4 \pi} \cl \legl(\cos{\theta_{ij}}) \beamlwmap^2
\end{equation}
if the patch is processed at full resolution, and 
\begin{equation}
S_{ij} = \sum_\ell \frac{2 \ell + 1}{4 \pi} \cl \legl(\cos{\theta_{ij}}) \beamlwmap^2 \beamlsmooth^2 \winldeg^2
\end{equation}
if the patch must be processed at degraded resolution. Here, $\legl(x)$ are the Legendre polynomials, $\winldeg$ is the $\nsidedeg$ pixel window function, and the $B_\ell$ are the WMAP and anti-aliasing Gaussian beams, represented in harmonic space as 
\begin{equation}
B_\ell = \exp \left( \frac{ -\ell (\ell + 1) f^2}{16 \log 2} \right),
\end{equation}
where $f$ is the relevant FWHM.

\subsubsection{Instrumental noise covariance}

At full resolution, the WMAP noise is uncorrelated in pixel space (see, e.g., Ref.~\cite{Hinshaw_etal:2003}). The noise covariance matrix is therefore diagonal, and can be written as
\begin{equation}
N_{ij} = \frac{\sigma_{\rm WMAP}^2}{N_{{\rm obs, }i}} \delta_{ij},
\end{equation}
where $\sigmawmap = 6.549$ mK is the RMS noise of the W-band detectors, $N_{{\rm obs, }i}$ is the number of times pixel $i$ has been observed, and $\delta_{ij}$ is the Kronecker delta.

The anti-aliasing smoothing applied as part of the degradation process induces correlations between the noise measured in each pixel. Coupled with the variations in sky coverage represented by the $N_{\rm obs}$ values, this makes the exact pixel-space covariance more difficult to write down. Progress can be made by separating the full-resolution noise model into isotropic white noise, calculated from the power spectrum in harmonic space, which is modulated by a map encoding the variations in sky coverage, which belongs entirely in pixel space. The effects of the smoothing and degradation can then be applied individually to each component, and recombined in the covariance matrix.

Uncorrelated pixel noise is identical to the noise generated by white (i.e., flat) noise in harmonic space. To produce the correct RMS noise, the amplitude of the white noise power spectrum, $\nl$, should be set such that
\begin{equation}
\nl = \frac{ \sigmawmap^2 \Omega_{\rm pix} }{ \bar{N}_{\rm obs} },
\end{equation}
where $\Omega_{\rm pix}$ is the pixel area, and we have incorporated the mean number of observations per pixel, $\bar{N}_{\rm obs}$, into the definition. After smoothing and degradation, the noise power spectrum is no longer flat, having been multiplied by both the anti-aliasing beam and the degraded pixel window function (both squared).

Having absorbed the mean number of pixel observations into the isotropic noise, to include the effects of varying sky coverage we need only consider relative changes in the number of observations in each pixel. These are captured by generating a map of $\sqrt{ \bar{N}_{\rm obs} / N_{{\rm obs,} i}}$. After degradation, this map will have been smoothed and degraded exactly as the input temperature map; its mean at any resolution is one. The noise covariance matrix at degraded resolution is then 
\begin{equation}
N_{ij} = \frac{\bar{N}_{\rm obs}}{\sqrt{N_{{\rm obs}, i} N_{{\rm obs}, j}}} \sum_\ell \frac{2 \ell + 1}{4 \pi} \nl \legl(\cos{\theta_{ij}}) \beamlsmooth^2 \winldeg^2,
\end{equation}
where the separation into components residing in pixel and harmonic space is clear. Note that this expression does not include the instrumental beam, as this is {\em detector} noise.

\subsubsection{Full covariance}

\begin{figure}
\includegraphics[width=8.5cm]{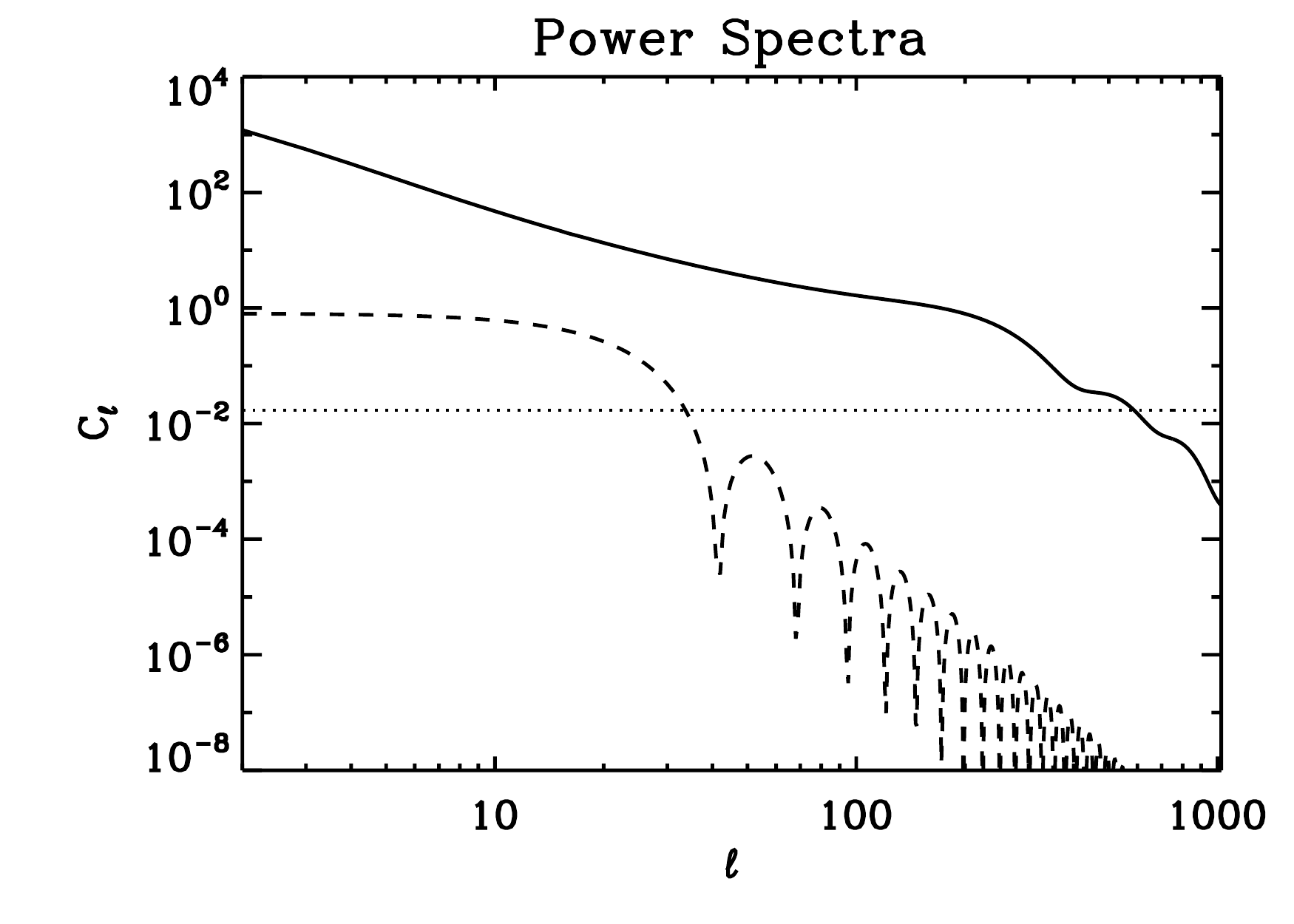}
\includegraphics[width=8.5cm]{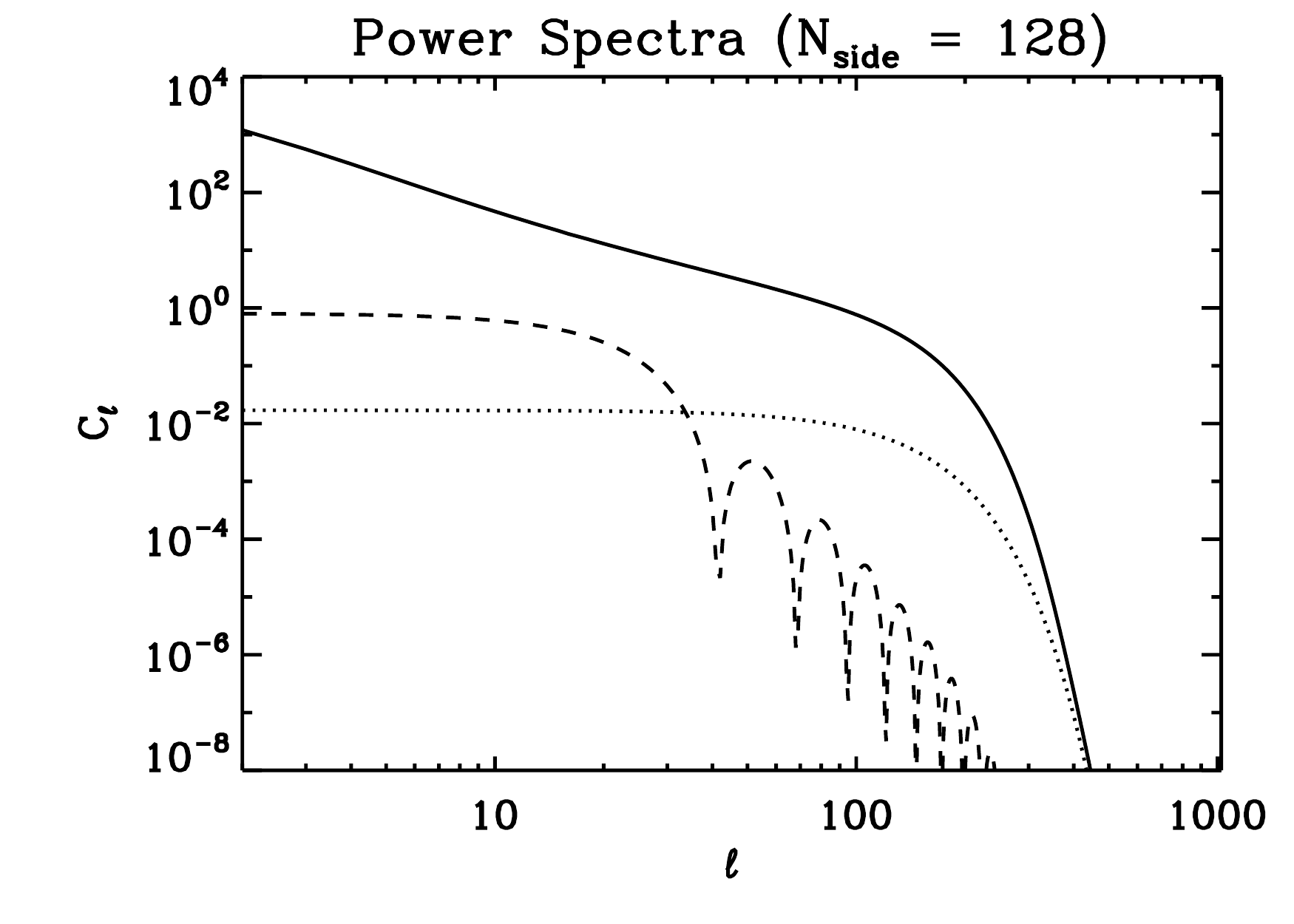}
\caption{Power spectra of the WMAP7 best-fit CMB signal (solid lines), WMAP7 noise (dotted) and a single simulated bubble collision (dashed) with amplitude $z_0 = 5\times10^{-5}$ and angular radius $\thetac = 7^\circ$. The power spectra are plotted at full WMAP resolution, $\nside = 512$ (left), and after smoothing and degradation to $\nsidedeg = 128$ (right).
\label{fig:cls_full_vs_deg}
}
\end{figure}

The sum over $\ell$ in the covariance matrix calculations is strictly a sum to infinity, but is approximated with a sum from $0 \le \ell \le \lmaxcov$, truncated at a point where all significant contributions have been included. At full resolution, the data are noise-dominated at high-$\ell$ (see Fig.~\ref{fig:cls_full_vs_deg}), so the covariance is not particularly sensitive to the precise value of $\lmaxcov$, provided it is larger than $\sim 1000$. We use $\lmaxcov = 1024$.\footnote{
The WMAP7 observations~\cite{Larson_etal:2011} extend the power spectrum measurements to $\ell = 1200$, but the CMB signal-to-noise ratio for the W-band is below one for $\ell \gtrsim 600$. Setting $\lmaxcov = 1024$ ensures the CMB contribution is characterized well into the noise-dominated regime.} After degradation, the CMB and noise power spectra have been additionally damped by the extra smoothing and pixel window function. These effects combine to reduce the map power to $\sim 0.7\%$ of its maximum at $\lmax = 2\nsidedeg$, but the kernel's long tail means that there is still information at higher-$\ell$. If this is ignored, the maps contain more information than is accounted for in the covariance matrix. The signal-to-noise ratio of any relic present is therefore artificially boosted, and the evidence is over-estimated: truncating at $2 \nsidedeg$ yields over-estimates of $\sim 5$ in the log-evidence at $\nsidedeg = 64$. Tests reveal that evidence ratios are stable provided multipoles $\ell \gtrsim 3 \nsidedeg$ are included in the covariance matrix. We choose to be conservative and push this to $\lmaxcov = 4 \nsidedeg$.

The two components of the covariance are added to form the full covariance matrix.
\begin{equation}
\matr{C} = \matr{S} + \matr{N}.
\end{equation}
This matrix must be inverted for use in the likelihood function; the inversion is carried out using the Cholesky decomposition implemented in the \lapack\ software package. The full-resolution covariance matrix has a prominent diagonal due to the uncorrelated pixel noise, and hence is readily invertable. However, at lower resolutions the additional smoothing can make the covariance matrix nearly singular. Diagonal regularizing noise is therefore added at the level of $2\,\muk^2$ to all degraded-resolution covariance matrices to aid inversion: we have checked that this does not affect the results of the likelihood calculation.

\subsection{Calculation of patch-based evidence ratios with MultiNest}

To form the full posterior (Eq.~\ref{equation:posteriorfinalapp}), we must calculate the patch-based evidence ratios (Eq.~\ref{equation:rho_int_over_local_params}) for each blob. Our pipeline uses the \multinest\ sampler~\cite{Feroz_Hobson:2008,Feroz_Hobson_Bridges:2009}, which outputs the normalized posterior and the evidence for the data in each patch under a specific model.  This is not precisely the information required in Eq.~\ref{equation:rho_int_over_local_params}, which integrates the product of the likelihood and the full-sky prior. However, we can use Bayes' theorem in the patch to convert the output posterior into a likelihood according to
\begin{equation}\label{eq:patchbasedbayes}
\prob_b (\model_0, \model_1 | \data) = \frac{ \prob (\model_0) \prob_b (\model_1 ) \prob_b (\data | \model_0, \model_1) }{ \prob_b (\data) }.
\end{equation}
Here, the subscript $b$ indicates that the probabilities are formed using only the data for a single blob. In particular, we identify
\begin{equation}
\prob_b (\data | \model_0, \model_1) \propto e^{- (\data - \template)
      {\matr{C}^{(bb)}}^{-1}
    (\data - \template)^{\rm{T}} /2} 
\end{equation}
as the quantity necessary for the patch-based evidence ratio. The normalization assumed in Eq.~\ref{eq:patchbasedbayes} is
\begin{equation}
\int_b \diff \model_0 \diff \model_1 \prob (\model_0) \prob_b (\model_1) = 1,
\end{equation}
which implies that $\prob_b (\model_1)$ is related to the prior $\prob (\model_1)$ in Eq.~\ref{equation:rho_int_over_local_params} by an overall normalization:
\begin{equation}\label{eq:blob_to_full_normalization}
F_b = \frac{\prob (\model_1)}{\prob_b (\model_1)} = 1 - \int_{\bar{b} } \diff \model_0 \diff \model_1 \prob (\model_0) \prob (\model_1).
\end{equation}
We also have
\begin{equation}
\prob_b (\data) = \int_b \diff \model_0 \diff \model_1 \prob (\model_0) \prob_b (\model_1) \prob_b (\data| \model_0, \model_1).
\end{equation}
Using these expressions, we can solve for the patch-based evidence in terms of known quantities:
\begin{eqnarray}\label{eq:multinest-evidence-ratios}
\rho_b (\model_0) = F_b \times \frac{ \prob_b (\data)}{ \prob_b (\data | 0,0 )} \times \frac{ 1 }{ \prob (\model_0) } \int_b \diff \model_1 \prob_b (\model_0, \model_1 | \data),
\end{eqnarray}
where $ \prob_b (\data | 0,0 )$ is the likelihood in the patch with no sources.

All steps of the algorithm are parallelised wherever possible using OpenMP and MPI to take advantage of both shared- and distributed-memory clusters.


\section{Adaptive-resolution tests}\label{sec:adaptive_res_tests}

\subsection{Stability of degraded evidence values}

The adaptive-resolution analysis pipeline was tested thoroughly to estimate the effect the degrading has on the calculated evidence values. Na\"{\i}vely, we would expect the evidence not to change (on average) for resolutions at which the feature is well-sampled, i.e., for which the pixel scale is much smaller than the feature itself. Once the resolution has decreased enough for the feature -- and hence the template used to fit it -- to appear pixelated, the evidence should begin to drop off sharply. In harmonic space this is equivalent to requiring that enough modes are left intact by the pre-degradation smoothing that the template power spectrum can be discerned.

The ideal test in this situation is to create a simulation containing a sufficiently large template to be un-pixelated at the lowest resolution at which a feature can be processed: in this case, $\nside = 64$. This feature should then be processed at all resolutions considered, from the highest (WMAP-resolution) to the lowest, to determine how the evidence behaves. Unfortunately, the very nature of the problem makes this is impossible: such templates would require enormous covariance matrices at full resolution.

Progress can be made by breaking the test into parts. Simulations containing templates -- strong bubble collision signatures, in this case -- are generated on a range of angular scales. Each simulation is processed twice: first at the highest resolution possible, then again at one resolution lower. This indicates how the evidence changes when a patch is not processed at its ``ideal'' resolution, which could occur if the maximum angular size, $\thetac$, is overestimated. The lower-resolution evidence values are calculated with four times fewer pixels: if the evidences returned by the two runs do not differ greatly, we can be confident that small reductions in resolution do not affect the evidence values returned.

\begin{figure}
\includegraphics[width=5.5cm]{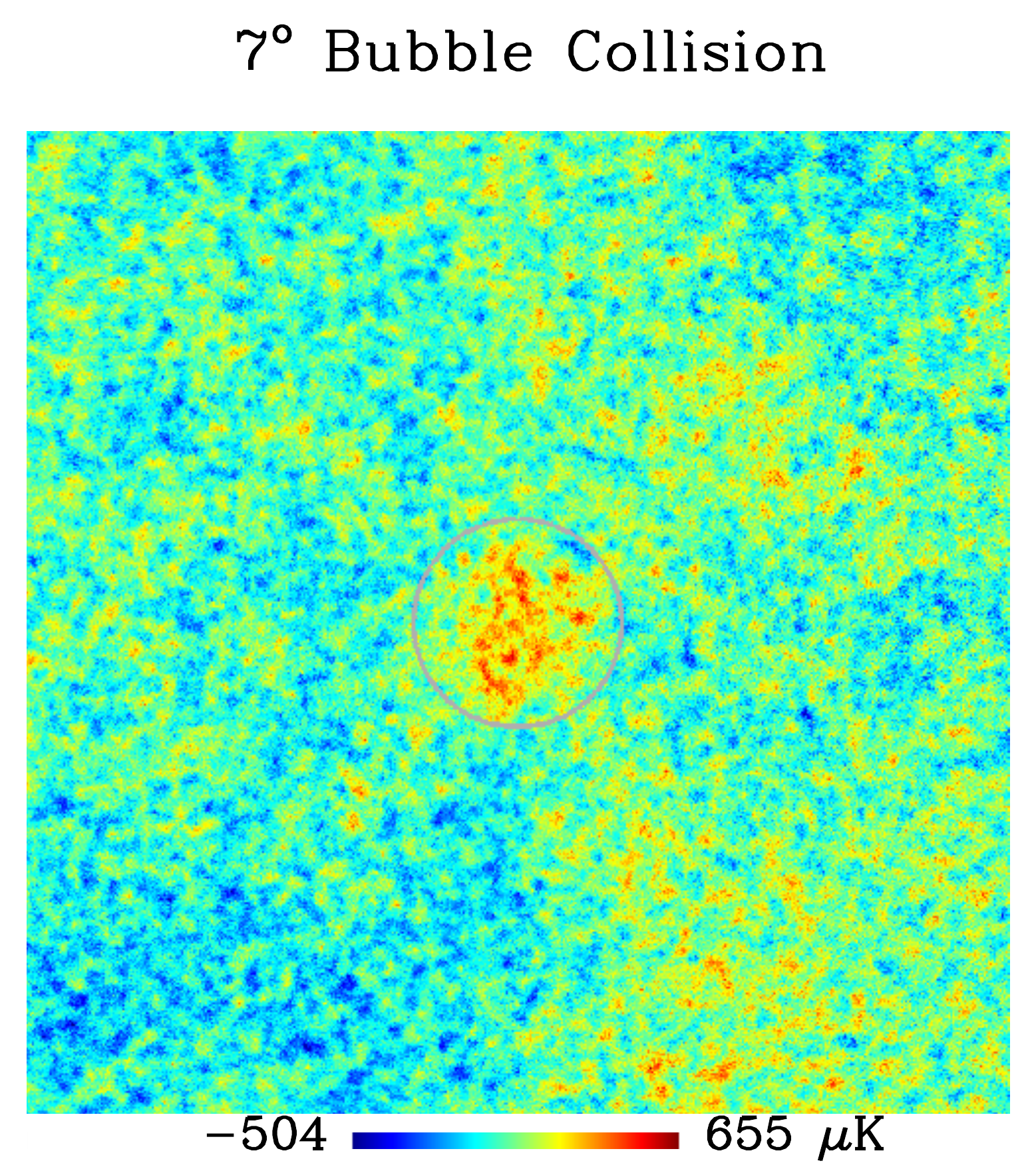}
\includegraphics[width=5.5cm]{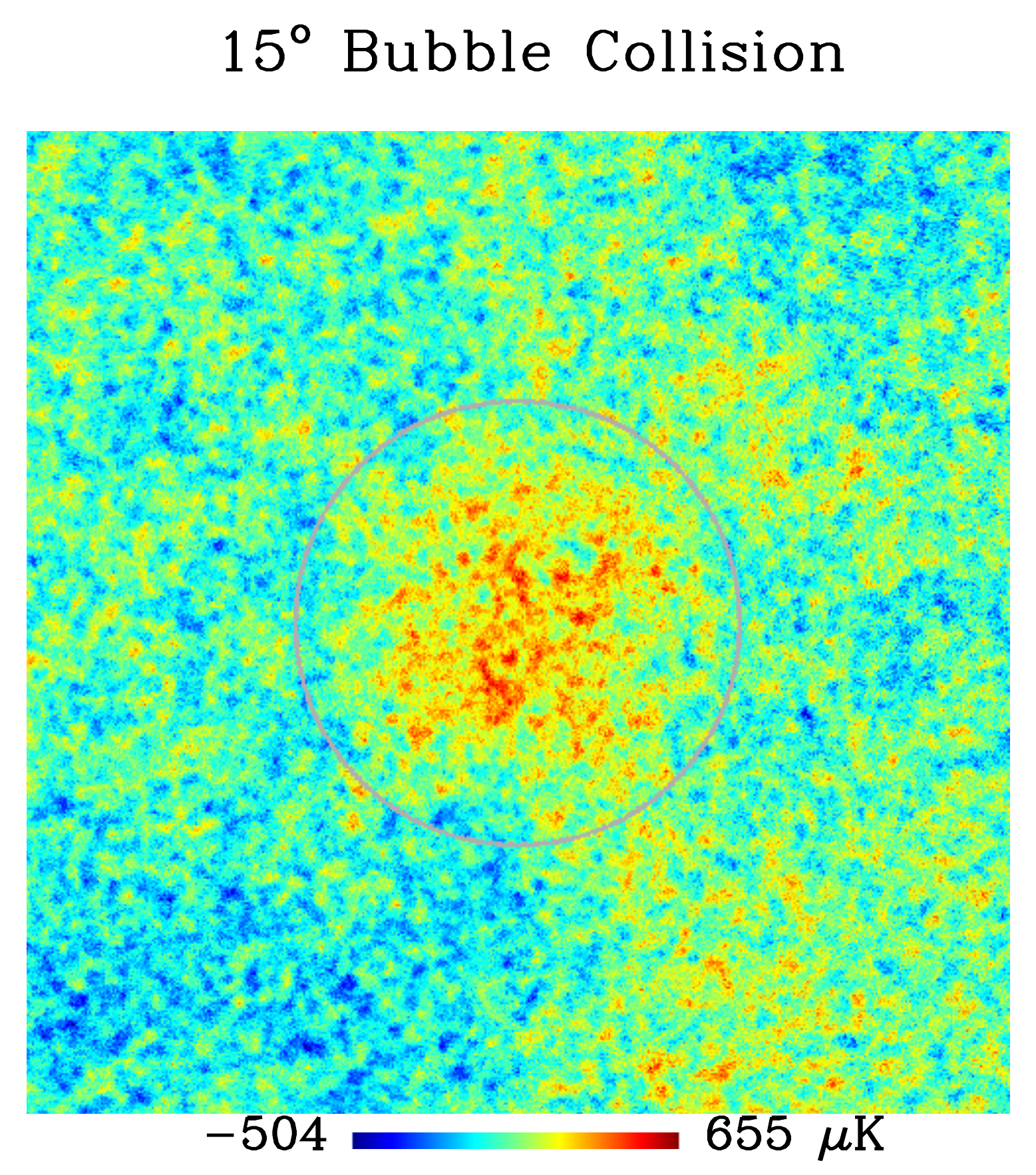}
\includegraphics[width=5.5cm]{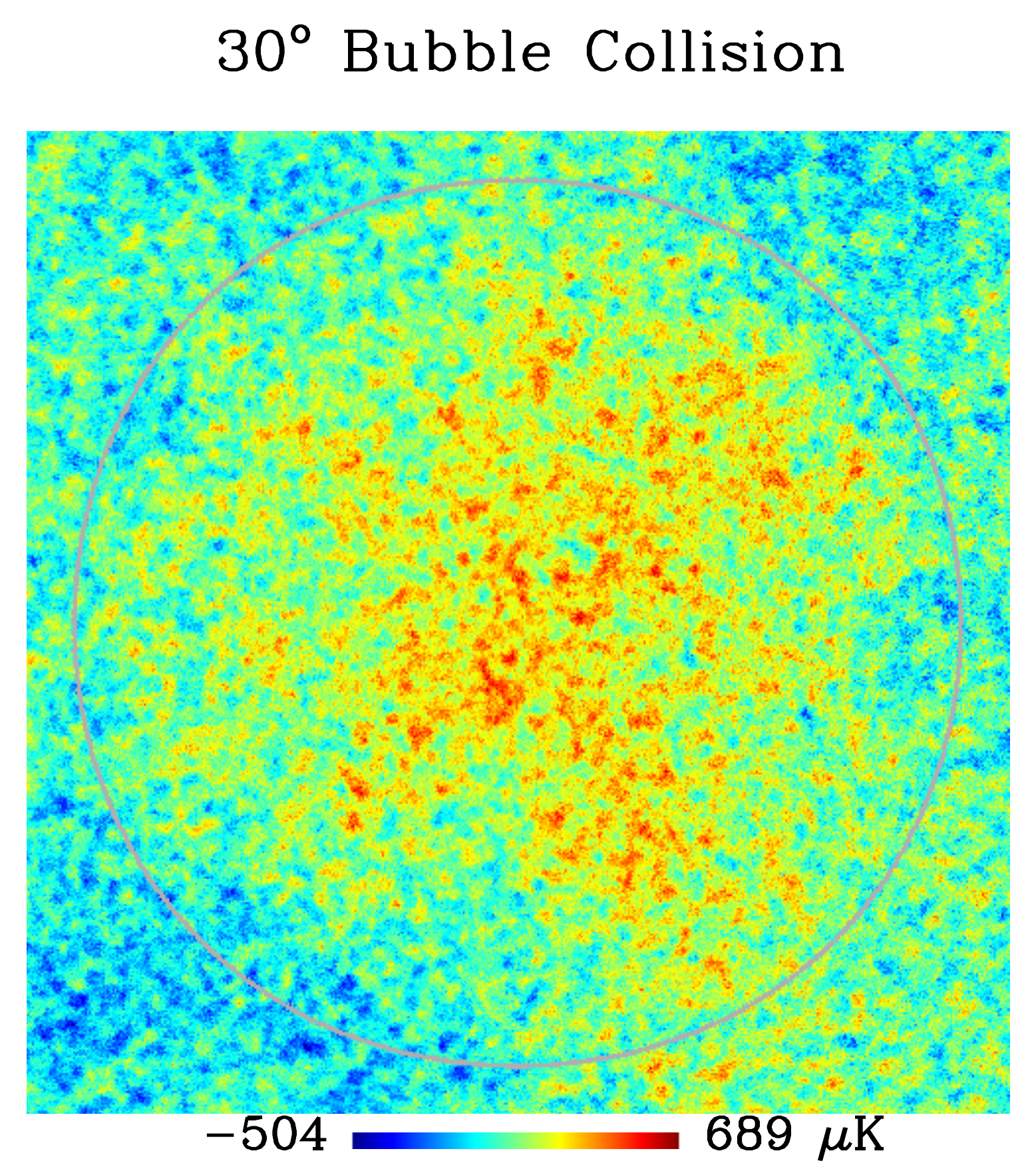}
\caption{The maps used to test the adaptive-resolution analysis. Three maps are generated, each containing a single bubble collision of radius $\thetac = 7^\circ$, $15^\circ$ or $30^\circ$. The maps contain identical CMB and noise realizations, and in each case the bubble collision is placed at $(\theta_0, \phi_0) =  (45^\circ, 45^\circ)$ with amplitude  $z_0 = 8.0\times10^{-5}$. The $7^\circ$ and $30^\circ$ collision maps are also used to test the effects of neglecting correlations with data outside the patch and the restriction of template locations to the regions highlighted by the candidate detection stage. All plots are shown at the same scale, and are $67^\circ$ on a side.
\label{fig:degrade_tests}
}
\end{figure}

Three maps are generated containing single bubble collision signatures of angular scale $\thetac$ equal to $7^\circ$, $15^\circ$ and $30^\circ$, each of which contains approximately the same number of pixels at $\nside = 512$, 256 and 128, respectively. In each case the signatures have the same amplitude, $z_0 = 8 \times 10^{-5}$, and position, $(\theta_0, \phi_0) = (45^\circ, 45^\circ)$; the maps also contain the same CMB and noise realizations, and are plotted in Fig.~\ref{fig:degrade_tests}. The evidence is calculated once at the highest resolution possible -- $\nside$ of 512, 256 and 128 for the $7^\circ$, $15^\circ$ and $30^\circ$ collisions, respectively -- and once again after degrading one step in resolution -- i.e., for $\nside$ of 256, 128 and 64. In each pair of tests the same ranges in size and position are sampled, and no mask is used. The only small difference comes in the pixels included in the patch, as the lower-resolution patches sample a slightly larger region than the high-resolution patches.\footnote{When defining the patches using the \healpix\ query\_disc subroutine, the ``inclusive'' option is set to include all pixels which fall even partly within the radius we sample.}

The results of the tests are shown in Table~\ref{tab:degrade_tests}. In each case, the evidence calculated at high resolution matches the evidence at low resolution to within \multinest\ precision. We conclude that the adaptive-resolution analysis produces stable evidence ratios for the resolutions considered in this work. Note that the evidence does not always decrease: in fact it can increase. This is because the realization noise (i.e., the combined CMB and noise signal) is different after degradation: the bubble collision signature is compared to larger-scale modes at lower $\nsidedeg$. Further, note that the log-evidence errors tabulated are the typical random variations due to sampling, estimated by repeatedly testing an individual patch with initial conditions set from different random seeds. \multinest\ also provides a statistical estimate of the error in an evidence calculation, derived from the relative entropy of the samples (see Refs.~\cite{Skilling:2004,Feroz_Hobson:2008}). We find these estimates (typically $\sim 0.1$ in log-evidence) to be subdominant to the variations due to sampling.

\begin{table*}
\begin{tabular}{c c c c}
\hline
\hline
\ Bubble $\thetac$ ($^\circ$) \ & \ $\nsidedeg$ \ & \ $\npix$ \ & \ $\log \rho$ \ \\
\hline
$7$ & $512$ & $41618$ & $12.6 \pm 0.3$ \ \\
$7$ & $256$ & $10544$ & $12.8 \pm 0.3$ \ \\
$15$ & $256$ & $40715$ & $14.2 \pm 0.3$ \ \\
$15$ & $128$ & $10314$ & $14.0 \pm 0.3$ \ \\
$30$ & $128$ & $49894$ & $13.2 \pm 0.3$ \ \\
$30$ & $64$ & $12619$ & $13.1 \pm 0.3$ \ \\
\hline
\hline
 \end{tabular}
 \begin{center}
 \caption{Tests of the stability of the degraded evidence values. Three maps, each containing a small, medium or large simulated bubble collision, are used to examine how the evidence ratio changes when a patch is degraded from $\nside = 512$ to 256, 256 to 128 and 128 to 64. 
   \label{tab:degrade_tests}}
 \end{center}
\end{table*}

\subsection{Robustness to smoothing-induced contamination}

The simulations used to test the adaptive-resolution pipeline contain only a CMB realization, a noise realization and a bubble collision template; no mask was used. Real CMB datasets also contain foregrounds (or foreground residuals after component separation), the worst of which are masked. Due to the need to smooth prior to degradation, there is a potential for contaminants to leak from behind the mask.\footnote{The smoothing procedure is fastest in harmonic space, where it is a multiplication rather than a convolution. Inclusion of the mask in this procedure is complex, and it is simplest to perform the smoothing on full-sky data.} While it is possible to mitigate this effect by extending the mask (and potentially using a smoothing kernel that is localized in pixel space)~\cite{Feeney_Pontzen_Peiris:2011}, it is highly undesirable to discard hard-won data. The likely scale and amplitude of any smoothing-induced contamination is therefore investigated to determine whether the mask should be extended.

The masks recommended by the WMAP team~\cite{Gold:2010fm} comprise two components: a Galactic cut (of varying conservatism) and a point-source mask. The point-source mask is created from a range of external and internal catalogs (as listed in Ref.~\cite{Bennett:2003ca}), and is updated with each data release. The point-sources, which appear in the data as approximately Gaussian peaks with the same FWHM ($0.22^\circ$) as the instrumental beam, are masked by excising a region of radius $\sim0.6^\circ$ centered on each point-source. A small number of the strongest sources are more aggressively masked, out to a $\sim1.2^\circ$ radius. The cut made to remove the extended emission of the Milky Way is much larger, and forms an irregular band $\sim20^\circ-40^\circ$ in width centered on the Galactic Plane. For clarity, the effects of the two components of the mask are investigated individually.

\begin{figure}
\includegraphics[width=6.0cm]{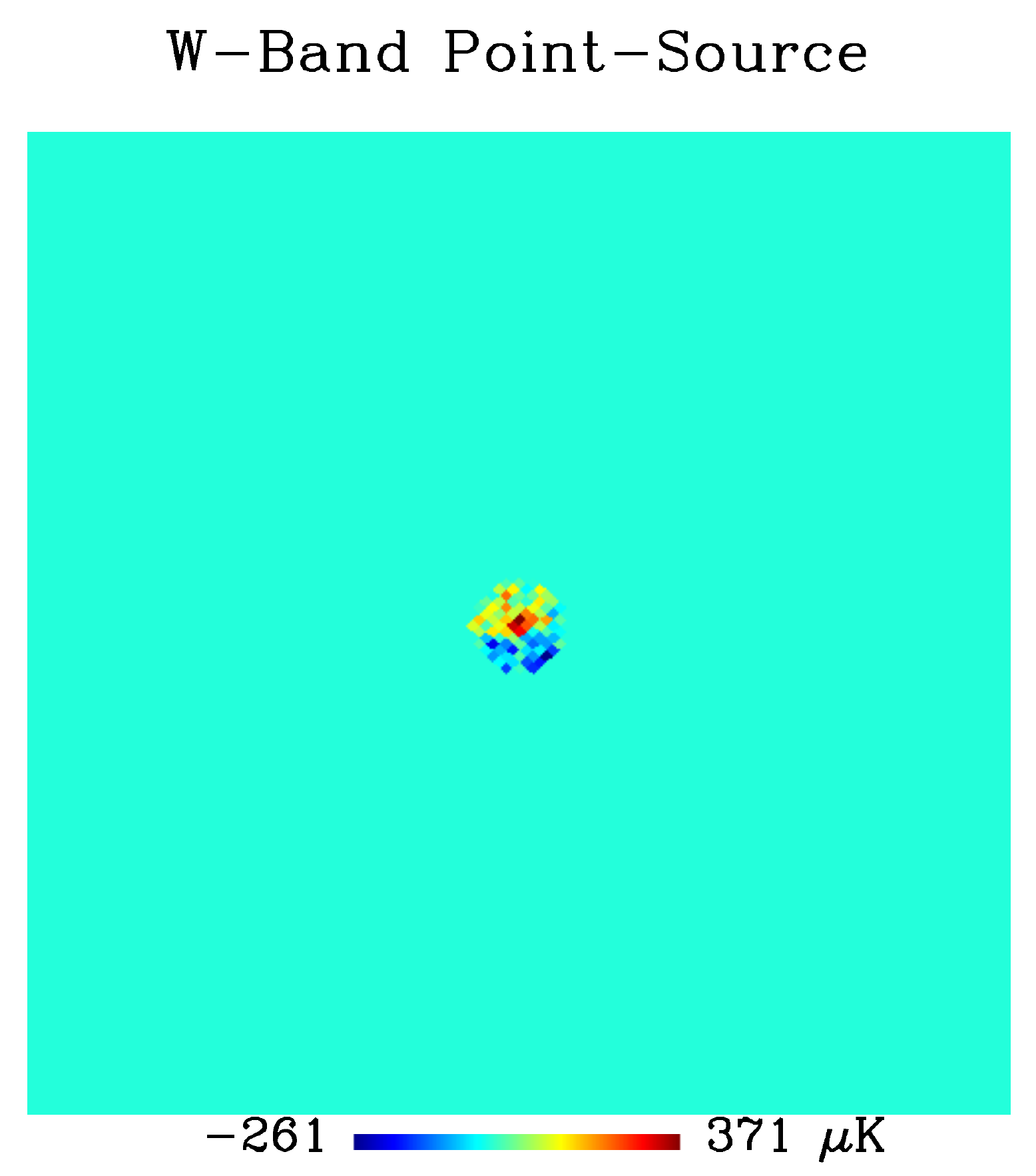}
\includegraphics[width=6.0cm]{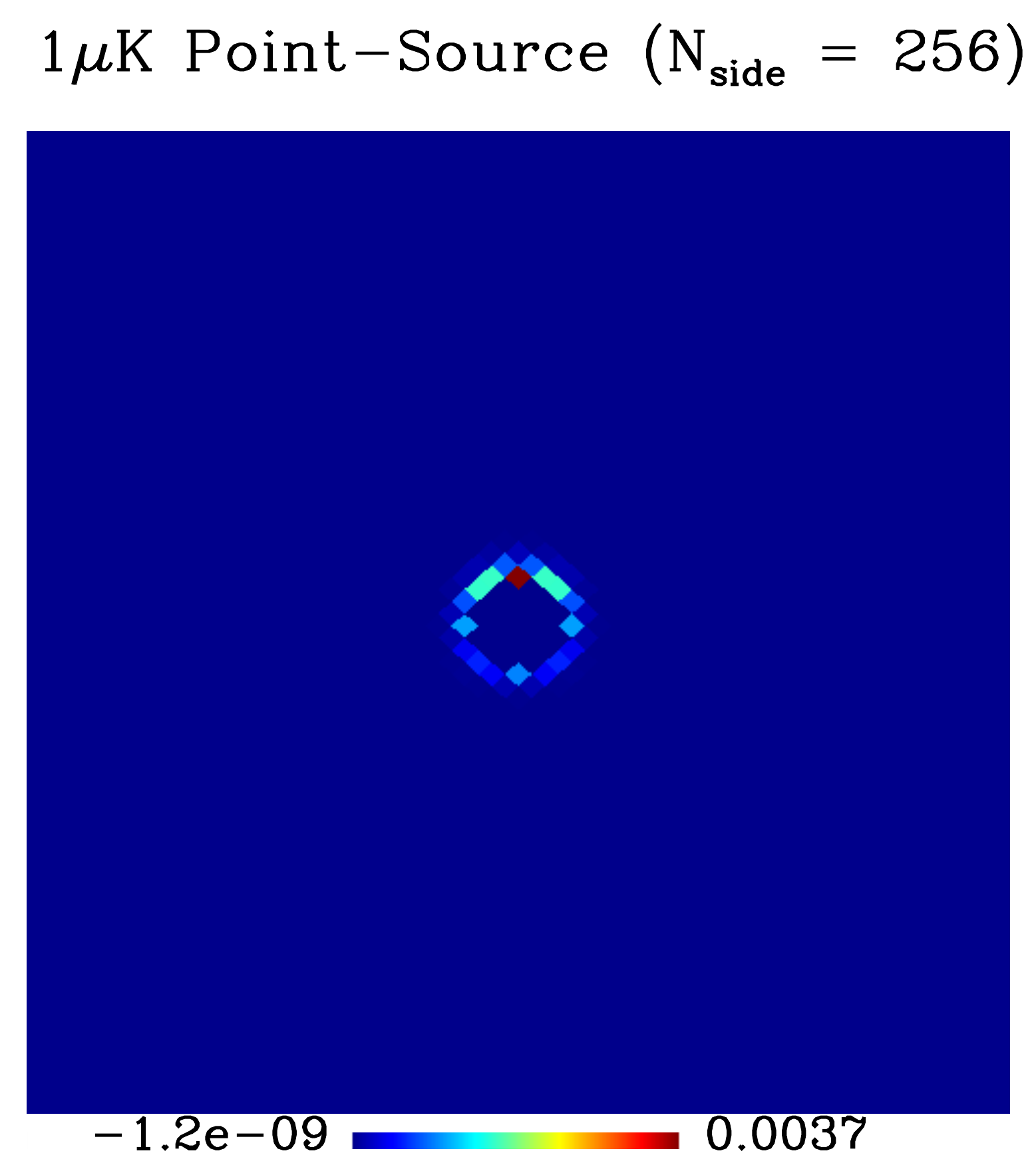}
\includegraphics[width=6.0cm]{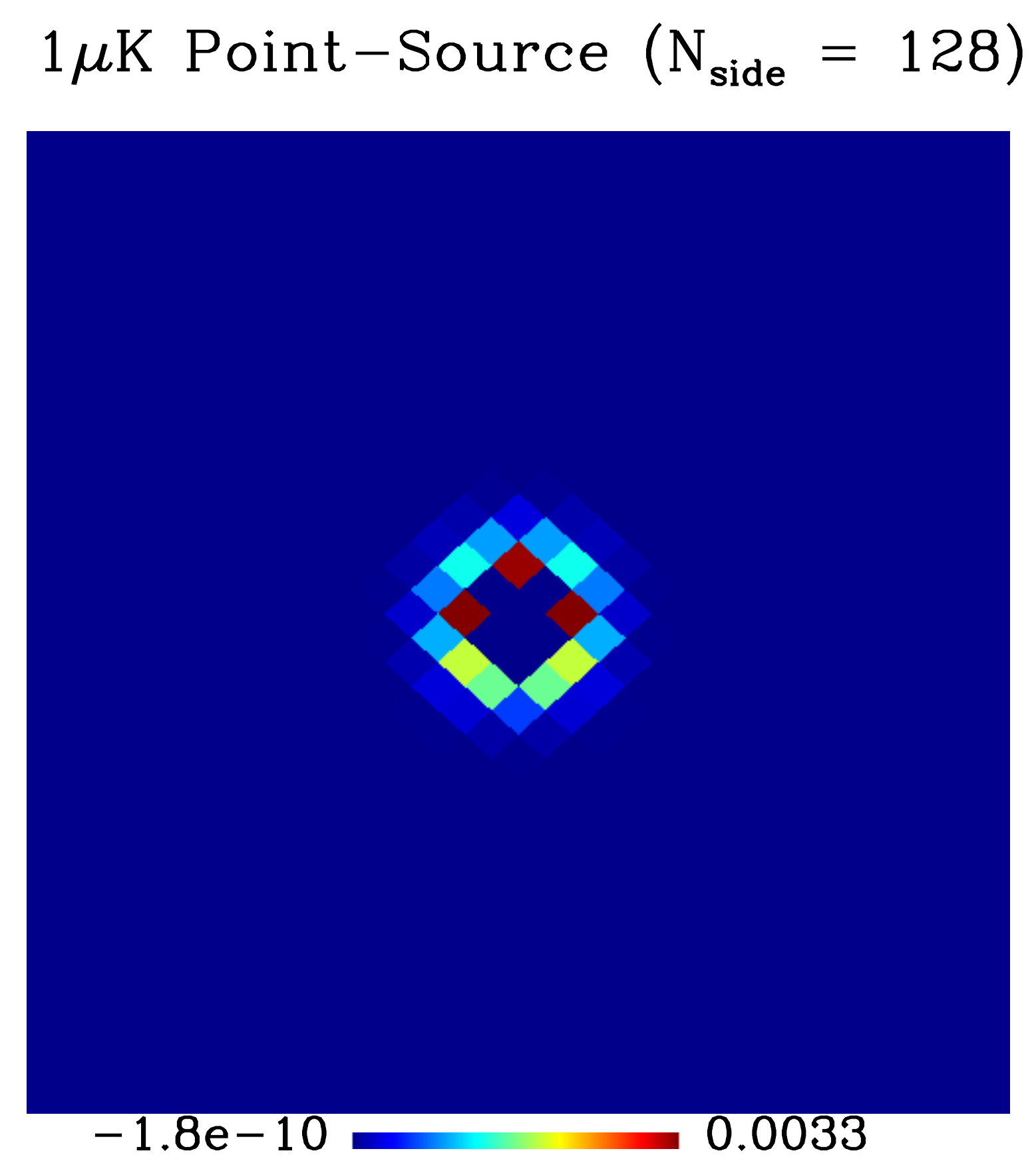}
\includegraphics[width=6.0cm]{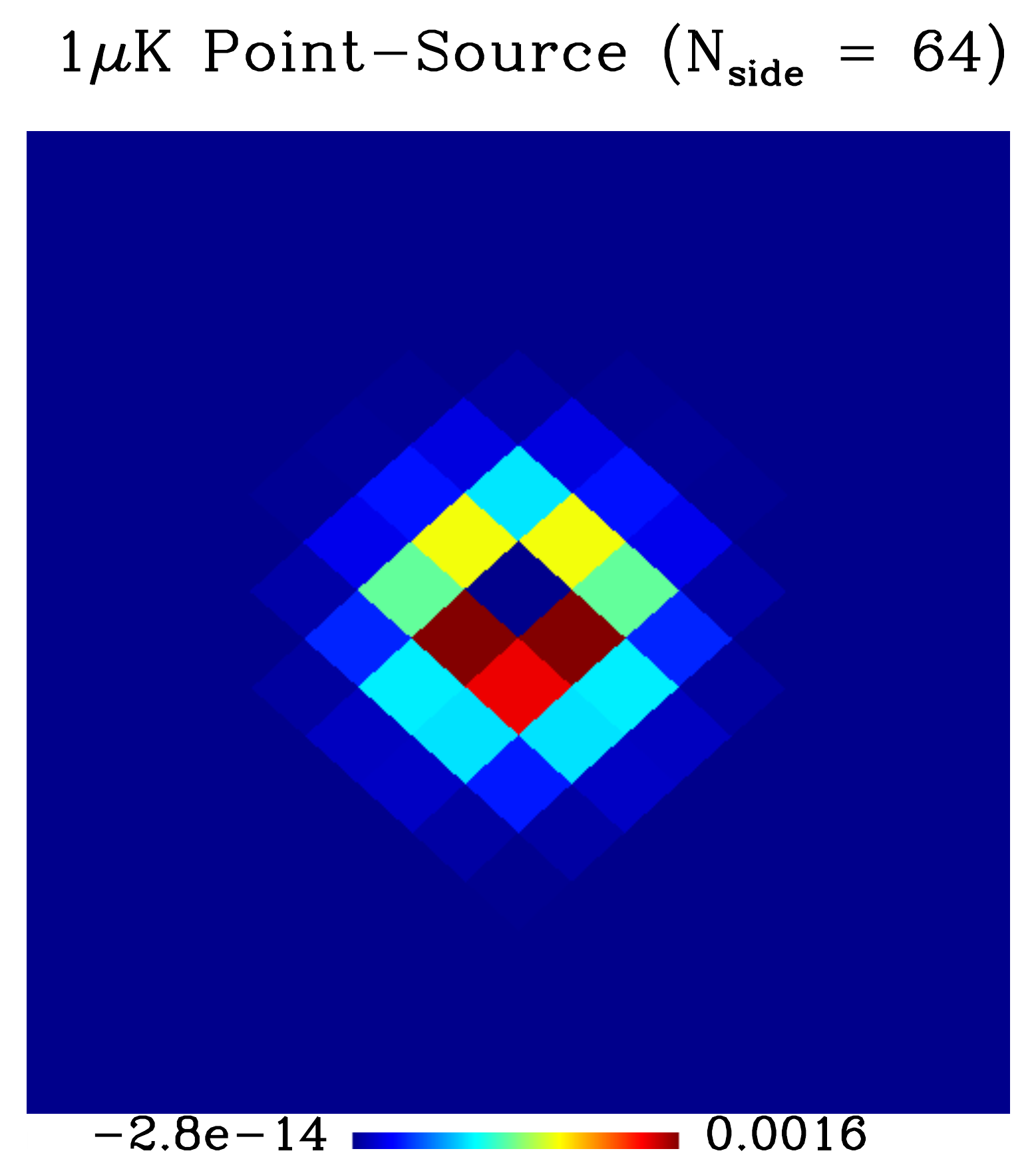}
\caption{A typical WMAP W-band point-source contaminant (top-left, plotted using the inverted point-source mask), and the effects of smoothing a unit-amplitude simulated point-source and degrading to $\nsidedeg = 256$, 128 and 64. All low-resolution plots are shown with a degraded mask applied, which limits any residuals to $\mathcal{O}(\muk)$. All plots are $12.5^\circ$ on a side.
\label{fig:point_source_tests}
}
\end{figure}

We can estimate the effects of point-source smoothing using a very simple test. A W-band point-source is simulated by placing a normalized $1\,\muk$ delta-function at a position taken from the WMAP point-source catalog~\cite{Gold:2010fm}, then convolving it with a Gaussian of $0.22^\circ$ FWHM. Since smoothing is a linear process, this can then be scaled to investigate the effects of sources of different amplitudes. This map is then smoothed and degraded to each scale used in the analysis (i.e., $\nside = \{ 256, 128, 64\}$), and masked using the degraded point-source mask at each resolution. The resulting maps, plotted in Fig.~\ref{fig:point_source_tests}, can then be scaled to mimic a source of a given temperature. The plots show that, at all resolutions considered, the maximum contamination injected into a single pixel is a few thousandths of the point-source's amplitude. Assuming that such sources have amplitudes of $100 - 1000\,\muk$~\cite{Wright:2008ib}, these results suggest that our degradation technique induces contaminants of at most $1 - 2\,\muk$ into fewer than 10 pixels. This level is completely subdominant to the CMB signal, and so should not affect the analysis. We therefore need not extend the point-source mask when smoothing and degrading.

\begin{figure}
\includegraphics[width=8.0cm]{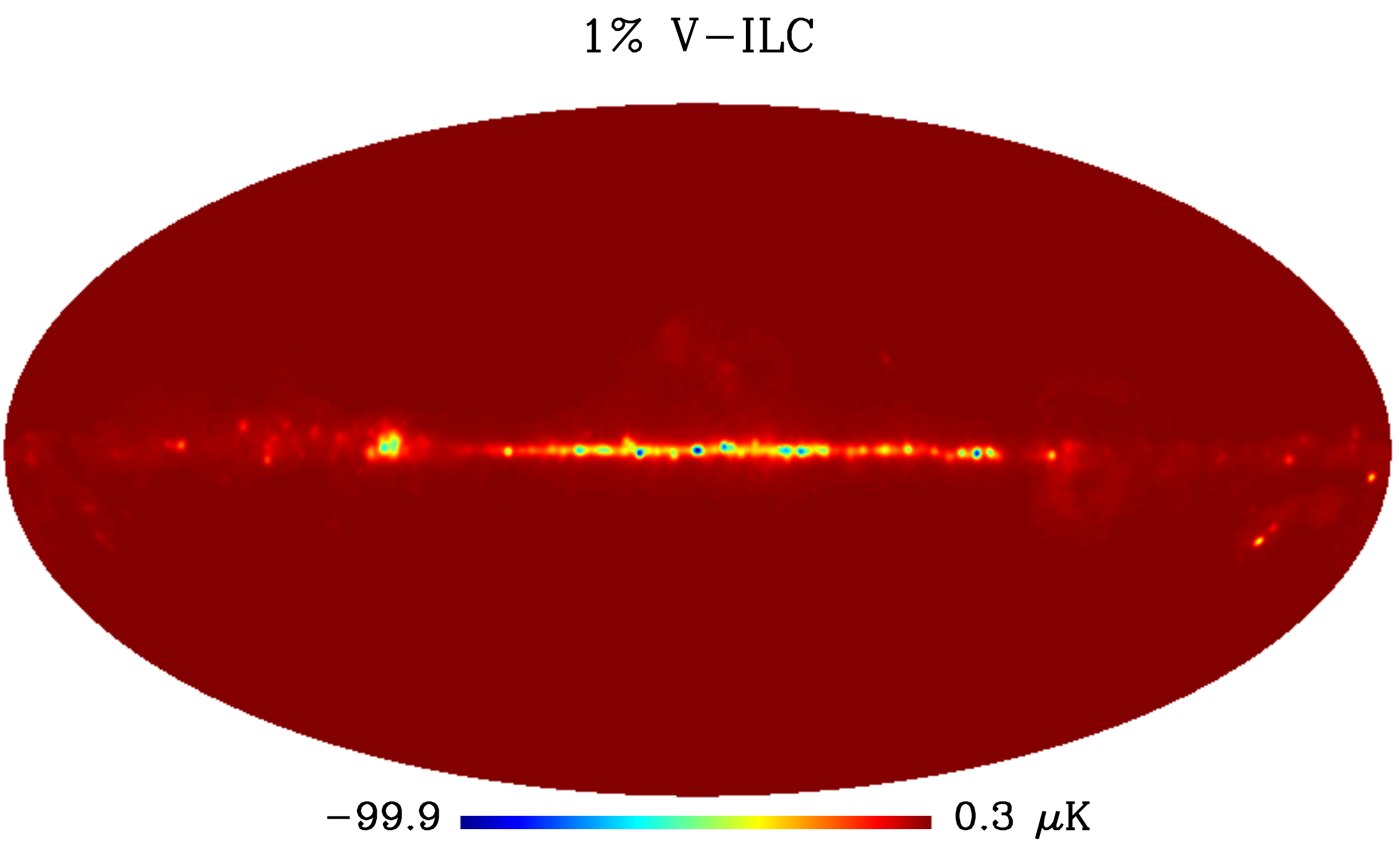}
\includegraphics[width=8.0cm]{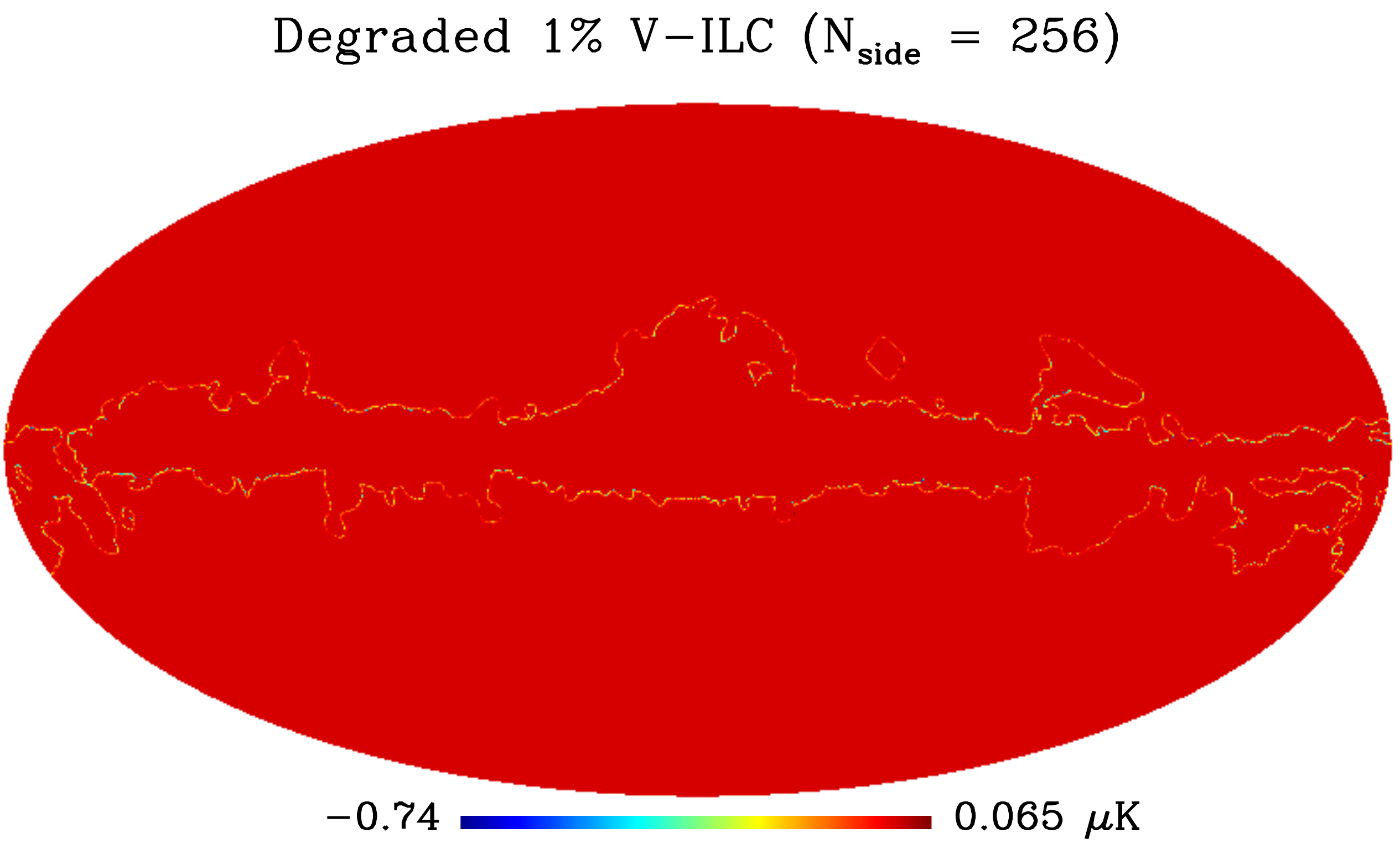}
\includegraphics[width=8.0cm]{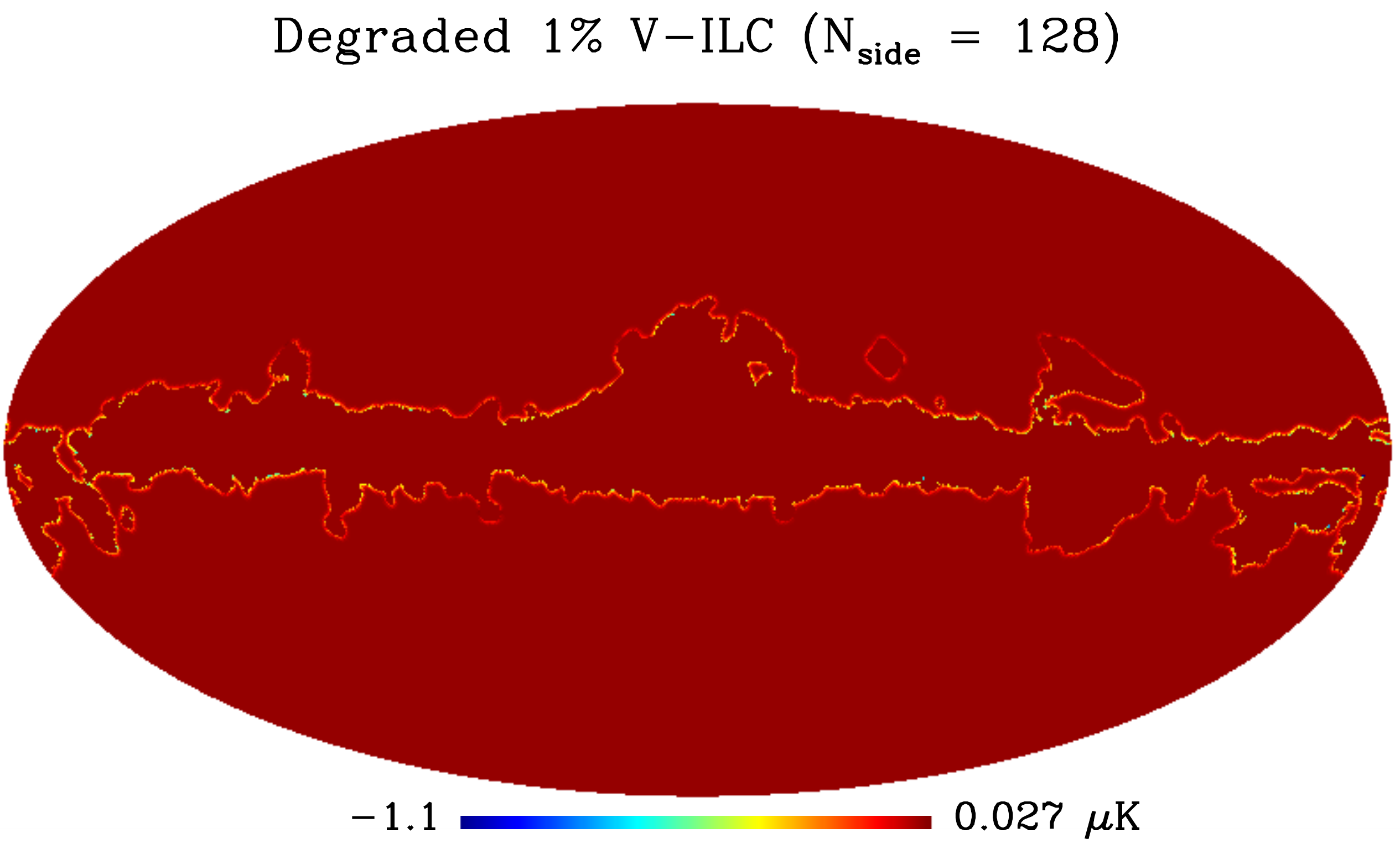}
\includegraphics[width=8.0cm]{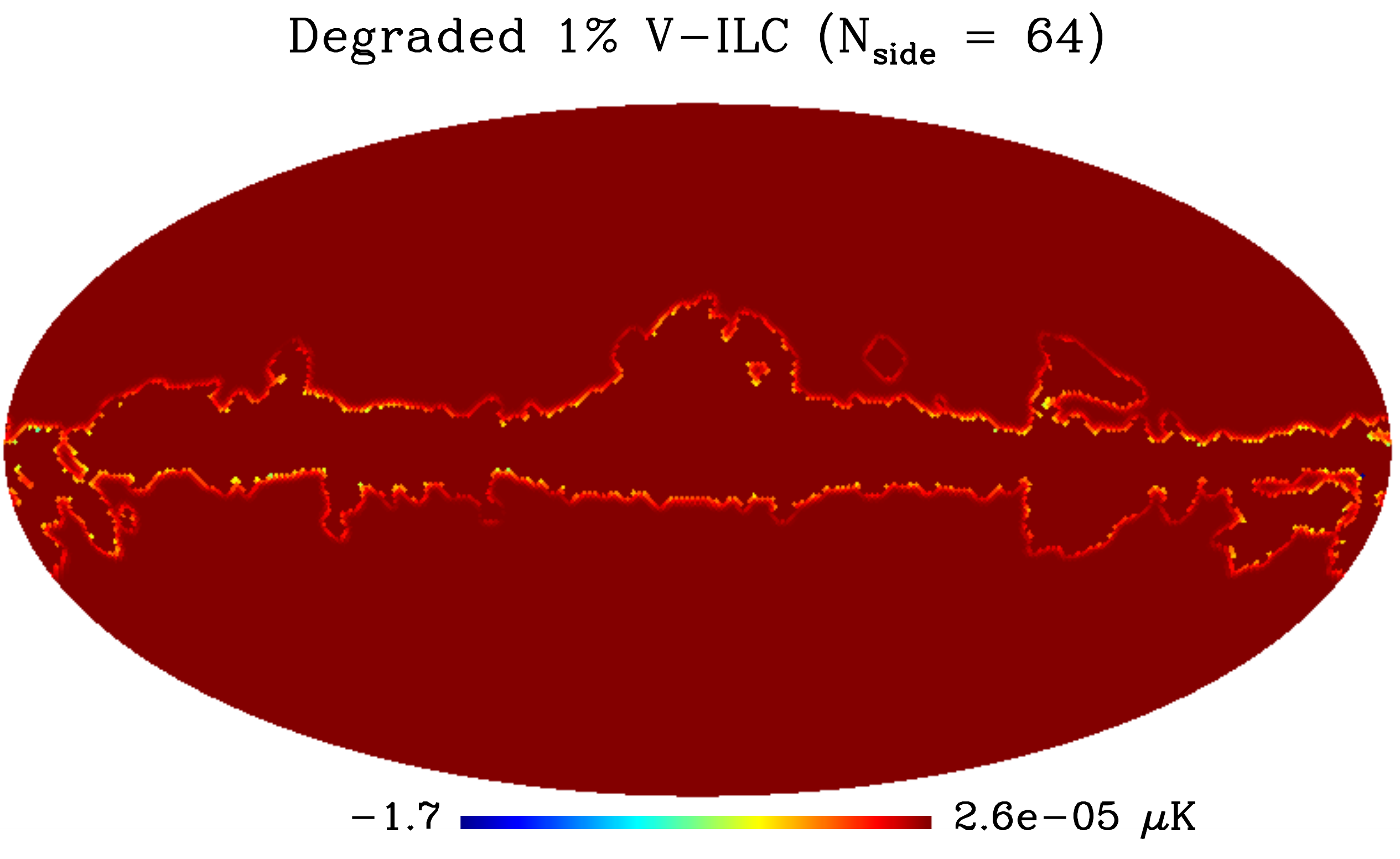}
\caption{One percent of the difference between the WMAP V-band and ILC maps provides an estimate of the Galactic contamination present in WMAP data products. This map is plotted here (top-left), alongside the residuals produced by smoothing and degrading it to $\nsidedeg = 256$, 128 and 64. All low-resolution plots are shown with a degraded KQ85 Galaxy-only mask applied.
\label{fig:gal_cont_tests}
}
\end{figure}

An estimate of the contaminants leaked from the Galactic cut can be obtained in a similar fashion. In place of the simulated point source, we must make an estimate of the Galactic foreground residuals. Modeling this precisely is difficult, so, following Ref.~\cite{Pontzen_Peiris:2010}, 1\% of the difference between the V-band signal and the WMAP7 Internal Linear Combination (ILC) map~\cite{Gold:2010fm} is used for illustrative purposes. This combination is indicative of the morphology and amplitude of residuals within the ILC: there are contaminants of around 50 times the amplitude visible to the eye in the WMAP foreground-reduced maps. As with the simulated point-source, we take this map, smooth and degrade it to $\nside$ values of 256, 128 and 64, and then mask using only the Galactic portion of the 7-year KQ85 mask. The resultant maps are plotted in Fig.~\ref{fig:gal_cont_tests}, along with the input. The extra smoothing creates a strip, a few pixels wide, of contamination around the Galactic mask, typically at a level of $0.3 - 0.4\,\muk$. Scaling this up by a factor of 50 yields contaminants of $\sim 20\,\muk$. Although this is an order of magnitude higher than that created by point-sources, it is extremely localized, and does not mimic any of the target signals. We conclude that, as with the point-source mask, there is no need to extend the Galactic cut when degrading. Our adaptive-resolution algorithm should be robust to smoothing-induced contamination, a hypothesis that will be further tested at a later point by processing a null simulation containing realistic foreground residuals.

\section{Testing the formalism's approximations}\label{sec:approximation_tests}

The formalism set out in Section~\ref{sec:formalism} includes three main approximations: that the parameters of $\Lambda$CDM can be set to the best-fit measured values, that the likelihood need only be integrated over ranges of $\{ \theta_0, \phi_0 \}$ (and, indeed, $\thetac$) corresponding to patches containing candidate sources, and that correlations between data inside and outside these patches can be neglected. The error made in the latter two approximations will depend on the nature of the source templates. In particular, the accuracy of the second approximation will depend on how well the candidate regions cover the underlying sources, and how the likelihood falls off as a function of the location and size of the feature. It is also important to note that the approximations will most likely improve as a function of the signal-to-noise for the source: if the likelihood is not peaked in any region of parameter space, we are not justified in choosing to integrate only over particular regions.

\begin{figure}
\includegraphics[width=10.0cm]{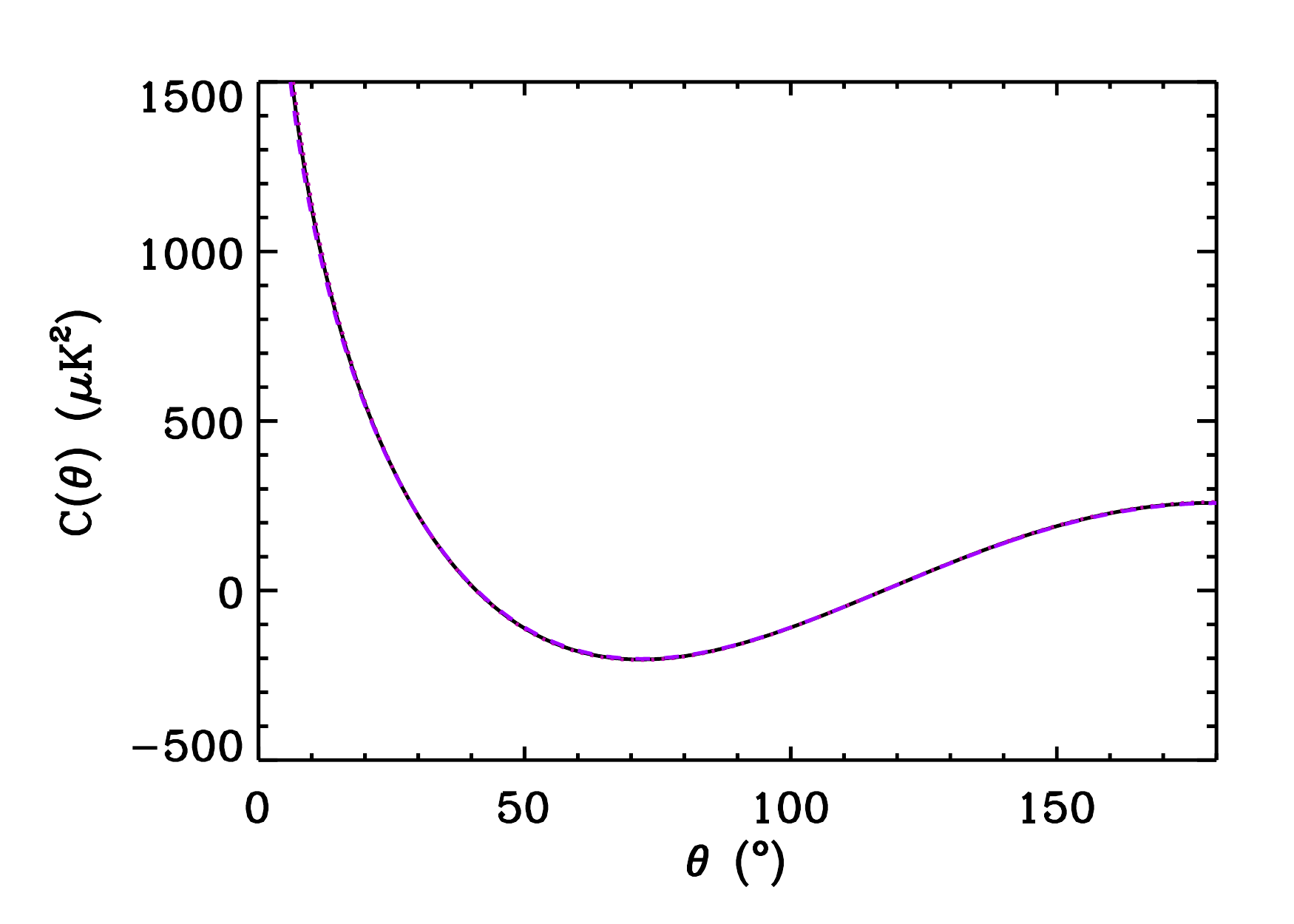}
\caption{A plot of the angular correlation function of the CMB for three cosmological models: the best-fit cosmological parameters from WMAP 7-year CMB, BAO and ${\rm H_0}$~\cite{Larson_etal:2011} (solid), a cosmological model where the dark energy density $\Omega_\Lambda$ is two standard deviations too high (dashed), and a cosmological model where $\Omega_\Lambda$ is two standard deviations too low (dotted). The three curves are nearly indistinguishable.
  \label{fig-cosvarplot}
}
\end{figure}

We begin by discussing the assumption that the parameters of $\lcdm$ can be set to the best-fit measured values. The only place these parameters enter the evidence calculation is through the pixel-pixel covariance matrix used to formulate the likelihood function. To get an idea for how this might vary with the parameters of $\lcdm$, in Fig.~\ref{fig-cosvarplot} we show the pixel-pixel covariance as a function of angle for three cosmologies: the fiducial best-fit WMAP7$+$BAO$+{\rm H_0}$ cosmology and two cosmological models where the dark energy density, $\Omega_\Lambda$, is two standard deviations higher or lower than the best-fit value (chosen because varying $\Omega_\Lambda$ significantly changes power on the largest scales, which are most relevant for the signatures considered). The difference at the level of the covariance is very small, and the difference in the inverse covariance appearing in Eq.~\ref{eq:likelihood} is correspondingly small. We can write the inverse covariance as
\begin{equation}
\matr{C}(\model_\lcdm)^{-1} = \matr{C}(\bar{\model}_\lcdm)^{-1} + \Delta \matr{C}(\model_\lcdm)^{-1},
\end{equation}
where $\Delta \matr{C}(\model_\lcdm)^{-1}$ is the small difference in the inverse covariance resulting from varying cosmological parameters. Neglecting the effect of changing the cosmological parameters amounts to requiring
\begin{equation}
\data \Delta \matr{C}(\model_\lcdm)^{-1} \data^{\rm{T}} \gg \data \Delta \matr{C}(\model_\lcdm)^{-1} \template \gg \template \Delta \matr{C}(\model_\lcdm)^{-1} \template,
\end{equation}
which is natural for small-amplitude templates that have compact support on the sky. In this case, the $\ns = 0$ and $\ns = 1$ terms in the expansion Eq.~\ref{equation:modellikelihood} would take the form
\begin{equation}
 \prob (\data | \ns=0) = \frac{1}{(2 \pi)^{\npix / 2} |\matr{C}|} e^{- \data  \matr{C}(\bar{\model}_\lcdm)^{-1} \data^{\rm{T}} /2} \int \diff \model_\lcdm \prob (\model_\lcdm)  e^{- \data \Delta \matr{C}(\model_\lcdm)^{-1} \data^{\rm{T}} /2}
\end{equation}
 \begin{eqnarray}
\prob (\data | \ns=1) &\simeq&  \prob (\data | \ns=0) \rho_b,
 \end{eqnarray}
with the patch-based evidence ratio $\rho_b$ defined as before (i.e. with the covariance defined for fixed cosmological parameters and confined to pixels in the blob). Higher order terms would follow suit. Therefore, in this approximation all of the dependence on cosmological parameters would be absorbed into an overall renormalization of $\prob (\data | \ns=0)$, and would not affect our final answer for the posterior of the global model parameters.

\begin{figure}
\includegraphics[width=7.5cm]{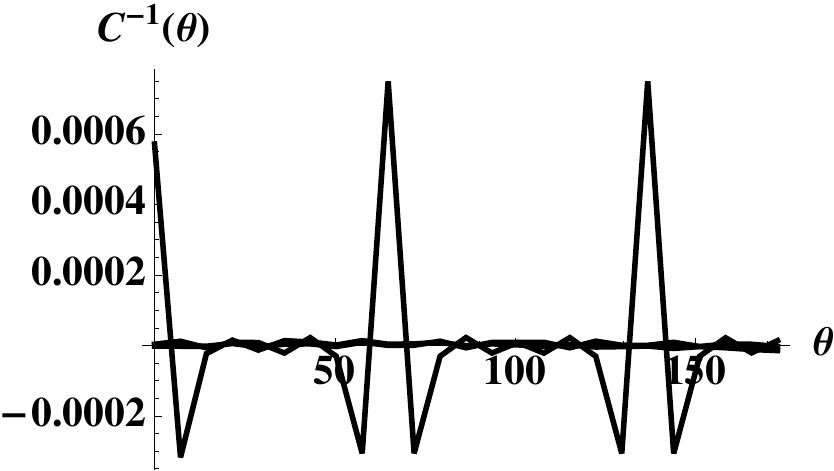}
\caption{ The inverse of the pixel-space $\Lambda$CDM correlation function between $\theta = 0$ and all $\theta$, $\theta = 72^{\circ}$ and all $\theta$, and $\theta = 144^{\circ}$ and all $\theta$ (left to right). It can be seen that $\matr{C}^{-1}$ is largest in magnitude over a $\sim 30^{\circ}$ window around the pixel being correlated.  
  \label{fig-ictheta}
}
\end{figure}

\begin{figure}
\includegraphics[width=6 cm]{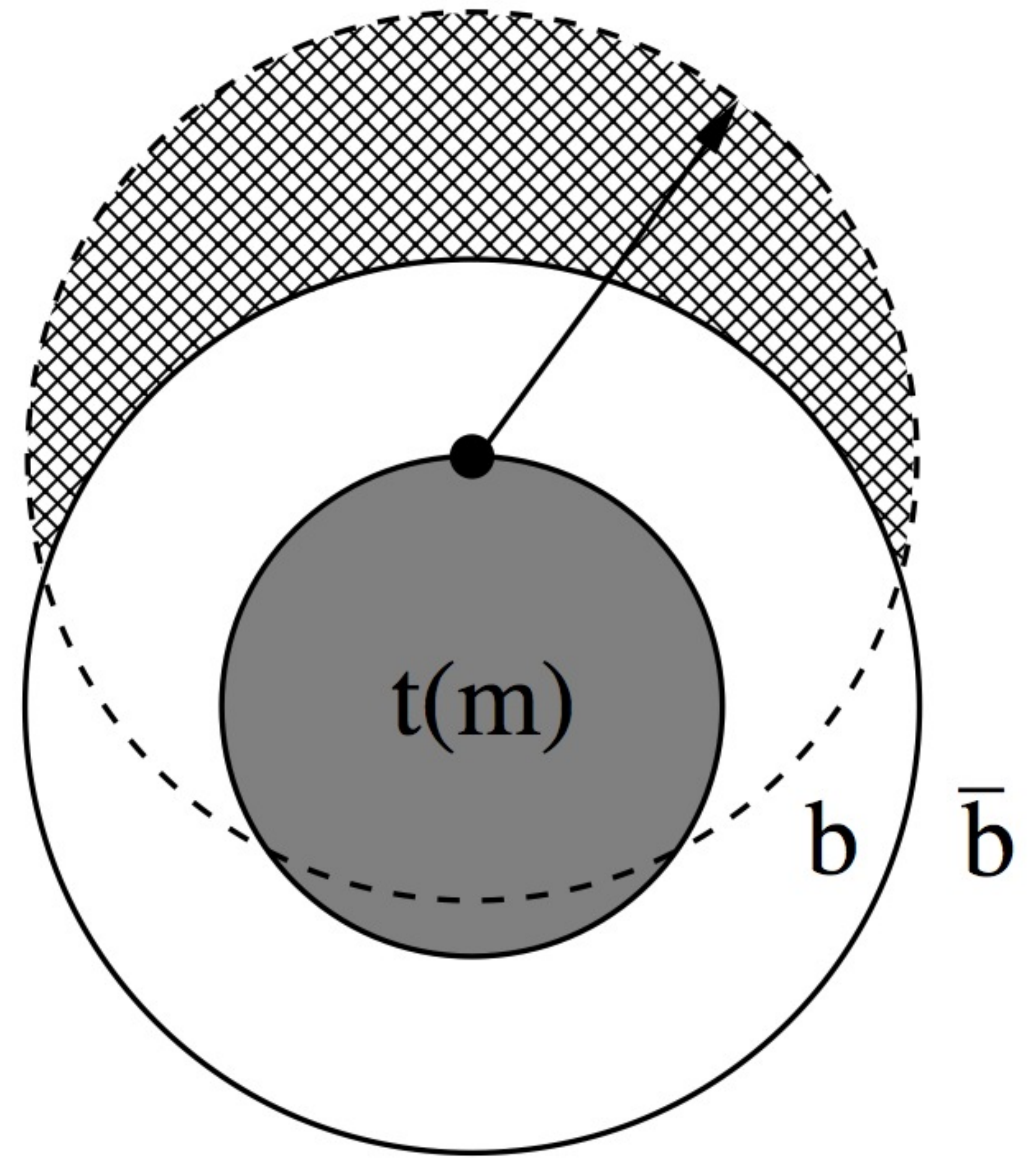}
\caption{The case where the source is well-contained within the blob. There is a clearly peaked likelihood for templates contained within the shaded disk. For the pixel denoted by the black dot, the inverse covariance is significant only within the dashed circle. Our approximation neglects correlations with pixels in the hatched region. This approximation is worst for pixels on the edge of the template.
  \label{fig-bnotb}
}
\end{figure}

To explore the latter two assumptions, consider the single-source contribution to the posterior Eq.~\ref{eq:margm1}. Neglecting correlations between regions inside and outside the patch (see Eq.~\ref{eq:bbbarneglected}) corresponds to the assumption that 
\begin{equation}
\template \matr{C}^{-1} \data_{\bar{b}}^{\rm{T}} \ll 1 \, , 
\end{equation}
where the template has support in a blob $b$, and the data consist of pixels in region $\bar{b}$ outside the blob. In Fig.~\ref{fig-ictheta}, we plot the inverse covariance between several positions and the rest of the sky (e.g., a set of rows of the inverse covariance matrix) in $\Lambda$CDM using the best-fit WMAP 7-year cosmological parameters, keeping only the first 50 multipole moments. It can be seen that the inverse covariance is only significant within a disk of radius $\sim 15^{\circ}$ around each of the template pixels. Therefore, we need not retain all of the pixels on the sky. In Fig.~\ref{fig-bnotb}, we depict the case where the likelihood is peaked for templates inside a region well-contained within the blob (shaded disk). The inverse covariance will be significant within a disk (dashed circle) of $\sim 15^{\circ}$ around each pixel where the template is non-zero (black dot). Our approximation neglects correlations between the template and the pixels contained within the dashed circle, but outside the blob (i.e., in region $\bar{b}$). The exponential will clearly yield a decreasing correction to the integral as the size of the blob is increased, becoming vanishingly small when the radius of the blob is $\sim 15^{\circ}$ larger than the size of templates near the maximum of the likelihood. For terms of higher order in $\ns$, the correction is slightly more complicated, but for blobs separated by a distance greater than $15^{\circ}$, the argument is the same. Another assumption we have made is encoded in Eq.~\ref{eq:correlationneglect2}: that the pixels in region $\bar{b}$ do not contribute to the inverse of the covariance inside region $b$. Further, if the actual source is well-contained within the blob, the likelihood will presumably peak in a region well-contained within the blob (which is the assumption behind our first approximation of integrating over region $b$ alone).  

The reasoning set out above provides qualitative support for the approximations that make this analysis feasible. To determine quantitatively how good these approximations are, we have performed three numerical tests. The first two tests are designed to determine whether correlations with pixels outside the blobs can indeed be discounted; the third determines the effects of restricting the position integral to within our candidate blobs.

\subsection{Tests of neglected correlations}

The ideal test of the effect of neglected correlations would be to perform evidence integrals for simulated collisions of varying sizes using covariance matrices ranging from patch-sized to full-sky. Unfortunately, memory restrictions mean we can only hold the full-sky covariance matrix in memory for \healpix\ resolutions smaller than $\nside = 64$. At this resolution each pixel is $\sim1^\circ$ across, so only large collisions are faithfully represented. As with the tests of the adaptive resolution analysis, we therefore split the test up, performing two tests: one checking the effect of neglected correlations on the largest scales, and the other at smaller scales.

The first test uses the map simulated for the largest-scale degradation tests, containing a $30^\circ$ bubble collision template placed at $(\theta_0, \phi_0) = (45.0^\circ, 45.0^\circ)$ with amplitude $z_0 = 8 \times 10^{-5}$ (as plotted in Fig.~\ref{fig:degrade_tests}). The simulation is passed to the candidate detection stage, and the patch corresponding to the bubble collision is singled out. The evidence ratio is then calculated twice at $\nsidedeg = 64$, first using the covariance matrix calculated for the patch, and second using the covariance matrix calculated on the full-sky. For clarity, the test is carried out without masking, and the integration limits on the template parameters are kept constant for both runs. The difference between evidence ratios returned will indicate the scale of any error induced by neglecting correlations at the largest scales.

The second test reuses the smallest map considered in the degradation tests (again, plotted in Fig.~\ref{fig:degrade_tests}), containing a $7^\circ$ bubble collision template with the same amplitude and position as in test one. As in the first test, the candidate detection algorithm is applied, and the feature containing the template is extracted. The evidence is calculated first using the standard patch size $15.4^\circ$ in radius, then using progressively larger patches of sky until the patch is $30^\circ$ in radius. At this point, Fig.~\ref{fig-ictheta} implies that the covariance matrix contains all pixels significantly correlated with those in the candidate collision region. The difference between evidence ratios will therefore indicate the errors associated with neglecting correlations on smaller scales. As with the first test, the integration limits used in each case are the same, as is the resolution ($\nsidedeg = 256$) at which the calculation is performed, and no mask is used.

The results of the two tests are presented in Tables~\ref{tab:assumption_test_1} and~\ref{tab:assumption_test_2}. In each case, increasing the patch size does not change the evidence value obtained beyond \multinest\ precision. This supports that the assumption that correlations outside of the patch can be neglected, and indicates that doing so does not add a significant source of systematic error, given the scale of the variations induced by the nested sampler.

\begin{table*}
\begin{tabular}{c c c}
\hline
\hline
\ $\theta_{\rm patch} (^\circ)$ \ & \ $\npix$ \ & \ $\log \rho$ \ \\
\hline
$60.1$ & $12619$ & $13.1 \pm 0.3$ \ \\
$180.0$ & $49149$ & $13.3 \pm 0.3$ \ \\
\hline
\hline
 \end{tabular}
 \begin{center}
 \caption{The evidence ratios obtained when a patch covariance matrix is used versus the full-sky covariance matrix. Note that the full-sky covariance matrix does not quite cover the entire sky: three pixels are left out. This is a consequence of the patch-based nature of the algorithm, and does not affect the conclusions. 
   \label{tab:assumption_test_1}}
 \end{center}
\end{table*}

\begin{table*}
\begin{tabular}{c c c}
\hline
\hline
\ $\theta_{\rm patch} (^\circ)$ \ & \ $\npix$ \ & \ $\log \rho$ \ \\
\hline
$15.4$ & $14670$ & $13.0 \pm 0.3$ \ \\
$20.0$ & $24167$ & $13.1 \pm 0.3$ \ \\
$25.0$ & $37388$ & $13.2 \pm 0.3$ \ \\
$30.0$ & $53338$ & $13.0 \pm 0.3$ \ \\
\hline
\hline
 \end{tabular}
 \begin{center}
 \caption{The evidence ratios obtained when the size of the patch covariance matrix is incrementally increased until all correlations are included. 
   \label{tab:assumption_test_2}}
 \end{center}
\end{table*}

\subsection{Test of localization of likelihood peaks}

The third assumption test is designed to assess the claim that the likelihood is peaked in position space, and that the evidence values obtained are unchanged when only the peaks are considered; this amounts to changing the limits of integration in Eq.~\ref{equation:rho_int_over_local_params}. This test uses the same $7^\circ$-bubble collision map as in the second test. This time, the patch radius is held constant at $30^\circ$, but the set of central positions sampled is incrementally increased until the template can be centered anywhere within the entire patch. The nested sampler therefore has access to greater portions of the collision environs -- by the fourth run it can sample central positions placing the templates entirely outside the simulated collision region -- and can provide an estimate of how much evidence is discarded by restricting the template position to lie within the blob.

\begin{figure}
\includegraphics[width=8.5cm]{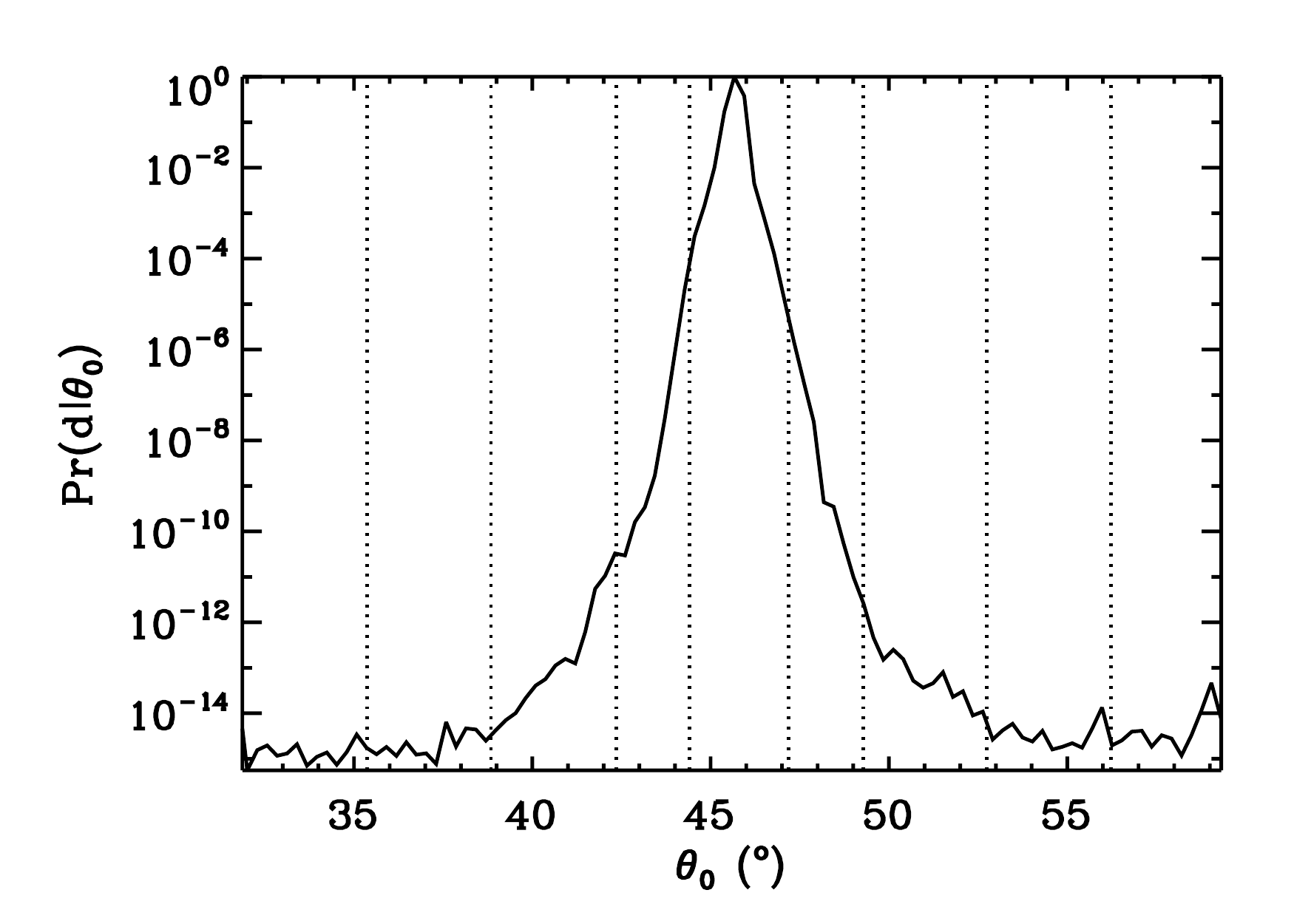}
\includegraphics[width=8.5cm]{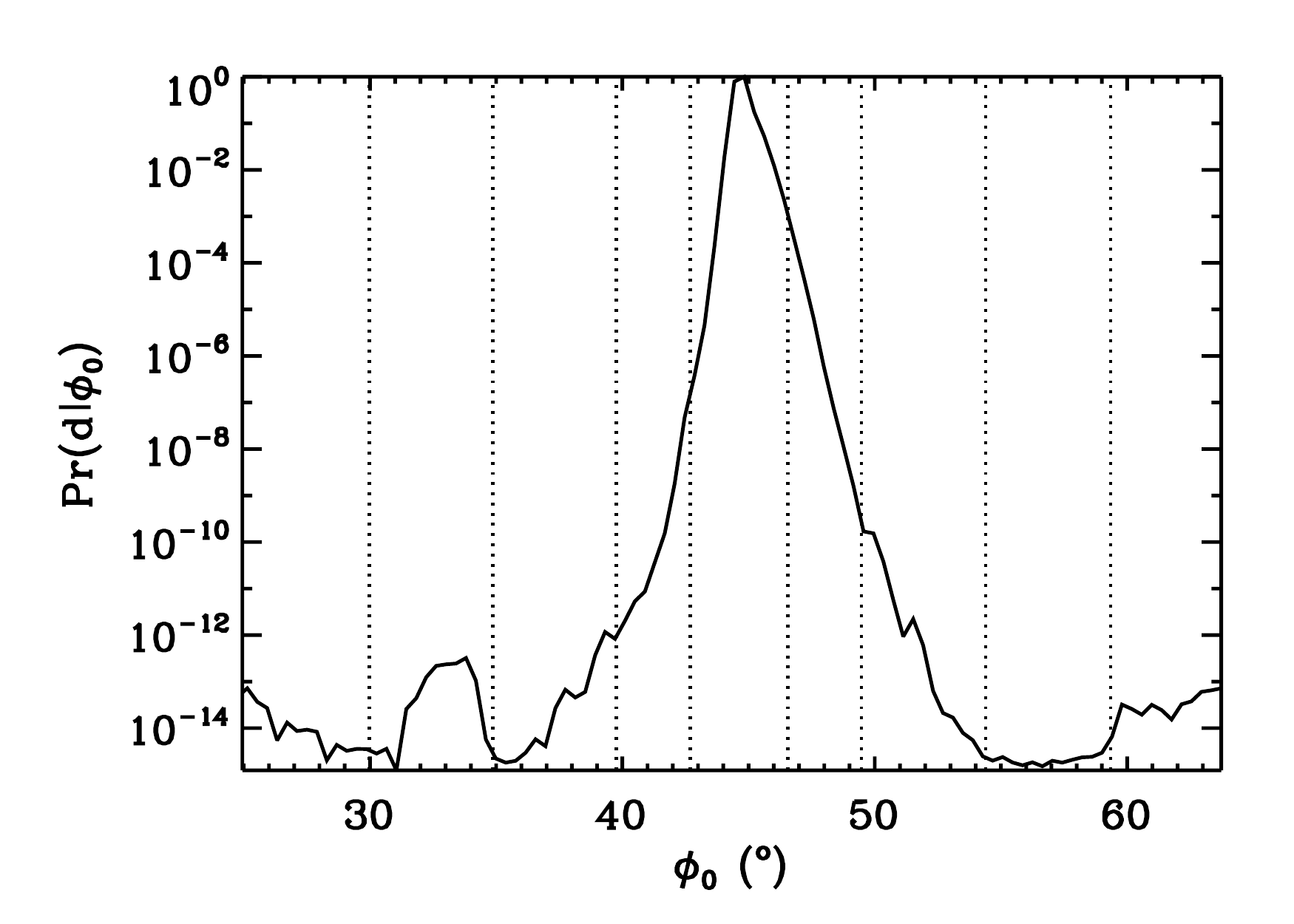}
\includegraphics[width=8.5cm]{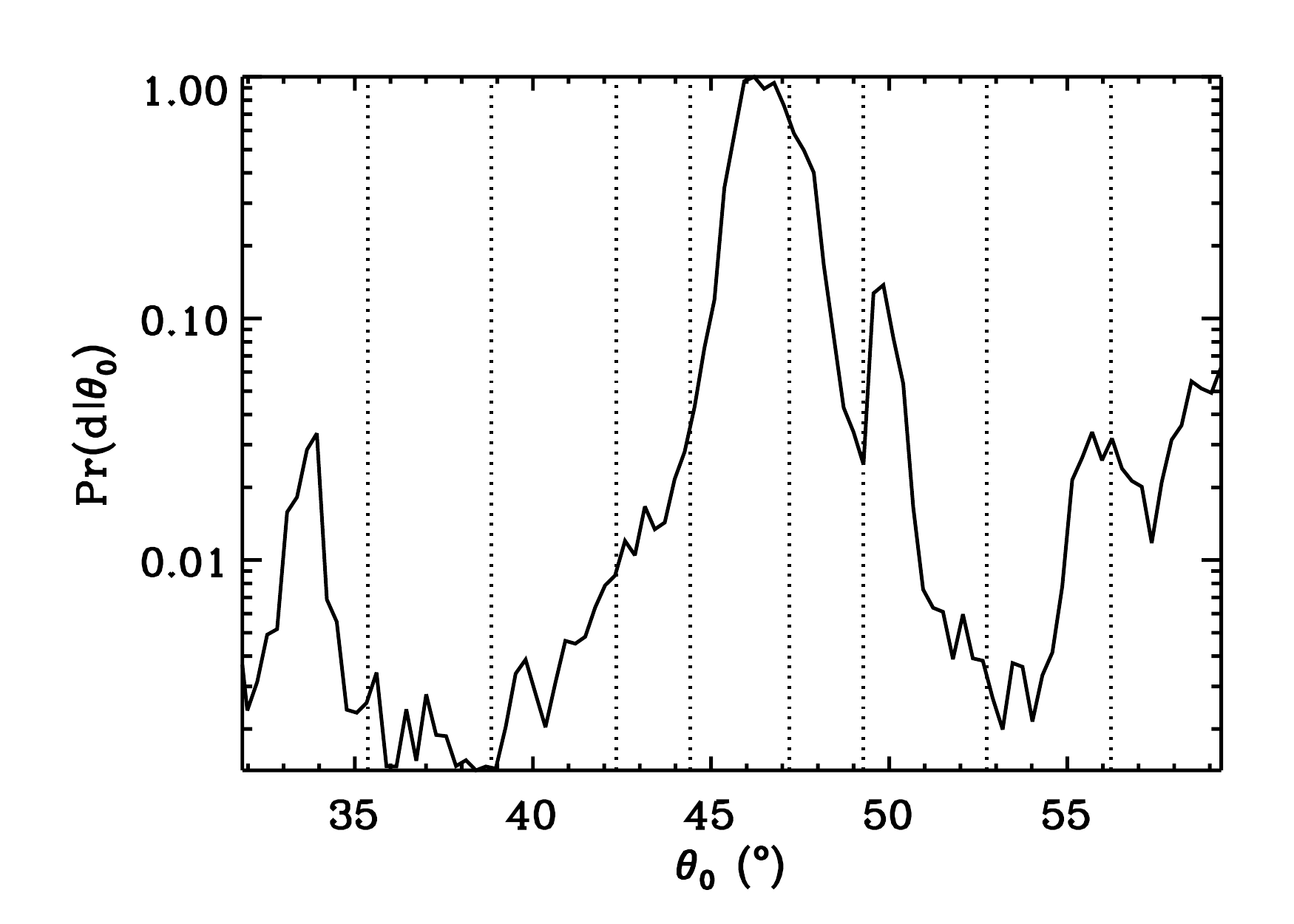}
\includegraphics[width=8.5cm]{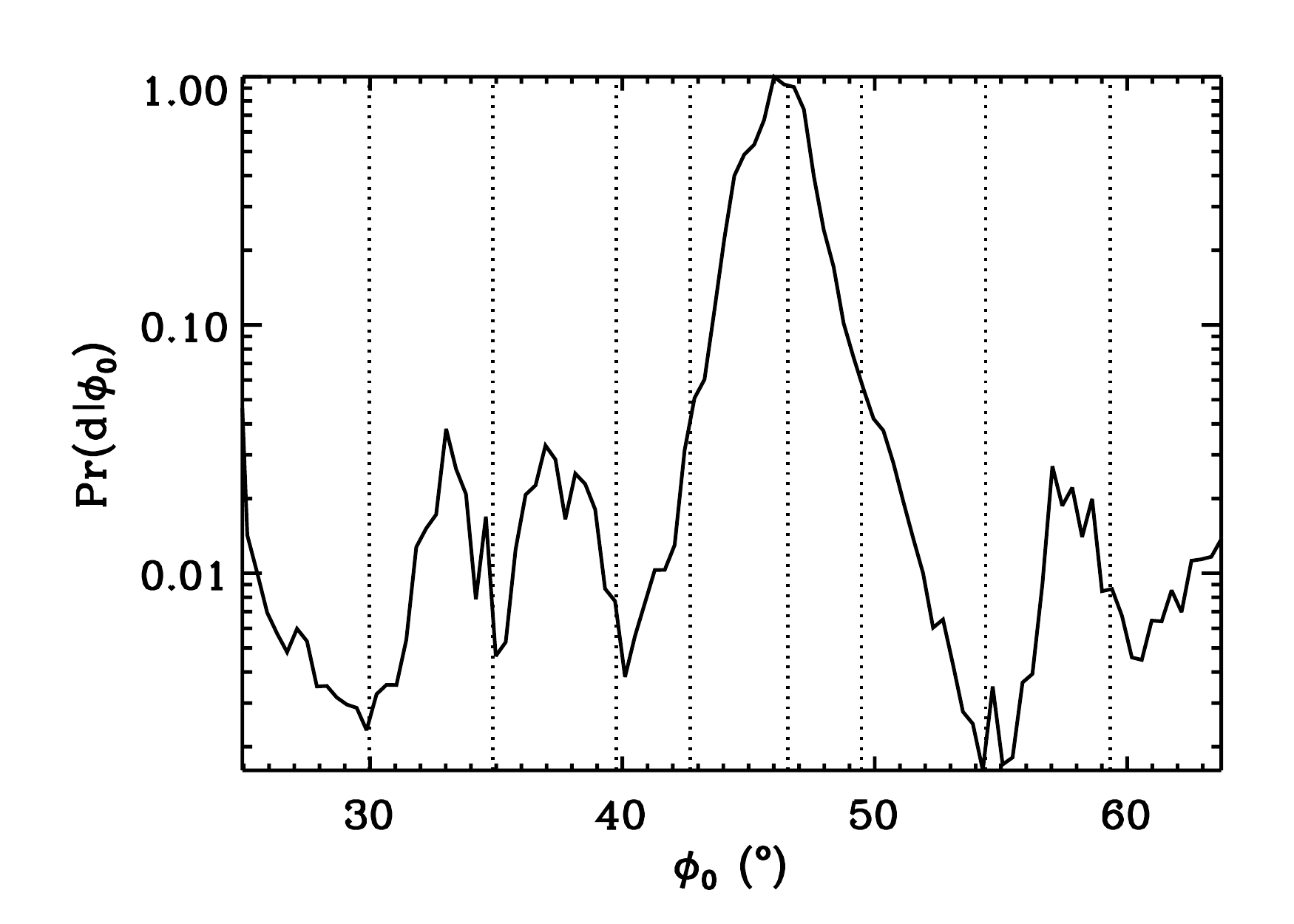}
\caption{The likelihood as a function of co-latitude $\theta_0$ (left) and longitude $\phi_0$ (right), plotted on a logarithmic scale. The top row is for a collision with $z_0 = 8 \times 10^{-5}$ (as used to calculate the localization-test evidence ratios); the bottom row is for a collision with $z_0 = 2.5 \times 10^{-5}$ (shown for comparison). Overplotted as dashed lines are the integration limits used in each of the five runs testing how the evidence changes as the central positions are sampled from larger regions. These limits correspond to sampling central positions from 10\%, 25\%, 50\%, 75\% and 100\% of the patch by angular radius. The likelihood is strongly peaked in both angular coordinates.
\label{fig:localization_test}
}
\end{figure}

The results of the test are reported in Table~\ref{tab:assumption_test_3}. The evidence ratio is stable, indicating that sampling from a larger range of central positions does not affect the outcome. This implies that the likelihood is indeed well-localized, supporting the assumption that the evidence integration need only be carried out over a restricted range of positions. The likelihood is plotted as a function of the two angular coordinates in Fig.~\ref{fig:localization_test}, and indeed is very strongly peaked about the position highlighted by the candidate detection stage. For comparison, we also plot a collision with the same $\theta_{\rm crit}$ but with an amplitude of $z_0 = 2.5 \times 10^{-5}$, which is near the lower-limit of our sensitivity (from Fig.~\ref{fig:mf_sensitivity} only $1\%$ of such sources would be flagged by the candidate-detection algorithm). Even in this case, the likelihood is strongly peaked near the centre of the collision.

\begin{table*}
\begin{tabular}{c c c}
\hline
\hline
\ Center Range $/ \,\theta_{\rm patch}$ \ & \ Center Range $(^\circ)$ \ & \ $\log \rho$ \ \\
\hline
$10\%$ & $1.4$ & $13.0 \pm 0.3$ \ \\
$25\%$ & $3.5$ & $13.0 \pm 0.3$ \ \\
$50\%$ & $7.0$ & $13.0 \pm 0.3$ \ \\
$75\%$ & $10.5$ & $13.1 \pm 0.3$ \ \\
$100\%$ & $14.0$ & $12.8 \pm 0.3$ \ \\
\hline
\hline
 \end{tabular}
 \begin{center}
 \caption{The evidence ratios obtained when the range of collision centers sampled is incrementally increased until the collision can be placed anywhere within the patch. 
   \label{tab:assumption_test_3}}
 \end{center}
\end{table*}

\section{Null test: Analysis of WMAP end-to-end simulation}\label{sec:null_test}

The WMAP data contain a number of components that cannot be included in the pixel-space covariance matrix; in particular, the foregrounds are not known precisely, and so their subtraction leaves behind unknown, highly anisotropic residuals. It is therefore important to apply the Bayesian analysis pipeline on a null dataset containing estimated foreground residuals -- and any other potential systematic effects -- to determine whether they generate false-positive results. The full end-to-end simulation of the WMAP experiment used to set the optimal-filter detection thresholds is perfectly suited to the task, as it is created from simulated time-ordered data, including foreground contamination, and is processed by exactly the same pipeline as the WMAP observations. 

The raw optimal filter analysis of the WMAP 7-year W-band end-to-end simulation (with the KQ85 mask applied) generates 19 bubble collision candidates and 10 texture candidates. Any candidates which are heavily masked are discarded, as are candidates whose estimated range of sizes has no overlap with the relevant prior on $\thetac$. Finally, any candidates which clearly correspond to a single feature are merged, leaving a set of 12 and 4 bubble collision and texture candidates, respectively. The sizes and locations of these candidates as estimated by the optimal filters are tabulated in Table~\ref{tab:e2e_bayes_results}.

\begin{table*}
\begin{tabular}[t]{c c c c c}
\hline
\hline
\ ID \ & \ $\thetac$ range $(^\circ)$ \ & \ $\theta_0 (^\circ)$ \ & \ $\phi_0 (^\circ)$ \ & \ $\log \rho$ \ \\
\hline
1 & $2$-$4$ & 63.1 & 142.0 & $-9.68 \pm 0.3$ \\
2 & $3$-$5$ & 50.2 & 221.5 & $-10.87 \pm 0.3$ \\
3 & $6$-$8$ & 126.4 & 214.8 & $-8.75 \pm 0.3$ \\
4 & $8$-$12$ & 43.8 & 152.7 & $-8.57 \pm 0.3$ \\
5 & $10$-$14$ & 19.1 & 326.3 & $-9.21 \pm 0.3$ \\
6 & $14$-$18$ & 139.7 & 11.3 & $-7.43 \pm 0.3$ \\
7 & $20$-$26$ & 137.0 & 42.7 & $-6.74 \pm 0.3$ \\
8 & $30$-$40$ & 56.8 & 154.0 & $-6.01 \pm 0.3$ \\
9 & $35$-$45$ & 118.6 & 232.4 & $-5.83 \pm 0.3$ \\
10 & $60$-$70$ & 102.0 & 326.6 & $-6.43 \pm 0.3$ \\
11 & $80$-$90$ & 86.1 & 305.9 & $-6.64 \pm 0.3$ \\
12 & $85$-$90$ & 53.9 & 19.7 & $-5.43 \pm 0.3$ \\
\hline
\hline
\end{tabular}
\
\
\
\
\
\begin{tabular}[t]{c c c c c}
\hline
\hline
\ ID \ & \ $\thetac$ range ($^\circ$)\ & \ $\theta_0 (^\circ)$ \ & \ $\phi_0 (^\circ)$ \ & \ $\log \rho$ \ \\
\hline
1 & $2$-$5$ & 50.2 & 221.5 & $-4.41 \pm 0.3$ \\
2 & $4$-$9$ & 45.8 & 153.1 & $-6.01 \pm 0.3$ \\
3 & $4$-$15$ & 143.2 & 3.2 & $-3.99 \pm 0.3$ \\
4 & $7$-$50$ & 38.7 & 94.8 & $-6.10 \pm 0.3$ \\
\hline
\hline
 \end{tabular}
 \begin{center}
 \caption{The size ranges and locations of the final 12 candidate bubble collisions (left) and 4 candidate textures (right) found in the end-to-end simulation of the WMAP experiment, along with their patch evidence ratios. The angular positions are related to Galactic longitude, $l$, and latitude, $b$, by the transformations $l = \phi$ and $b =90^{\circ} - \theta$.
   \label{tab:e2e_bayes_results}}
 \end{center}
\end{table*}

\begin{figure}
\includegraphics[width=8.0cm]{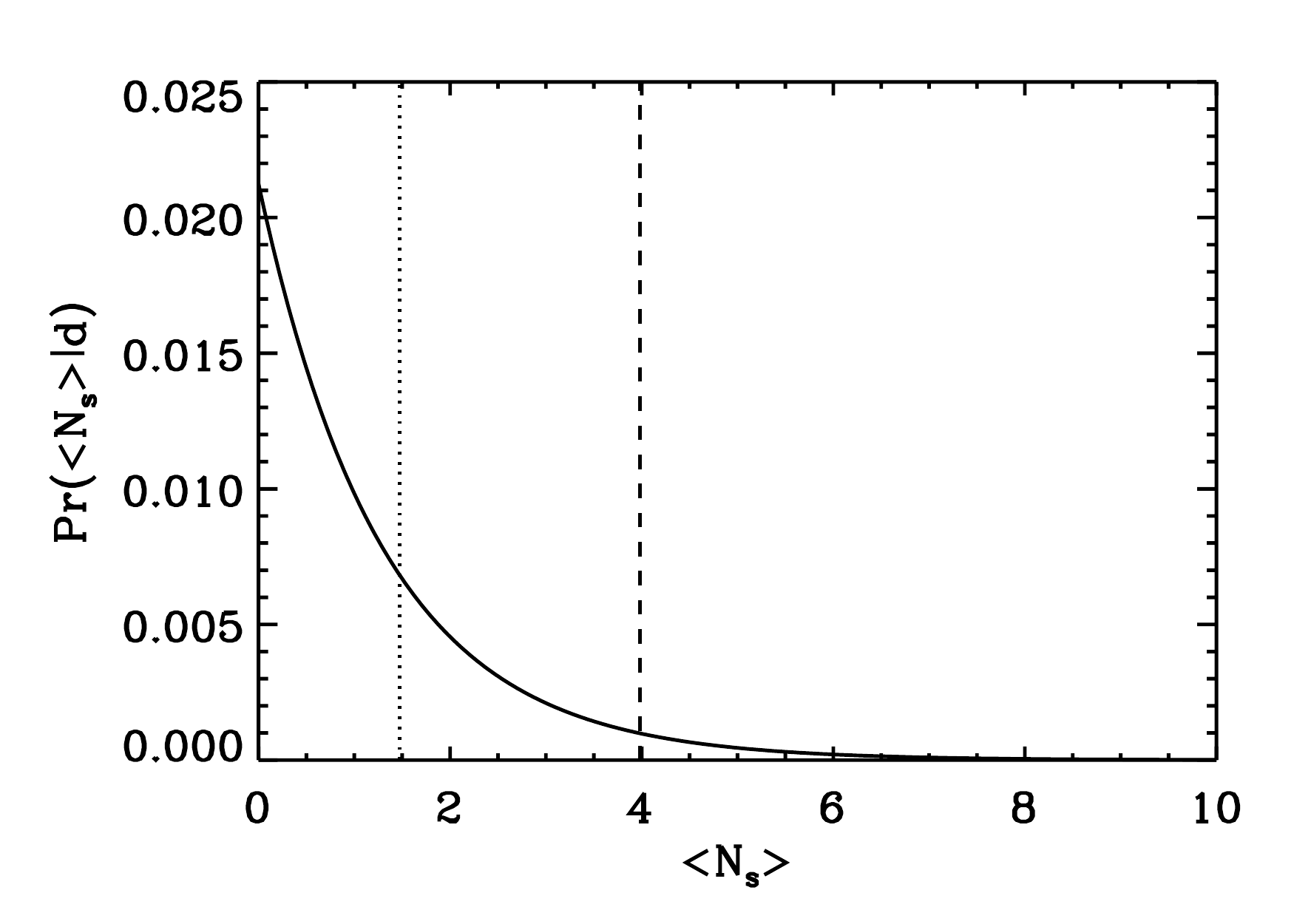}
\includegraphics[width=8.0cm]{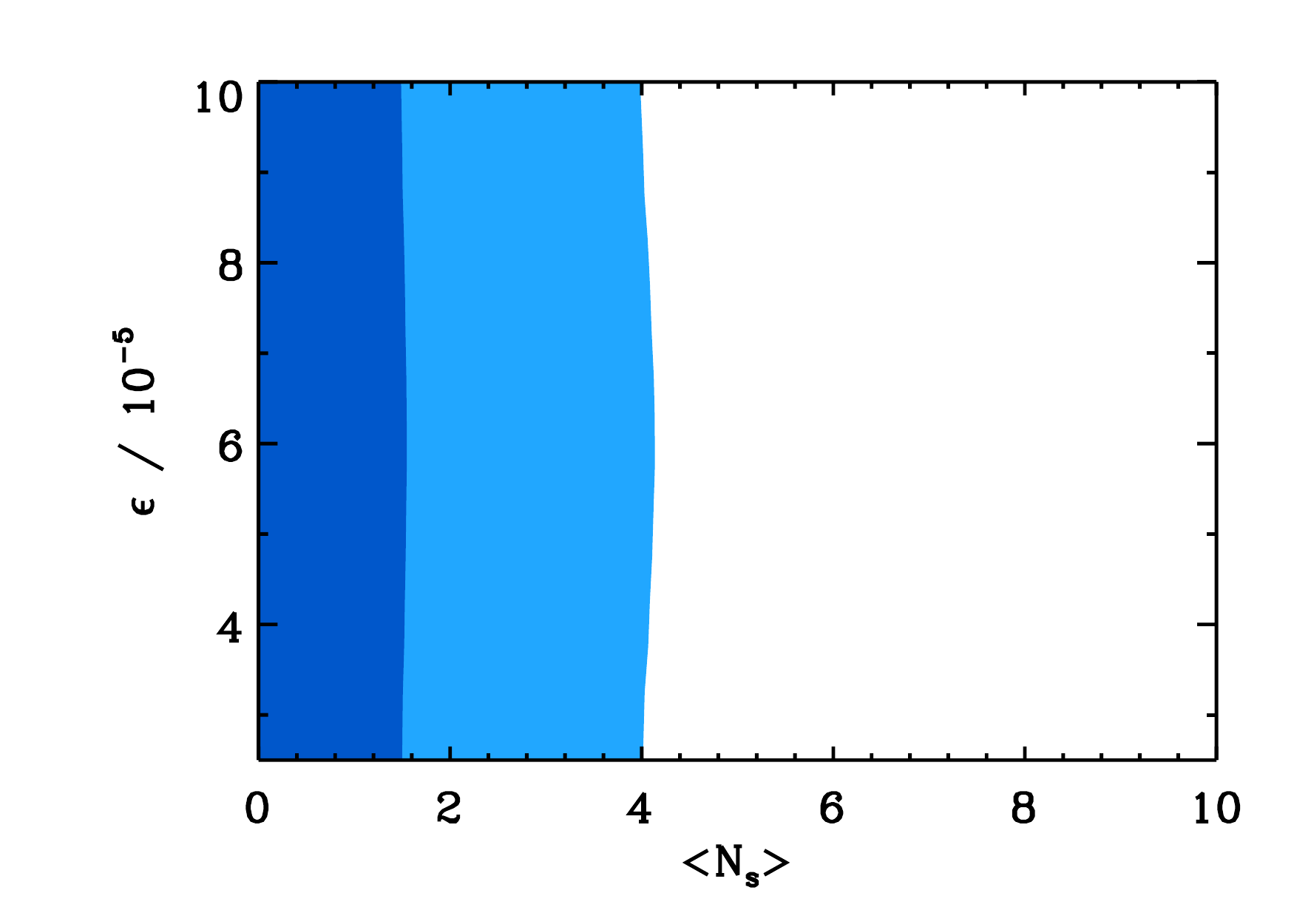}
\caption{The posterior probabilities of the global parameters of the bubble collision (left) and texture (right) models, given the end-to-end simulation of the WMAP 7-year W-band. The posterior is plotted as a function of one parameter, $\nsavge$, for bubble collisions, and two parameters, $\nsavge$ and $\epsilon$, for textures. The regions containing 68\% and 95\% of the posterior probability are indicated by the dotted and dashed lines in the bubble collision plot, and as dark and light regions in the texture plot. Both posteriors are strongly peaked at zero sources.
\label{fig:e2e_global_posteriors}
}
\end{figure}

The patch evidence ratios obtained by applying the adaptive-resolution Bayesian evidence calculation to these candidates are also presented in Table~\ref{tab:e2e_bayes_results}. As expected, no candidate exhibits strong evidence in favor of either the bubble collision or texture model: the maximum evidence ratios are $e^{-5.4}$ and $e^{-4.0}$, respectively. Merging these results produces the posteriors on the global model parameters as plotted in Fig.~\ref{fig:e2e_global_posteriors}. The posteriors for both models are clearly peaked at $\nsavge = 0$, confirming our prior knowledge that there are no bubble collisions or textures in the end-to-end simulation of the WMAP W-band data. The null test therefore indicates that we should not expect un-modeled foreground residuals and unknown systematics to generate false positives in the WMAP data. Further, the maximum evidence ratios obtained provide indicators of the level of response foregrounds and systematics can produce: any features found in the data should exceed these values in order to be considered interesting.

\section{Analysis of WMAP 7-Year data}\label{sec:wmap_analysis}

\begin{figure}
\includegraphics[width=8.0cm]{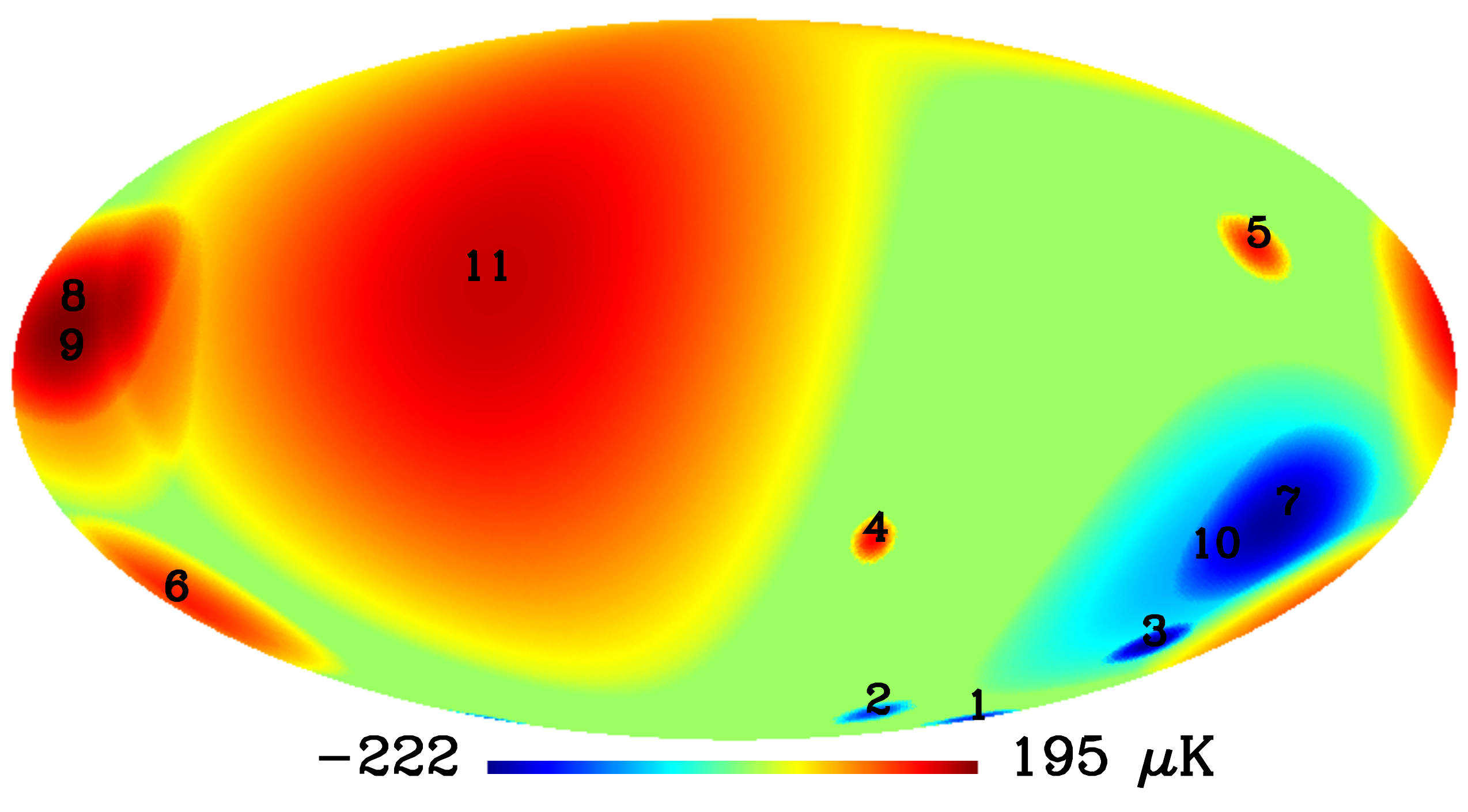}
\includegraphics[width=8.0cm]{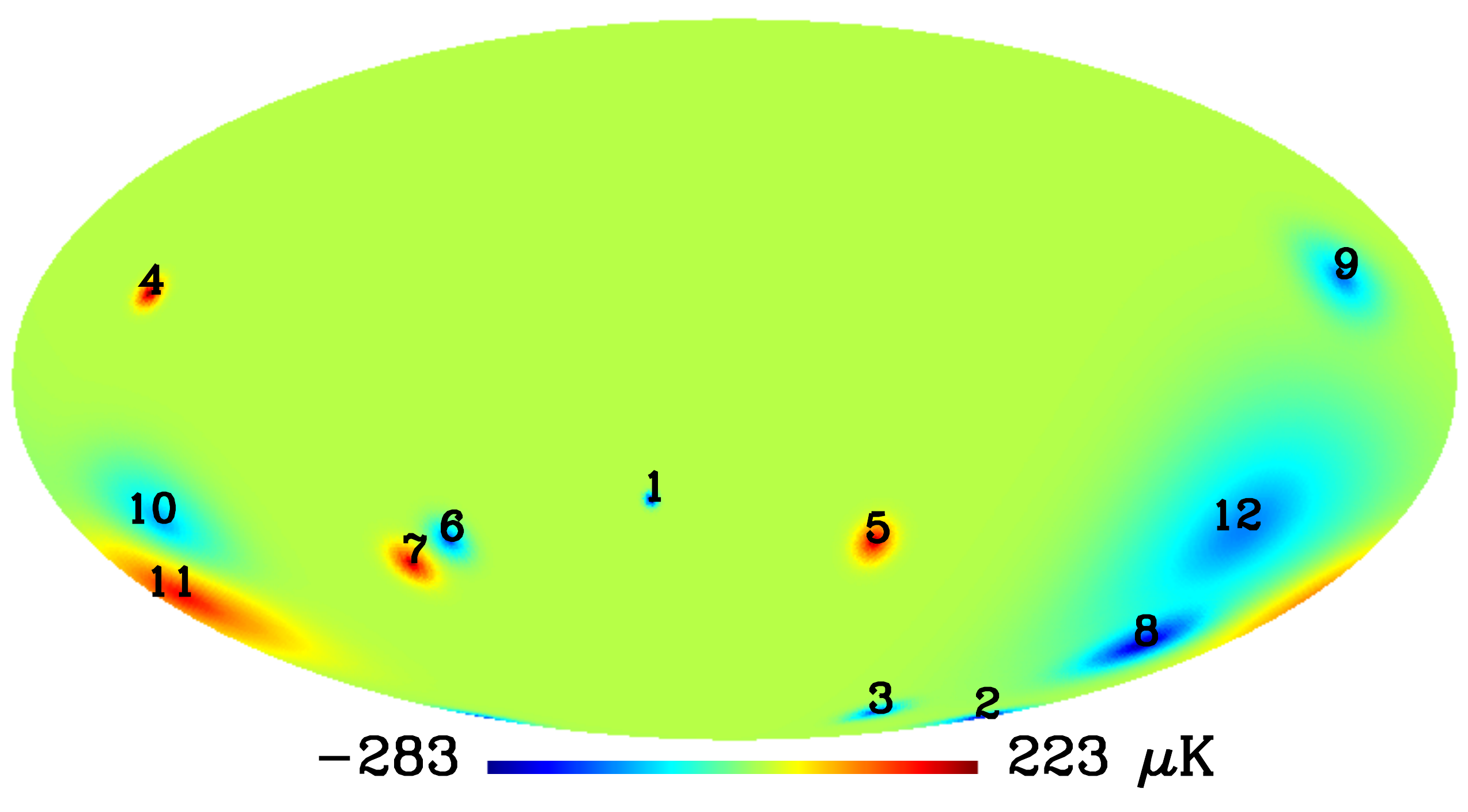}
\includegraphics[width=8.0cm]{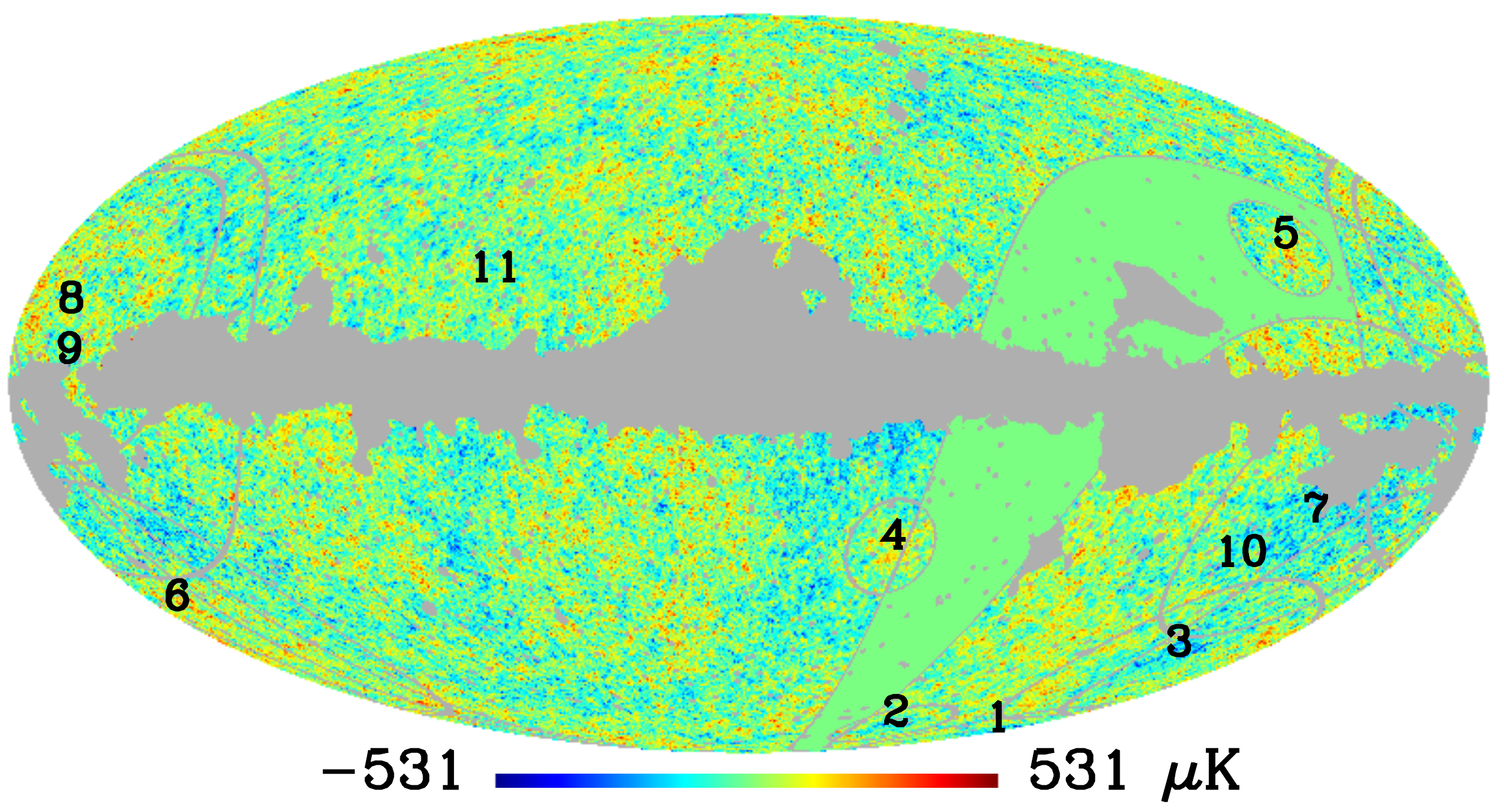}
\includegraphics[width=8.0cm]{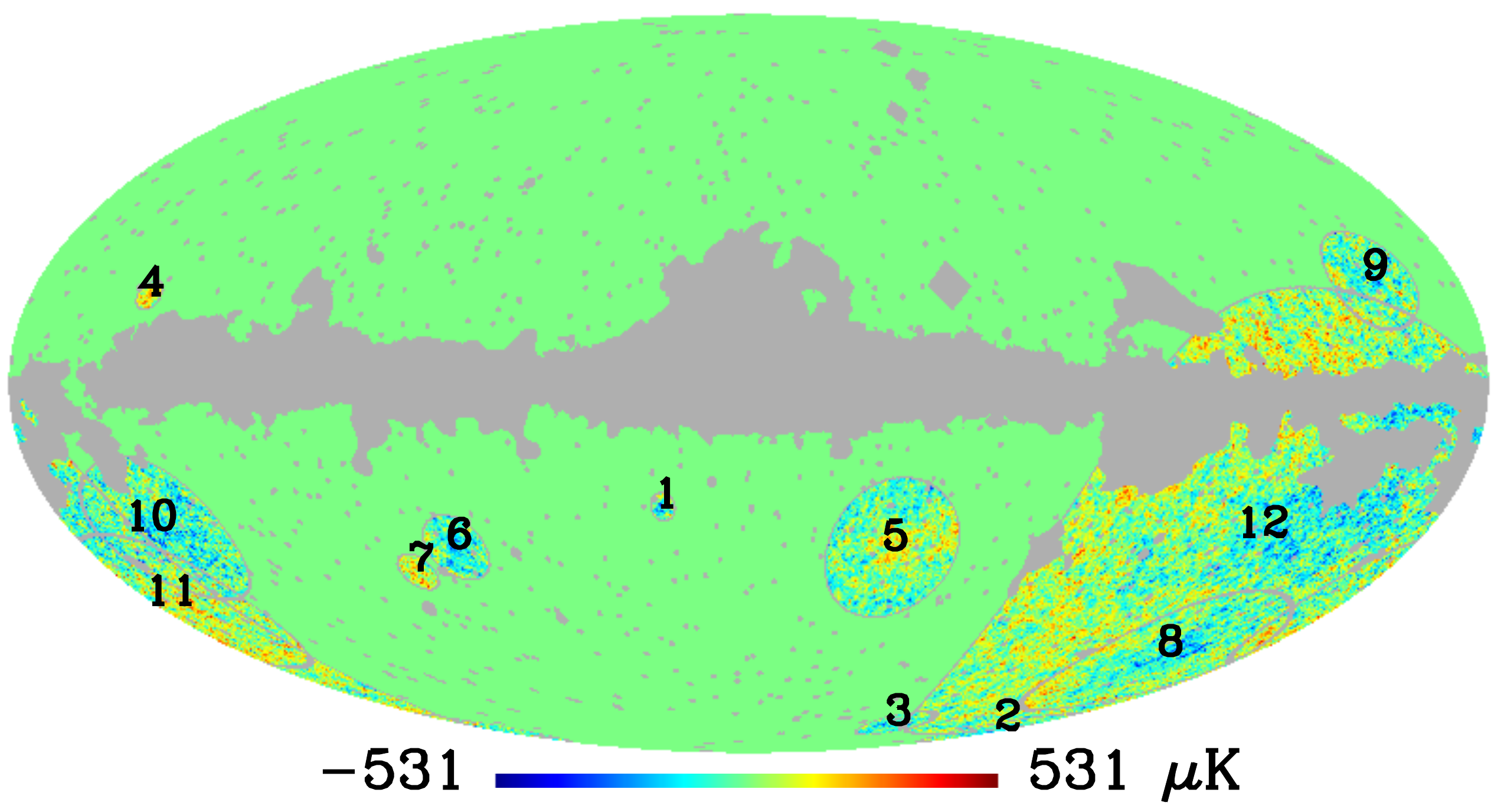}
\caption{Top: the estimated sizes and amplitudes of the bubble collision (left) and texture (right) candidates located in the WMAP 7-year data by the optimal filters. Bottom: the patches of WMAP 7-year data passed to the Bayesian evidence calculation for each of these candidates. The bubble collision plot shows all of the data involved in the evidence calculation for each candidate; for clarity, the texture plot only shows the core region of each patch. Note that the plots of the optimal filter candidates (top) contain only the estimated contributions due to additional sources, and the temperature ranges therefore differ from the plots of the WMAP data (bottom).
\label{fig:wmap7_candidates}
}
\end{figure}

Our analysis of the WMAP 7-year W-band foreground-reduced temperature map (with the KQ85 mask applied) produces a total of 32 bubble collision candidates and 33 texture candidates. The candidates' locations are visually inspected, and those which are mostly obscured by the mask are discarded; candidates found to have no overlap with the relevant priors on $\thetac$ are likewise cut. Any candidates which are obviously coincident are merged at this point. The remaining candidates are then required to also be significant in an optimal filter analysis of the WMAP 7-year V-band foreground-reduced temperature map. This simple check requires that each feature is interesting across a range of frequencies, indicating that it is not due to foregrounds. This final cut leaves a set of 11 and 12 bubble collision and texture candidates respectively. The most probable sizes and locations of these candidates are tabulated in Table~\ref{tab:wmap7_bc_results} and plotted in Fig.~\ref{fig:wmap7_candidates}.

\begin{table*}
\begin{tabular}[t]{c c c c c}
\hline
\hline
\ ID \ & \ $\thetac$ range ($^\circ$) \ & \ $\theta_0 (^\circ)$ \ & \ $\phi_0 (^\circ)$ \ & \ $\log \rho$ \ \\
\hline
1 & $2$-$6$ & 169.0 & 187.5 & $-8.82 \pm 0.3$ \\
2 & $3$-$5$ & 167.2 & 268.7 & $-7.79 \pm 0.3$ \\
3 & $4$-$14$ & 147.4 & 207.1 & $-5.80 \pm 0.3$ \\
4 & $4$-$8$ & 123.2 & 321.3 & $-7.17 \pm 0.3$ \\
5 & $6$-$8$ & 62.7 & 220.4 & $-9.33 \pm 0.3$ \\
6 & $14$-$22$ & 136.6 & 176.5 & $-5.96 \pm 0.3$ \\
7 & $12$-$20$ & 118.0 & 212.0 & $-7.39 \pm 0.3$ \\
8 & $20$-$24$ & 75.8 & 168.8 & $-9.46 \pm 0.3$ \\
9 & $28$-$35$ & 85.8 & 166.3 & $-7.04 \pm 0.3$ \\
10 & $22$-$40$ & 126.8 & 220.1 & $-5.67 \pm 0.3$ \\
11 & $80$-$90$ & 69.6 & 61.9 & $-6.66 \pm 0.3$ \\
\hline
\hline
\end{tabular}
\
\
\
\
\
\begin{tabular}[t]{c c c c c}
\hline
\hline
\ ID \ & \ $\thetac$ range ($^\circ$)\ & \ $\theta_0 (^\circ)$ \ & \ $\phi_0 (^\circ)$ \ & \ $\log \rho$ \ \\
\hline
1 & $1$-$3$ & 114.6 & 22.1 & $-6.08 \pm 0.3$ \\
2 & $1.5$-$5$ & 168.7 & 184.4 & $-3.12 \pm 0.3$ \\
3 & $1.5$-$4$ & 166.8 & 268.8 & $-4.58 \pm 0.3$ \\
4 & $1.5$-$3$ & 72.4 & 150.1 & $-5.57 \pm 0.3$ \\
5 & $2$-$15$ & 123.5 & 321.0 & $-3.31 \pm 0.3$ \\
6 & $2$-$7$ & 123.2 & 79.5 & $-4.73 \pm 0.3$ \\
7 & $2$-$4$ & 128.3 & 93.5 & $-5.56 \pm 0.3$ \\
8 & $2$-$15$ & 147.4 & 210.2 & $-2.06 \pm 0.3$ \\
9 & $5$-$10$ & 69.3 & 202.1 & $-8.04 \pm 0.3$ \\
10 & $8$-$15$ & 119.3 & 155.7 & $-7.71 \pm 0.3$ \\
11 & $4$-$15$ & 135.4 & 172.1 & $-5.48 \pm 0.3$\\
12 & $7$-$50$ & 120.7 & 220.4 & $-5.84 \pm 0.3$ \\
\hline
\hline
 \end{tabular}
 \begin{center}
 \caption{The size ranges and locations of the final 11 candidate bubble collisions (left) and 12 candidate textures (right), along with their patch evidence ratios. The size ranges tabulated are those derived from the optimal filter analysis; where they extend beyond the relevant priors they are truncated before the evidence calculation. The angular positions are related to Galactic longitude, $l$, and latitude, $b$, by the transformations $l = \phi$ and $b =90^{\circ} - \theta$. Note that, although bubble collision candidates 8 and 9 overlap considerably, their positions and size ranges are sufficiently discrepant to justify processing individually.
   \label{tab:wmap7_bc_results}}
 \end{center}
\end{table*}

\begin{figure}
\includegraphics[width=8.0cm]{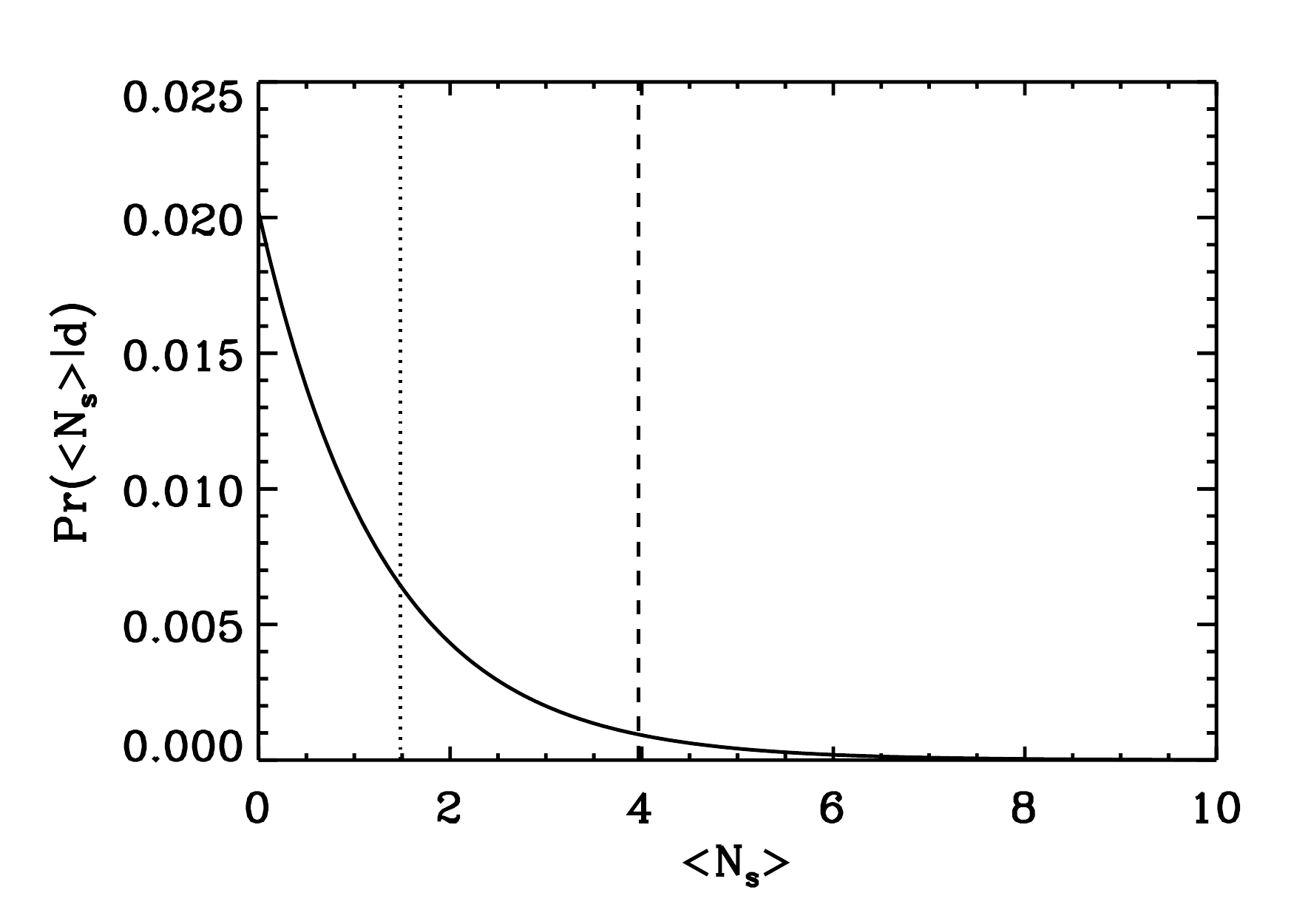}
\includegraphics[width=8.0cm]{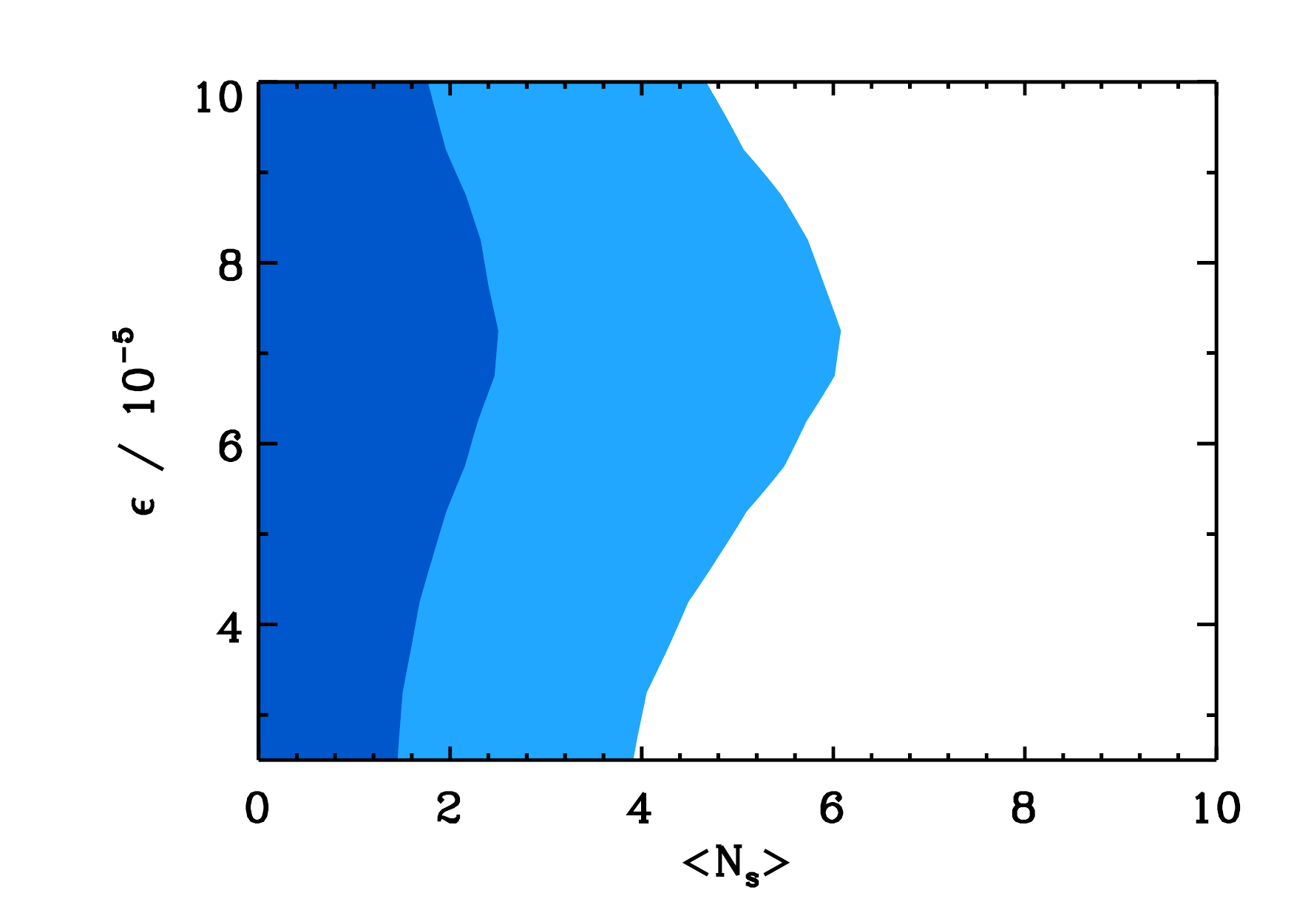}
\caption{The posterior probabilities of the global parameters of the bubble collision (left) and texture (right) models, given the WMAP 7-year data. The posterior is plotted as a function of one parameter, $\nsavge$, for bubble collisions, and two parameters, $\nsavge$ and $\epsilon$, for textures. The most probable regions containing 68\% and 95\% of the posterior probability are indicated by the dotted and dashed lines in the bubble collision plot, and as dark and light regions in the texture plot. Both posteriors are strongly peaked at $\nsavge = 0$.
\label{fig:wmap7_global_posteriors}
}
\end{figure}

Applying the adaptive-resolution evidence calculation to the candidates produces the patch evidence ratios also reported in Table~\ref{tab:wmap7_bc_results}. No single candidate is strong enough to claim a detection on its own. However, as demonstrated in Refs.~\cite{Feeney_etal:2010dd,Feeney_etal:2012jf}, it is possible for a number of weak candidates to favor the addition of relics to $\Lambda$CDM even if their individual evidence ratios are less than one: only by combining the results obtained for all candidates can the overall predictive power of the underlying model be revealed. The posteriors on the global parameters of the bubble collision and texture models, derived by combining the results from the candidates, are plotted in Fig.~\ref{fig:wmap7_global_posteriors}: both posteriors are peaked at zero sources. The texture model's dimensionless scale of symmetry breaking is constrained to be $2.6 \times 10^{-5} \le \epsilon \le 1.0 \times 10^{-4}$ (at 95\% confidence), which, as the prior is defined only within the range $2.5 \times 10^{-5} \le \epsilon \le 1.0 \times 10^{-4}$, indicates that the WMAP data do not provide any interesting constraint on this parameter.

The WMAP 7-year data do not favor the addition of either bubble collisions or textures to $\Lambda$CDM. As none of the candidates exhibits significant evidence for the addition of sources to $\Lambda$CDM, we do not check the candidates for foreground residuals.

\section{Discussion}\label{sec:new_vs_old}

In Refs~\cite{Feeney_etal:2010jj,Feeney_etal:2010dd} and~\cite{Feeney_etal:2012jf}, searches for bubble collisions and textures using earlier versions of the Bayesian source detection pipeline were published. Each previous analysis shares a number of candidate features in common with the current analysis, allowing consistency checks to be carried out between versions of the pipeline. Comparing results between versions is non-trivial, and must take into account each change made to the algorithm. In particular:

\begin{enumerate}

\item The prior on the bubble collision size has changed from uniform in the range $2$-$11.25^\circ$ to being proportional to $\sin{\thetac}$ in the range $2-90^\circ$. {\em Ceteris paribus}, this will reduce evidence ratios previously reported for bubble collision candidates, particularly those at small scales.

\item The bubble collision template previously allowed for a discontinuity at the template boundary with amplitude $\zc$. This parameter is now set to zero due to updated theoretical results~\cite{Gobbetti_Kleban:2012,Kleban_Levi_Sigurdson:2011}, and the bubble collision model considered in this analysis is consequently nested within the model considered previously. The effects of removing the edge can be determined exactly using the Savage-Dickey Density Ratio~\cite{Dickey:1971}: the change in evidence will be the ratio of the posterior and prior probabilities of the edge amplitude, evaluated at $\zc = 0$ using the results of the previous analysis, i.e.,
\begin{equation*}
\Delta \log \rho = \log \frac{ {\rm Pr}(\zc | {\bf d}, {\rm old}) }{ {\rm Pr}(\zc, {\rm old}) } \bigg |_{\zc = 0}.
\end{equation*}
As there was little evidence to support the edge parameter in the earlier analysis, the ratio of posterior to prior at $\zc = 0$ is typically $\sim 10$, and the new evidence ratios are boosted accordingly.

\item The new analysis replaces the WMAP 7-year KQ75 mask used in all prior analyses with the KQ85 mask, revealing $\sim 8 \%$ more of the sky. The change in the fraction of the sky available to the algorithm increases the prior volume on observable source positions by $\sim 8 \%$ as well, and the log evidence ratios hence decrease by a similar amount.

\item The candidate detection method has changed from needlets to optimal filters, and the ranges of size and position deemed significant therefore also change. The integration limits for the patch evidences can differ by small amounts if the new ranges either reveal or truncate regions of non-zero likelihood.

\end{enumerate}
In addition, since the first bubble collision analysis we have used an increased number of \multinest\ live points and tighter tolerance and efficiency settings. The current settings were chosen to ensure accurate calculation of the evidence; however, tests show that there is no significant difference in our results due to the new settings.

Table~\ref{tab:old_vs_new} shows the expected and observed changes in evidence ratio obtained for the four best bubble collision and texture candidates processed in the previous analyses. In the majority of cases the differences between the observed and expected changes in evidence ratios are consistent to \multinest\ precision, but there are two bubble collision cases where the new results show a significant difference.

\begin{table*}
\begin{tabular}{c c c c c}
\hline
\hline
source \ & \ new ID \ & \ old ID \ & \ expected $\Delta \log \rho$ \ & \ observed $\Delta \log \rho$ \ \\
\hline
bubble & 3 & 2 & $-2.1$ & $-1.2$ \\
bubble & 4 & 3 & $-4.1$ & $-3.1$ \\
bubble & 1 & 7 & $-4.0$ & $-3.4$ \\
bubble & 2 & 10 & $-3.8$ & $-4.0$ \\
texture & 2 & 6 & $-0.1$ & $-0.5$ \\
texture & 3 & 8 & $-0.1$ & $-0.4$ \\
texture & 8 & 9 & $-0.1$ & $-0.2$ \\
texture & 5 & 10 & $-0.1$ & $-0.2$ \\
\hline
\hline
 \end{tabular}
 \begin{center}
 \caption{The expected and observed differences in evidence ratio found for the four best bubble collision and texture candidates between the previous incarnation of the pipeline and the present analysis. Included in the expected change are the new form of the $\thetac$ prior for the bubble collisions, the removal of the edge from the bubble collision template, and the change in the mask. Note that the evidence ratios and IDs of the texture candidates were not previously published.
   \label{tab:old_vs_new}}
 \end{center}
\end{table*}

The first case is the Cold Spot, candidate 3. The memory restrictions of the previous analysis~\cite{Feeney_etal:2010jj,Feeney_etal:2010dd} required that the $\thetac$ range sampled be truncated (compare Tables VII in Ref.~\cite{Feeney_etal:2010dd} and~\ref{tab:wmap7_bc_results} in the current analysis). The new adaptive-resolution algorithm allows the full range of $\thetac$ estimated by the candidate detection stage to be sampled, revealing an additional peak in the posterior and boosting the evidence accordingly. The second case is candidate 4. The patches of data used to calculate the evidence for this candidate contain a $\sim4^\circ\times3^\circ$ region masked by the KQ75 mask but not by the KQ85 mask. The improvement in evidence ratio most likely derives from uncovering these extra pixels, which produce an extra hot contribution to an already hot feature.

The only differences between the current texture analysis and that of Ref.~\cite{Feeney_etal:2012jf} are the candidate detection algorithm and the mask employed. The candidates that contributed most significantly to the posterior in the previous analysis are all still detected by the optimal filters, and the changes in evidence observed are consistent within \multinest\ precision, given the new mask and small differences in integration limits from changing the candidate detection stage. Indeed, the posteriors produced by the two analyses (Fig. 2 of Ref.~\cite{Feeney_etal:2012jf} and the right-hand plot of Fig.~\ref{fig:wmap7_global_posteriors}) are almost indistinguishable by eye.


\section{Conclusions}\label{sec:conclusions}

We have presented a hierarchical Bayesian algorithm for the detection of spatially-localized sources in high-resolution CMB datasets. The algorithm uses the posterior over the global parameters describing the population of sources to determine whether their presence is warranted by the data and prior theoretical knowledge. To cope with the volume of data available, a conservative approximation to the posterior is calculated by selecting the most likely candidate sources using optimal filters and assuming that the remaining data do not contribute to the likelihood. Candidates are processed at the highest data resolution possible, given their size and the computing power available. The effects of the approximations and adaptive-resolution analysis have been quantified using a suite of tests, and are found to be comparable to the typical variance in sampling from the un-approximated posterior.

As a demonstration, the pipeline has been applied to search for evidence of bubble collisions and cosmic textures in the WMAP 7-year data. This work removes the size restriction imposed by memory constraints on the previous bubble collision analysis~\cite{Feeney_etal:2010dd,Feeney_etal:2010jj}, as well as optimising the candidate detection stages of previous bubble collision and texture analyses~\cite{Feeney_etal:2010dd,Feeney_etal:2010jj,Feeney_etal:2012jf}. The WMAP data do not favor the addition of either bubble collisions or cosmic textures to the $\Lambda$CDM model: even though such sources provide higher-likelihood fits, they are not sufficiently predictive to overcome the extra model complexity. In the context of these models our results also place limits on the average numbers of bubble collisions and textures within our detection thresholds on the CMB sky, which are constrained to be fewer than 4.0 and 5.2 at 95\% confidence, respectively. The WMAP data do not place any significant constraint on the dimensionless scale of symmetry breaking for textures, $\epsilon$.

The Planck satellite~\cite{Tauber2010} will soon release temperature data with a factor of 2-3 improvement in resolution and $\sim10$ in pixel noise over WMAP. Further, it will extract essentially all of the information from the temperature power spectrum, providing a near-ideal characterization of the dominant source of noise in the analysis. These facts strongly motivate performing the texture and bubble collision analyses on the Planck data when they become available. In addition, high-quality CMB polarization data are being gathered by experiments such as Planck, ACTPol~\cite{Niemack_etal:2010}, SPTPol~\cite{McMahon_etal:2009} and Spider~\cite{Crill_etal:2008}. Textures do not induce a polarization signal~\cite{2011MNRAS.410...33V}, but bubble collisions are expected to create characteristic imprints~\cite{Czech:2010rg}, complementary to those in the temperature data. Extension of the hierarchical Bayesian analysis pipeline to process polarization data, either in isolation or by cross-correlating with temperature maps, therefore represents a promising avenue for future tests of these models. Further, such an upgraded pipeline could be readily applied to other localized signatures in the CMB, such as the Sunyaev-Zel'dovich effect produced by clusters of galaxies.


\acknowledgements
This work was partially supported by a grant from the Foundational Questions Institute (FQXi) Fund, a donor-advised fund of the Silicon Valley Community Foundation on the basis of proposal FQXi-RFP3-1015 to the Foundational Questions Institute. SMF is supported by the Perren Fund and STFC. Research at Perimeter Institute is supported by the Government of Canada through Industry Canada and by the Province of Ontario through the Ministry of Research and Innovation. JDM is supported by a Newton International Fellowship from the Royal Society and the British Academy and, during this work, was also supported by a Leverhulme Early Career Fellowship from the Leverhulme Trust. HVP is supported by STFC and the Leverhulme Trust. We acknowledge use of the \healpix\ package and the Legacy Archive for  Microwave Background Data Analysis (LAMBDA).  Support for LAMBDA is provided by the NASA Office of Space Science.


\bibliography{texvsbub1}

\begin{thebibliography}{55}%
\makeatletter
\providecommand \@ifxundefined [1]{%
 \@ifx{#1\undefined}
}%
\providecommand \@ifnum [1]{%
 \ifnum #1\expandafter \@firstoftwo
 \else \expandafter \@secondoftwo
 \fi
}%
\providecommand \@ifx [1]{%
 \ifx #1\expandafter \@firstoftwo
 \else \expandafter \@secondoftwo
 \fi
}%
\providecommand \natexlab [1]{#1}%
\providecommand \enquote  [1]{``#1''}%
\providecommand \bibnamefont  [1]{#1}%
\providecommand \bibfnamefont [1]{#1}%
\providecommand \citenamefont [1]{#1}%
\providecommand \href@noop [0]{\@secondoftwo}%
\providecommand \href [0]{\begingroup \@sanitize@url \@href}%
\providecommand \@href[1]{\@@startlink{#1}\@@href}%
\providecommand \@@href[1]{\endgroup#1\@@endlink}%
\providecommand \@sanitize@url [0]{\catcode `\\12\catcode `\$12\catcode
  `\&12\catcode `\#12\catcode `\^12\catcode `\_12\catcode `\%12\relax}%
\providecommand \@@startlink[1]{}%
\providecommand \@@endlink[0]{}%
\providecommand \url  [0]{\begingroup\@sanitize@url \@url }%
\providecommand \@url [1]{\endgroup\@href {#1}{\urlprefix }}%
\providecommand \urlprefix  [0]{URL }%
\providecommand \Eprint [0]{\href }%
\providecommand \doibase [0]{http://dx.doi.org/}%
\providecommand \selectlanguage [0]{\@gobble}%
\providecommand \bibinfo  [0]{\@secondoftwo}%
\providecommand \bibfield  [0]{\@secondoftwo}%
\providecommand \translation [1]{[#1]}%
\providecommand \BibitemOpen [0]{}%
\providecommand \bibitemStop [0]{}%
\providecommand \bibitemNoStop [0]{.\EOS\space}%
\providecommand \EOS [0]{\spacefactor3000\relax}%
\providecommand \BibitemShut  [1]{\csname bibitem#1\endcsname}%
\let\auto@bib@innerbib\@empty
\bibitem [{\citenamefont {Turok}(1989)}]{Turok:1989ai}%
  \BibitemOpen
  \bibfield  {author} {\bibinfo {author} {\bibfnamefont {N.}~\bibnamefont
  {Turok}},\ }\href {\doibase 10.1103/PhysRevLett.63.2625} {\bibfield
  {journal} {\bibinfo  {journal} {Phys. Rev. Lett.}\ }\textbf {\bibinfo
  {volume} {63}},\ \bibinfo {pages} {2625} (\bibinfo {year}
  {1989})}\BibitemShut {NoStop}%
\bibitem [{\citenamefont {Turok}\ and\ \citenamefont
  {Spergel}(1991)}]{Turok:1991qq}%
  \BibitemOpen
  \bibfield  {author} {\bibinfo {author} {\bibfnamefont {N.}~\bibnamefont
  {Turok}}\ and\ \bibinfo {author} {\bibfnamefont {D.~N.}\ \bibnamefont
  {Spergel}},\ }\href {\doibase 10.1103/PhysRevLett.66.3093} {\bibfield
  {journal} {\bibinfo  {journal} {Phys. Rev. Lett.}\ }\textbf {\bibinfo
  {volume} {66}},\ \bibinfo {pages} {3093} (\bibinfo {year}
  {1991})}\BibitemShut {NoStop}%
\bibitem [{\citenamefont {Spergel}\ \emph {et~al.}(1991)\citenamefont
  {Spergel}, \citenamefont {Turok}, \citenamefont {Press},\ and\ \citenamefont
  {Ryden}}]{Spergel:1990ee}%
  \BibitemOpen
  \bibfield  {author} {\bibinfo {author} {\bibfnamefont {D.~N.}\ \bibnamefont
  {Spergel}}, \bibinfo {author} {\bibfnamefont {N.}~\bibnamefont {Turok}},
  \bibinfo {author} {\bibfnamefont {W.~H.}\ \bibnamefont {Press}}, \ and\
  \bibinfo {author} {\bibfnamefont {B.~S.}\ \bibnamefont {Ryden}},\ }\href
  {\doibase 10.1103/PhysRevD.43.1038} {\bibfield  {journal} {\bibinfo
  {journal} {Phys. Rev.}\ }\textbf {\bibinfo {volume} {D43}},\ \bibinfo {pages}
  {1038} (\bibinfo {year} {1991})}\BibitemShut {NoStop}%
\bibitem [{\citenamefont {Pen}\ \emph {et~al.}(1994)\citenamefont {Pen},
  \citenamefont {Spergel},\ and\ \citenamefont {Turok}}]{Pen:1993nx}%
  \BibitemOpen
  \bibfield  {author} {\bibinfo {author} {\bibfnamefont {U.-L.}\ \bibnamefont
  {Pen}}, \bibinfo {author} {\bibfnamefont {D.~N.}\ \bibnamefont {Spergel}}, \
  and\ \bibinfo {author} {\bibfnamefont {N.}~\bibnamefont {Turok}},\ }\href
  {\doibase 10.1103/PhysRevD.49.692} {\bibfield  {journal} {\bibinfo  {journal}
  {Phys. Rev.}\ }\textbf {\bibinfo {volume} {D49}},\ \bibinfo {pages} {692}
  (\bibinfo {year} {1994})}\BibitemShut {NoStop}%
\bibitem [{\citenamefont {Turok}\ and\ \citenamefont
  {Spergel}(1990)}]{Turok:1990gw}%
  \BibitemOpen
  \bibfield  {author} {\bibinfo {author} {\bibfnamefont {N.}~\bibnamefont
  {Turok}}\ and\ \bibinfo {author} {\bibfnamefont {D.}~\bibnamefont
  {Spergel}},\ }\href {\doibase 10.1103/PhysRevLett.64.2736} {\bibfield
  {journal} {\bibinfo  {journal} {Phys. Rev. Lett.}\ }\textbf {\bibinfo
  {volume} {64}},\ \bibinfo {pages} {2736} (\bibinfo {year}
  {1990})}\BibitemShut {NoStop}%
\bibitem [{\citenamefont {Aguirre}(2008)}]{Aguirre:2007gy}%
  \BibitemOpen
  \bibfield  {author} {\bibinfo {author} {\bibfnamefont {A.}~\bibnamefont
  {Aguirre}},\ }in\ \href@noop {} {\emph {\bibinfo {booktitle} {Beyond the Big
  Bang}}}\ (\bibinfo  {publisher} {Springer},\ \bibinfo {year}
  {2008})\BibitemShut {NoStop}%
\bibitem [{\citenamefont {Guth}(2007)}]{Guth:2007ng}%
  \BibitemOpen
  \bibfield  {author} {\bibinfo {author} {\bibfnamefont {A.~H.}\ \bibnamefont
  {Guth}},\ }\href {\doibase 10.1088/1751-8113/40/25/S25} {\bibfield  {journal}
  {\bibinfo  {journal} {J.Phys.}\ }\textbf {\bibinfo {volume} {A40}},\ \bibinfo
  {pages} {6811} (\bibinfo {year} {2007})},\ \Eprint
  {http://arxiv.org/abs/hep-th/0702178} {arXiv:hep-th/0702178 [HEP-TH]}
  \BibitemShut {NoStop}%
\bibitem [{\citenamefont {Aguirre}\ \emph {et~al.}(2007)\citenamefont
  {Aguirre}, \citenamefont {Johnson},\ and\ \citenamefont
  {Shomer}}]{Aguirre:2007an}%
  \BibitemOpen
  \bibfield  {author} {\bibinfo {author} {\bibfnamefont {A.}~\bibnamefont
  {Aguirre}}, \bibinfo {author} {\bibfnamefont {M.~C.}\ \bibnamefont
  {Johnson}}, \ and\ \bibinfo {author} {\bibfnamefont {A.}~\bibnamefont
  {Shomer}},\ }\href@noop {} {\bibfield  {journal} {\bibinfo  {journal} {Phys.
  Rev.}\ }\textbf {\bibinfo {volume} {D76}},\ \bibinfo {pages} {063509}
  (\bibinfo {year} {2007})}\BibitemShut {NoStop}%
\bibitem [{\citenamefont {Feeney}\ \emph
  {et~al.}(2011{\natexlab{a}})\citenamefont {Feeney}, \citenamefont {Johnson},
  \citenamefont {Mortlock},\ and\ \citenamefont {Peiris}}]{Feeney_etal:2010dd}%
  \BibitemOpen
  \bibfield  {author} {\bibinfo {author} {\bibfnamefont {S.~M.}\ \bibnamefont
  {Feeney}}, \bibinfo {author} {\bibfnamefont {M.~C.}\ \bibnamefont {Johnson}},
  \bibinfo {author} {\bibfnamefont {D.~J.}\ \bibnamefont {Mortlock}}, \ and\
  \bibinfo {author} {\bibfnamefont {H.~V.}\ \bibnamefont {Peiris}},\ }\href
  {\doibase 10.1103/PhysRevD.84.043507} {\bibfield  {journal} {\bibinfo
  {journal} {Phys. Rev.}\ }\textbf {\bibinfo {volume} {D84}},\ \bibinfo {pages}
  {043507} (\bibinfo {year} {2011}{\natexlab{a}})},\ \Eprint
  {http://arxiv.org/abs/1012.3667} {arXiv:1012.3667 [astro-ph.CO]} \BibitemShut
  {NoStop}%
\bibitem [{\citenamefont {Feeney}\ \emph
  {et~al.}(2011{\natexlab{b}})\citenamefont {Feeney}, \citenamefont {Johnson},
  \citenamefont {Mortlock},\ and\ \citenamefont {Peiris}}]{Feeney_etal:2010jj}%
  \BibitemOpen
  \bibfield  {author} {\bibinfo {author} {\bibfnamefont {S.~M.}\ \bibnamefont
  {Feeney}}, \bibinfo {author} {\bibfnamefont {M.~C.}\ \bibnamefont {Johnson}},
  \bibinfo {author} {\bibfnamefont {D.~J.}\ \bibnamefont {Mortlock}}, \ and\
  \bibinfo {author} {\bibfnamefont {H.~V.}\ \bibnamefont {Peiris}},\ }\href
  {\doibase 10.1103/PhysRevLett.107.071301} {\bibfield  {journal} {\bibinfo
  {journal} {Phys. Rev. Lett.}\ }\textbf {\bibinfo {volume} {107}},\ \bibinfo
  {pages} {071301} (\bibinfo {year} {2011}{\natexlab{b}})},\ \Eprint
  {http://arxiv.org/abs/1012.1995} {arXiv:1012.1995 [astro-ph.CO]} \BibitemShut
  {NoStop}%
\bibitem [{\citenamefont {Feeney}\ \emph {et~al.}(2012)\citenamefont {Feeney},
  \citenamefont {Johnson}, \citenamefont {Mortlock},\ and\ \citenamefont
  {Peiris}}]{Feeney_etal:2012jf}%
  \BibitemOpen
  \bibfield  {author} {\bibinfo {author} {\bibfnamefont {S.~M.}\ \bibnamefont
  {Feeney}}, \bibinfo {author} {\bibfnamefont {M.~C.}\ \bibnamefont {Johnson}},
  \bibinfo {author} {\bibfnamefont {D.~J.}\ \bibnamefont {Mortlock}}, \ and\
  \bibinfo {author} {\bibfnamefont {H.~V.}\ \bibnamefont {Peiris}},\ }\href
  {\doibase 10.1103/PhysRevLett.108.241301} {\bibfield  {journal} {\bibinfo
  {journal} {Phys.Rev.Lett.}\ }\textbf {\bibinfo {volume} {108}},\ \bibinfo
  {pages} {241301} (\bibinfo {year} {2012})},\ \Eprint
  {http://arxiv.org/abs/1203.1928} {arXiv:1203.1928 [astro-ph.CO]} \BibitemShut
  {NoStop}%
\bibitem [{\citenamefont {McEwen}\ \emph {et~al.}(2012)\citenamefont {McEwen},
  \citenamefont {Feeney}, \citenamefont {Johnson},\ and\ \citenamefont
  {Peiris}}]{McEwen:2012uk}%
  \BibitemOpen
  \bibfield  {author} {\bibinfo {author} {\bibfnamefont {J.~D.}\ \bibnamefont
  {McEwen}}, \bibinfo {author} {\bibfnamefont {S.~M.}\ \bibnamefont {Feeney}},
  \bibinfo {author} {\bibfnamefont {M.~C.}\ \bibnamefont {Johnson}}, \ and\
  \bibinfo {author} {\bibfnamefont {H.~V.}\ \bibnamefont {Peiris}},\ }\href
  {\doibase 10.1103/PhysRevD.85.103502} {\bibfield  {journal} {\bibinfo
  {journal} {Phys. Rev.}\ }\textbf {\bibinfo {volume} {D85}},\ \bibinfo {pages}
  {103502} (\bibinfo {year} {2012})},\ \Eprint {http://arxiv.org/abs/1202.2861}
  {arXiv:1202.2861 [astro-ph.CO]} \BibitemShut {NoStop}%
\bibitem [{\citenamefont {Bennett}\ \emph
  {et~al.}(2003{\natexlab{a}})\citenamefont {Bennett} \emph
  {et~al.}}]{Bennett:2003ba}%
  \BibitemOpen
  \bibfield  {author} {\bibinfo {author} {\bibfnamefont {C.~L.}\ \bibnamefont
  {Bennett}} \emph {et~al.} (\bibinfo {collaboration} {WMAP}),\ }\href
  {\doibase 10.1086/345346} {\bibfield  {journal} {\bibinfo  {journal}
  {Astrophys. J.}\ }\textbf {\bibinfo {volume} {583}},\ \bibinfo {pages} {1}
  (\bibinfo {year} {2003}{\natexlab{a}})},\ \Eprint
  {http://arxiv.org/abs/astro-ph/0301158} {arXiv:astro-ph/0301158} \BibitemShut
  {NoStop}%
\bibitem [{\citenamefont {{Sunyaev}}\ and\ \citenamefont
  {{Zeldovich}}(1972)}]{Sunyaev_Zeldovich:1972}%
  \BibitemOpen
  \bibfield  {author} {\bibinfo {author} {\bibfnamefont {R.~A.}\ \bibnamefont
  {{Sunyaev}}}\ and\ \bibinfo {author} {\bibfnamefont {Y.~B.}\ \bibnamefont
  {{Zeldovich}}},\ }\href@noop {} {\bibfield  {journal} {\bibinfo  {journal}
  {Comments on Astrophysics and Space Physics}\ }\textbf {\bibinfo {volume}
  {4}},\ \bibinfo {pages} {173} (\bibinfo {year} {1972})}\BibitemShut {NoStop}%
\bibitem [{\citenamefont {{Loredo}}(2012)}]{Loredo:2012}%
  \BibitemOpen
  \bibfield  {author} {\bibinfo {author} {\bibfnamefont {T.~J.}\ \bibnamefont
  {{Loredo}}},\ }\href@noop {} {\bibfield  {journal} {\bibinfo  {journal}
  {ArXiv e-prints}\ } (\bibinfo {year} {2012})},\ \Eprint
  {http://arxiv.org/abs/1208.3036} {arXiv:1208.3036 [astro-ph.IM]} \BibitemShut
  {NoStop}%
\bibitem [{\citenamefont {Freivogel}(2011)}]{Freivogel:2011eg}%
  \BibitemOpen
  \bibfield  {author} {\bibinfo {author} {\bibfnamefont {B.}~\bibnamefont
  {Freivogel}},\ }\href {\doibase 10.1088/0264-9381/28/20/204007} {\bibfield
  {journal} {\bibinfo  {journal} {Class.Quant.Grav.}\ }\textbf {\bibinfo
  {volume} {28}},\ \bibinfo {pages} {204007} (\bibinfo {year} {2011})},\
  \Eprint {http://arxiv.org/abs/1105.0244} {arXiv:1105.0244 [hep-th]}
  \BibitemShut {NoStop}%
\bibitem [{\citenamefont {Salem}(2012)}]{Salem:2012}%
  \BibitemOpen
  \bibfield  {author} {\bibinfo {author} {\bibfnamefont {M.~P.}\ \bibnamefont
  {Salem}},\ }\href {\doibase 10.1088/1475-7516/2012/01/021} {\bibfield
  {journal} {\bibinfo  {journal} {JCAP}\ }\textbf {\bibinfo {volume} {1201}},\
  \bibinfo {pages} {021} (\bibinfo {year} {2012})},\ \Eprint
  {http://arxiv.org/abs/1108.0040} {arXiv:1108.0040 [hep-th]} \BibitemShut
  {NoStop}%
\bibitem [{\citenamefont {{Planck Collaboration}}\ \emph
  {et~al.}(2011)\citenamefont {{Planck Collaboration}}, \citenamefont {{Ade}}
  \emph {et~al.}}]{Planck_Ade:2011}%
  \BibitemOpen
  \bibfield  {author} {\bibinfo {author} {\bibnamefont {{Planck
  Collaboration}}}, \bibinfo {author} {\bibfnamefont {P.~A.~R.}\ \bibnamefont
  {{Ade}}},  \emph {et~al.},\ }\href {\doibase 10.1051/0004-6361/201116464}
  {\bibfield  {journal} {\bibinfo  {journal} {Astron. Astrophys.}\ }\textbf
  {\bibinfo {volume} {536}},\ \bibinfo {eid} {A1} (\bibinfo {year} {2011})},\
  \Eprint {http://arxiv.org/abs/1101.2022} {arXiv:1101.2022 [astro-ph.IM]}
  \BibitemShut {NoStop}%
\bibitem [{\citenamefont {Hobson}\ and\ \citenamefont
  {McLachlan}(2003)}]{Hobson_McLachlan:2003}%
  \BibitemOpen
  \bibfield  {author} {\bibinfo {author} {\bibfnamefont {M.~P.}\ \bibnamefont
  {Hobson}}\ and\ \bibinfo {author} {\bibfnamefont {C.}~\bibnamefont
  {McLachlan}},\ }\href {\doibase 10.1046/j.1365-8711.2003.06094.x} {\bibfield
  {journal} {\bibinfo  {journal} {Mon. Not. R. Astron. Soc.}\ }\textbf
  {\bibinfo {volume} {338}},\ \bibinfo {pages} {765} (\bibinfo {year}
  {2003})},\ \Eprint {http://arxiv.org/abs/astro-ph/0204457}
  {arXiv:astro-ph/0204457} \BibitemShut {NoStop}%
\bibitem [{\citenamefont {{Hobson}}\ \emph {et~al.}(2010)\citenamefont
  {{Hobson}}, \citenamefont {{Rocha}},\ and\ \citenamefont
  {{Savage}}}]{Hobson_etal:2010}%
  \BibitemOpen
  \bibfield  {author} {\bibinfo {author} {\bibfnamefont {M.~P.}\ \bibnamefont
  {{Hobson}}}, \bibinfo {author} {\bibfnamefont {G.}~\bibnamefont {{Rocha}}}, \
  and\ \bibinfo {author} {\bibfnamefont {R.~S.}\ \bibnamefont {{Savage}}},\
  }\enquote {\bibinfo {title} {{Bayesian source extraction}},}\ in\ \href@noop
  {} {\emph {\bibinfo {booktitle} {Bayesian Methods in Cosmology}}},\ \bibinfo
  {editor} {edited by\ \bibinfo {editor} {\bibnamefont {{Hobson, M.~P., Jaffe,
  A.~H., Liddle, A.~R., Mukeherjee, P., \& Parkinson, D.}}}}\ (\bibinfo {year}
  {2010})\ pp.\ \bibinfo {pages} {167--+}\BibitemShut {NoStop}%
\bibitem [{\citenamefont {{Cruz}}\ \emph {et~al.}(2008)\citenamefont {{Cruz}},
  \citenamefont {{Mart{\'{\i}}nez-Gonzalez}}, \citenamefont {{Vielva}},
  \citenamefont {{Diego}}, \citenamefont {{Hobson}},\ and\ \citenamefont
  {{Turok}}}]{2008MNRAS.390..913C}%
  \BibitemOpen
  \bibfield  {author} {\bibinfo {author} {\bibfnamefont {M.}~\bibnamefont
  {{Cruz}}}, \bibinfo {author} {\bibfnamefont {E.}~\bibnamefont
  {{Mart{\'{\i}}nez-Gonzalez}}}, \bibinfo {author} {\bibfnamefont
  {P.}~\bibnamefont {{Vielva}}}, \bibinfo {author} {\bibfnamefont {J.~M.}\
  \bibnamefont {{Diego}}}, \bibinfo {author} {\bibfnamefont {M.}~\bibnamefont
  {{Hobson}}}, \ and\ \bibinfo {author} {\bibfnamefont {N.}~\bibnamefont
  {{Turok}}},\ }\href {\doibase 10.1111/j.1365-2966.2008.13812.x} {\bibfield
  {journal} {\bibinfo  {journal} {Mon. Not. R. Astron. Soc.}\ }\textbf
  {\bibinfo {volume} {390}},\ \bibinfo {pages} {913} (\bibinfo {year}
  {2008})},\ \Eprint {http://arxiv.org/abs/0804.2904} {arXiv:0804.2904}
  \BibitemShut {NoStop}%
\bibitem [{\citenamefont {Vielva}\ \emph {et~al.}(2011)\citenamefont {Vielva},
  \citenamefont {Martinez-Gonzalez}, \citenamefont {Cruz}, \citenamefont
  {Barreiro},\ and\ \citenamefont {Tucci}}]{2011MNRAS.410...33V}%
  \BibitemOpen
  \bibfield  {author} {\bibinfo {author} {\bibfnamefont {P.}~\bibnamefont
  {Vielva}}, \bibinfo {author} {\bibfnamefont {E.}~\bibnamefont
  {Martinez-Gonzalez}}, \bibinfo {author} {\bibfnamefont {M.}~\bibnamefont
  {Cruz}}, \bibinfo {author} {\bibfnamefont {R.~B.}\ \bibnamefont {Barreiro}},
  \ and\ \bibinfo {author} {\bibfnamefont {M.}~\bibnamefont {Tucci}},\ }\href
  {\doibase 10.1111/j.1365-2966.2010.17418.x} {\bibfield  {journal} {\bibinfo
  {journal} {Mon. Not. R. Astron. Soc.}\ }\textbf {\bibinfo {volume} {410}},\
  \bibinfo {pages} {33} (\bibinfo {year} {2011})},\ \Eprint
  {http://arxiv.org/abs/1002.4029} {arXiv:1002.4029 [astro-ph.CO]} \BibitemShut
  {NoStop}%
\bibitem [{\citenamefont {Cruz}\ \emph {et~al.}(2005)\citenamefont {Cruz},
  \citenamefont {Martinez-Gonzalez}, \citenamefont {Vielva},\ and\
  \citenamefont {Cayon}}]{Cruz:2004ce}%
  \BibitemOpen
  \bibfield  {author} {\bibinfo {author} {\bibfnamefont {M.}~\bibnamefont
  {Cruz}}, \bibinfo {author} {\bibfnamefont {E.}~\bibnamefont
  {Martinez-Gonzalez}}, \bibinfo {author} {\bibfnamefont {P.}~\bibnamefont
  {Vielva}}, \ and\ \bibinfo {author} {\bibfnamefont {L.}~\bibnamefont
  {Cayon}},\ }\href {\doibase 10.1111/j.1365-2966.2004.08419.x/abs/} {\bibfield
   {journal} {\bibinfo  {journal} {Mon. Not. R. Astron. Soc.}\ }\textbf
  {\bibinfo {volume} {356}},\ \bibinfo {pages} {29} (\bibinfo {year} {2005})},\
  \Eprint {http://arxiv.org/abs/astro-ph/0405341} {arXiv:astro-ph/0405341}
  \BibitemShut {NoStop}%
\bibitem [{\citenamefont {Cruz}\ \emph {et~al.}(2007)\citenamefont {Cruz},
  \citenamefont {Turok}, \citenamefont {Vielva}, \citenamefont
  {Martinez-Gonzalez},\ and\ \citenamefont {Hobson}}]{Cruz:2007pe}%
  \BibitemOpen
  \bibfield  {author} {\bibinfo {author} {\bibfnamefont {M.}~\bibnamefont
  {Cruz}}, \bibinfo {author} {\bibfnamefont {N.}~\bibnamefont {Turok}},
  \bibinfo {author} {\bibfnamefont {P.}~\bibnamefont {Vielva}}, \bibinfo
  {author} {\bibfnamefont {E.}~\bibnamefont {Martinez-Gonzalez}}, \ and\
  \bibinfo {author} {\bibfnamefont {M.}~\bibnamefont {Hobson}},\ }\href
  {\doibase 10.1126/science.1148694} {\bibfield  {journal} {\bibinfo  {journal}
  {Science}\ }\textbf {\bibinfo {volume} {318}},\ \bibinfo {pages} {1612}
  (\bibinfo {year} {2007})},\ \Eprint {http://arxiv.org/abs/0710.5737}
  {arXiv:0710.5737 [astro-ph]} \BibitemShut {NoStop}%
\bibitem [{\citenamefont {Cruz}\ \emph {et~al.}(2010)\citenamefont {Cruz},
  \citenamefont {Martinez-Gonzalez},\ and\ \citenamefont
  {Vielva}}]{Cruz:2009nd}%
  \BibitemOpen
  \bibfield  {author} {\bibinfo {author} {\bibfnamefont {M.}~\bibnamefont
  {Cruz}}, \bibinfo {author} {\bibfnamefont {E.}~\bibnamefont
  {Martinez-Gonzalez}}, \ and\ \bibinfo {author} {\bibfnamefont
  {P.}~\bibnamefont {Vielva}},\ }in\ \href@noop {} {\emph {\bibinfo {booktitle}
  {Highlights of Spanish Astrophysics V}}}\ (\bibinfo {year} {2010})\ \Eprint
  {http://arxiv.org/abs/0901.1986} {arXiv:0901.1986 [astro-ph]} \BibitemShut
  {NoStop}%
\bibitem [{\citenamefont {Aguirre}\ and\ \citenamefont
  {Johnson}(2011)}]{Aguirre:2009ug}%
  \BibitemOpen
  \bibfield  {author} {\bibinfo {author} {\bibfnamefont {A.}~\bibnamefont
  {Aguirre}}\ and\ \bibinfo {author} {\bibfnamefont {M.~C.}\ \bibnamefont
  {Johnson}},\ }\href {\doibase 10.1088/0034-4885/74/7/074901} {\bibfield
  {journal} {\bibinfo  {journal} {Rept.Prog.Phys.}\ }\textbf {\bibinfo {volume}
  {74}},\ \bibinfo {pages} {074901} (\bibinfo {year} {2011})},\ \Eprint
  {http://arxiv.org/abs/0908.4105} {arXiv:0908.4105 [hep-th]} \BibitemShut
  {NoStop}%
\bibitem [{\citenamefont {Chang}\ \emph {et~al.}(2009)\citenamefont {Chang},
  \citenamefont {Kleban},\ and\ \citenamefont {Levi}}]{Chang_Kleban_Levi:2009}%
  \BibitemOpen
  \bibfield  {author} {\bibinfo {author} {\bibfnamefont {S.}~\bibnamefont
  {Chang}}, \bibinfo {author} {\bibfnamefont {M.}~\bibnamefont {Kleban}}, \
  and\ \bibinfo {author} {\bibfnamefont {T.~S.}\ \bibnamefont {Levi}},\ }\href
  {\doibase 10.1088/1475-7516/2009/04/025} {\bibfield  {journal} {\bibinfo
  {journal} {JCAP}\ }\textbf {\bibinfo {volume} {0904}},\ \bibinfo {pages}
  {025} (\bibinfo {year} {2009})},\ \Eprint {http://arxiv.org/abs/0810.5128}
  {arXiv:0810.5128 [hep-th]} \BibitemShut {NoStop}%
\bibitem [{\citenamefont {Czech}\ \emph {et~al.}(2010)\citenamefont {Czech},
  \citenamefont {Kleban}, \citenamefont {Larjo}, \citenamefont {Levi},\ and\
  \citenamefont {Sigurdson}}]{Czech:2010rg}%
  \BibitemOpen
  \bibfield  {author} {\bibinfo {author} {\bibfnamefont {B.}~\bibnamefont
  {Czech}}, \bibinfo {author} {\bibfnamefont {M.}~\bibnamefont {Kleban}},
  \bibinfo {author} {\bibfnamefont {K.}~\bibnamefont {Larjo}}, \bibinfo
  {author} {\bibfnamefont {T.~S.}\ \bibnamefont {Levi}}, \ and\ \bibinfo
  {author} {\bibfnamefont {K.}~\bibnamefont {Sigurdson}},\ }\href {\doibase
  10.1088/1475-7516/2010/12/023} {\bibfield  {journal} {\bibinfo  {journal}
  {JCAP}\ }\textbf {\bibinfo {volume} {1012}},\ \bibinfo {pages} {023}
  (\bibinfo {year} {2010})},\ \Eprint {http://arxiv.org/abs/1006.0832}
  {arXiv:1006.0832 [astro-ph.CO]} \BibitemShut {NoStop}%
\bibitem [{\citenamefont {Gobbetti}\ and\ \citenamefont
  {Kleban}(2012)}]{Gobbetti_Kleban:2012}%
  \BibitemOpen
  \bibfield  {author} {\bibinfo {author} {\bibfnamefont {R.}~\bibnamefont
  {Gobbetti}}\ and\ \bibinfo {author} {\bibfnamefont {M.}~\bibnamefont
  {Kleban}},\ }\href {\doibase 10.1088/1475-7516/2012/05/025} {\bibfield
  {journal} {\bibinfo  {journal} {JCAP}\ }\textbf {\bibinfo {volume} {1205}},\
  \bibinfo {pages} {025} (\bibinfo {year} {2012})},\ \Eprint
  {http://arxiv.org/abs/1201.6380} {arXiv:1201.6380 [hep-th]} \BibitemShut
  {NoStop}%
\bibitem [{\citenamefont {Kleban}\ \emph {et~al.}(2011)\citenamefont {Kleban},
  \citenamefont {Levi},\ and\ \citenamefont
  {Sigurdson}}]{Kleban_Levi_Sigurdson:2011}%
  \BibitemOpen
  \bibfield  {author} {\bibinfo {author} {\bibfnamefont {M.}~\bibnamefont
  {Kleban}}, \bibinfo {author} {\bibfnamefont {T.~S.}\ \bibnamefont {Levi}}, \
  and\ \bibinfo {author} {\bibfnamefont {K.}~\bibnamefont {Sigurdson}},\
  }\href@noop {} {\  (\bibinfo {year} {2011})},\ \Eprint
  {http://arxiv.org/abs/1109.3473} {arXiv:1109.3473 [astro-ph.CO]} \BibitemShut
  {NoStop}%
\bibitem [{\citenamefont {Kozaczuk}\ and\ \citenamefont
  {Aguirre}(2012)}]{Kozaczuk:2012sx}%
  \BibitemOpen
  \bibfield  {author} {\bibinfo {author} {\bibfnamefont {J.}~\bibnamefont
  {Kozaczuk}}\ and\ \bibinfo {author} {\bibfnamefont {A.}~\bibnamefont
  {Aguirre}},\ }\href@noop {} {\  (\bibinfo {year} {2012})},\ \Eprint
  {http://arxiv.org/abs/1206.5038} {arXiv:1206.5038 [hep-th]} \BibitemShut
  {NoStop}%
\bibitem [{\citenamefont {Aguirre}\ and\ \citenamefont
  {Johnson}(2008)}]{Aguirre:2007wm}%
  \BibitemOpen
  \bibfield  {author} {\bibinfo {author} {\bibfnamefont {A.}~\bibnamefont
  {Aguirre}}\ and\ \bibinfo {author} {\bibfnamefont {M.~C.}\ \bibnamefont
  {Johnson}},\ }\href {\doibase 10.1103/PhysRevD.77.123536} {\bibfield
  {journal} {\bibinfo  {journal} {Phys. Rev.}\ }\textbf {\bibinfo {volume}
  {D77}},\ \bibinfo {pages} {123536} (\bibinfo {year} {2008})}\BibitemShut
  {NoStop}%
\bibitem [{\citenamefont {Freivogel}\ \emph {et~al.}(2009)\citenamefont
  {Freivogel}, \citenamefont {Kleban}, \citenamefont {Nicolis},\ and\
  \citenamefont {Sigurdson}}]{Freivogel_etal:2009it}%
  \BibitemOpen
  \bibfield  {author} {\bibinfo {author} {\bibfnamefont {B.}~\bibnamefont
  {Freivogel}}, \bibinfo {author} {\bibfnamefont {M.}~\bibnamefont {Kleban}},
  \bibinfo {author} {\bibfnamefont {A.}~\bibnamefont {Nicolis}}, \ and\
  \bibinfo {author} {\bibfnamefont {K.}~\bibnamefont {Sigurdson}},\ }\href
  {\doibase 10.1088/1475-7516/2009/08/036} {\bibfield  {journal} {\bibinfo
  {journal} {JCAP}\ }\textbf {\bibinfo {volume} {0908}},\ \bibinfo {pages}
  {036} (\bibinfo {year} {2009})}\BibitemShut {NoStop}%
\bibitem [{\citenamefont {Feroz}\ and\ \citenamefont
  {Hobson}(2008)}]{Feroz_Hobson:2008}%
  \BibitemOpen
  \bibfield  {author} {\bibinfo {author} {\bibfnamefont {F.}~\bibnamefont
  {Feroz}}\ and\ \bibinfo {author} {\bibfnamefont {M.~P.}\ \bibnamefont
  {Hobson}},\ }\href {\doibase 10.1111/j.1365-2966.2007.12353.x} {\bibfield
  {journal} {\bibinfo  {journal} {Mon. Not. R. Astron. Soc.}\ }\textbf
  {\bibinfo {volume} {384}},\ \bibinfo {pages} {449} (\bibinfo {year}
  {2008})},\ \Eprint {http://arxiv.org/abs/0704.3704} {arXiv:0704.3704
  [astro-ph]} \BibitemShut {NoStop}%
\bibitem [{\citenamefont {{Feroz}}\ \emph {et~al.}(2009)\citenamefont
  {{Feroz}}, \citenamefont {{Hobson}},\ and\ \citenamefont
  {{Bridges}}}]{Feroz_Hobson_Bridges:2009}%
  \BibitemOpen
  \bibfield  {author} {\bibinfo {author} {\bibfnamefont {F.}~\bibnamefont
  {{Feroz}}}, \bibinfo {author} {\bibfnamefont {M.~P.}\ \bibnamefont
  {{Hobson}}}, \ and\ \bibinfo {author} {\bibfnamefont {M.}~\bibnamefont
  {{Bridges}}},\ }\href {\doibase 10.1111/j.1365-2966.2009.14548.x} {\bibfield
  {journal} {\bibinfo  {journal} {Mon. Not. R. Astron. Soc.}\ }\textbf
  {\bibinfo {volume} {398}},\ \bibinfo {pages} {1601} (\bibinfo {year}
  {2009})},\ \Eprint {http://arxiv.org/abs/0809.3437} {arXiv:0809.3437}
  \BibitemShut {NoStop}%
\bibitem [{\citenamefont {Larson}\ \emph {et~al.}(2011)\citenamefont {Larson},
  \citenamefont {Dunkley}, \citenamefont {Hinshaw}, \citenamefont {Komatsu},
  \citenamefont {Nolta} \emph {et~al.}}]{Larson_etal:2011}%
  \BibitemOpen
  \bibfield  {author} {\bibinfo {author} {\bibfnamefont {D.}~\bibnamefont
  {Larson}}, \bibinfo {author} {\bibfnamefont {J.}~\bibnamefont {Dunkley}},
  \bibinfo {author} {\bibfnamefont {G.}~\bibnamefont {Hinshaw}}, \bibinfo
  {author} {\bibfnamefont {E.}~\bibnamefont {Komatsu}}, \bibinfo {author}
  {\bibfnamefont {M.~R.}\ \bibnamefont {Nolta}},  \emph {et~al.},\ }\href
  {\doibase 10.1088/0067-0049/192/2/16} {\bibfield  {journal} {\bibinfo
  {journal} {Astrophys. J. Suppl. Ser.}\ }\textbf {\bibinfo {volume} {192}},\
  \bibinfo {pages} {16} (\bibinfo {year} {2011})},\ \Eprint
  {http://arxiv.org/abs/1001.4635} {arXiv:1001.4635 [astro-ph.CO]} \BibitemShut
  {NoStop}%
\bibitem [{\citenamefont {McEwen}\ \emph {et~al.}(2008)\citenamefont {McEwen},
  \citenamefont {Hobson},\ and\ \citenamefont {Lasenby}}]{mcewen:2006:filters}%
  \BibitemOpen
  \bibfield  {author} {\bibinfo {author} {\bibfnamefont {J.~D.}\ \bibnamefont
  {McEwen}}, \bibinfo {author} {\bibfnamefont {M.~P.}\ \bibnamefont {Hobson}},
  \ and\ \bibinfo {author} {\bibfnamefont {A.~N.}\ \bibnamefont {Lasenby}},\
  }\href {\doibase 10.1109/TSP.2008.923198} {\bibfield  {journal} {\bibinfo
  {journal} {{IEEE Trans.\ Sig.\ Proc.}}\ }\textbf {\bibinfo {volume} {56}},\
  \bibinfo {pages} {3813} (\bibinfo {year} {2008})},\ \Eprint
  {http://arxiv.org/abs/astro-ph/0612688} {astro-ph/0612688} \BibitemShut
  {NoStop}%
\bibitem [{\citenamefont {{Marinucci}}\ \emph {et~al.}(2008)\citenamefont
  {{Marinucci}}, \citenamefont {{Pietrobon}}, \citenamefont {{Balbi}},
  \citenamefont {{Baldi}}, \citenamefont {{Cabella}}, \citenamefont
  {{Kerkyacharian}}, \citenamefont {{Natoli}}, \citenamefont {{Picard}},\ and\
  \citenamefont {{Vittorio}}}]{Marinucci:2007aj}%
  \BibitemOpen
  \bibfield  {author} {\bibinfo {author} {\bibfnamefont {D.}~\bibnamefont
  {{Marinucci}}}, \bibinfo {author} {\bibfnamefont {D.}~\bibnamefont
  {{Pietrobon}}}, \bibinfo {author} {\bibfnamefont {A.}~\bibnamefont
  {{Balbi}}}, \bibinfo {author} {\bibfnamefont {P.}~\bibnamefont {{Baldi}}},
  \bibinfo {author} {\bibfnamefont {P.}~\bibnamefont {{Cabella}}}, \bibinfo
  {author} {\bibfnamefont {G.}~\bibnamefont {{Kerkyacharian}}}, \bibinfo
  {author} {\bibfnamefont {P.}~\bibnamefont {{Natoli}}}, \bibinfo {author}
  {\bibfnamefont {D.}~\bibnamefont {{Picard}}}, \ and\ \bibinfo {author}
  {\bibfnamefont {N.}~\bibnamefont {{Vittorio}}},\ }\href@noop {} {\bibfield
  {journal} {\bibinfo  {journal} {Mon. Not. R. Astron. Soc.}\ }\textbf
  {\bibinfo {volume} {383}},\ \bibinfo {pages} {539} (\bibinfo {year}
  {2008})}\BibitemShut {NoStop}%
\bibitem [{\citenamefont {Scodeller}\ \emph {et~al.}(2011)\citenamefont
  {Scodeller}, \citenamefont {Rudjord}, \citenamefont {Hansen}, \citenamefont
  {Marinucci}, \citenamefont {Geller} \emph {et~al.}}]{Scodeller:2010mp}%
  \BibitemOpen
  \bibfield  {author} {\bibinfo {author} {\bibfnamefont {S.}~\bibnamefont
  {Scodeller}}, \bibinfo {author} {\bibfnamefont {O.}~\bibnamefont {Rudjord}},
  \bibinfo {author} {\bibfnamefont {F.}~\bibnamefont {Hansen}}, \bibinfo
  {author} {\bibfnamefont {D.}~\bibnamefont {Marinucci}}, \bibinfo {author}
  {\bibfnamefont {D.}~\bibnamefont {Geller}},  \emph {et~al.},\ }\href
  {\doibase 10.1088/0004-637X/733/2/121} {\bibfield  {journal} {\bibinfo
  {journal} {Astrophys. J.}\ }\textbf {\bibinfo {volume} {733}},\ \bibinfo
  {pages} {121} (\bibinfo {year} {2011})},\ \Eprint
  {http://arxiv.org/abs/1004.5576} {arXiv:1004.5576 [astro-ph.CO]} \BibitemShut
  {NoStop}%
\bibitem [{\citenamefont {McEwen}\ \emph {et~al.}(2007)\citenamefont {McEwen},
  \citenamefont {Hobson}, \citenamefont {Mortlock},\ and\ \citenamefont
  {Lasenby}}]{mcewen:2006:fcswt}%
  \BibitemOpen
  \bibfield  {author} {\bibinfo {author} {\bibfnamefont {J.~D.}\ \bibnamefont
  {McEwen}}, \bibinfo {author} {\bibfnamefont {M.~P.}\ \bibnamefont {Hobson}},
  \bibinfo {author} {\bibfnamefont {D.~J.}\ \bibnamefont {Mortlock}}, \ and\
  \bibinfo {author} {\bibfnamefont {A.~N.}\ \bibnamefont {Lasenby}},\ }\href
  {\doibase 10.1109/TSP.2006.887148} {\bibfield  {journal} {\bibinfo  {journal}
  {{IEEE Trans.\ Sig.\ Proc.}}\ }\textbf {\bibinfo {volume} {55}},\ \bibinfo
  {pages} {520} (\bibinfo {year} {2007})},\ \Eprint
  {http://arxiv.org/abs/astro-ph/0506308} {astro-ph/0506308} \BibitemShut
  {NoStop}%
\bibitem [{\citenamefont {Gorski}\ \emph {et~al.}(2005)\citenamefont {Gorski}
  \emph {et~al.}}]{Gorski:2004by}%
  \BibitemOpen
  \bibfield  {author} {\bibinfo {author} {\bibfnamefont {K.~M.}\ \bibnamefont
  {Gorski}} \emph {et~al.},\ }\href {\doibase 10.1086/427976} {\bibfield
  {journal} {\bibinfo  {journal} {Astrophys. J.}\ }\textbf {\bibinfo {volume}
  {622}},\ \bibinfo {pages} {759} (\bibinfo {year} {2005})},\ \Eprint
  {http://arxiv.org/abs/astro-ph/0409513} {arXiv:astro-ph/0409513} \BibitemShut
  {NoStop}%
\bibitem [{\citenamefont {Gold}\ \emph {et~al.}(2011)\citenamefont {Gold},
  \citenamefont {Odegard}, \citenamefont {Weiland}, \citenamefont {Hill},
  \citenamefont {Kogut} \emph {et~al.}}]{Gold:2010fm}%
  \BibitemOpen
  \bibfield  {author} {\bibinfo {author} {\bibfnamefont {B.}~\bibnamefont
  {Gold}}, \bibinfo {author} {\bibfnamefont {N.}~\bibnamefont {Odegard}},
  \bibinfo {author} {\bibfnamefont {J.~L.}\ \bibnamefont {Weiland}}, \bibinfo
  {author} {\bibfnamefont {R.~S.}\ \bibnamefont {Hill}}, \bibinfo {author}
  {\bibfnamefont {A.}~\bibnamefont {Kogut}},  \emph {et~al.},\ }\href {\doibase
  10.1088/0067-0049/192/2/15} {\bibfield  {journal} {\bibinfo  {journal}
  {Astrophys. J. Suppl. Ser.}\ }\textbf {\bibinfo {volume} {192}},\ \bibinfo
  {pages} {15} (\bibinfo {year} {2011})},\ \Eprint
  {http://arxiv.org/abs/1001.4555} {arXiv:1001.4555 [astro-ph.GA]} \BibitemShut
  {NoStop}%
\bibitem [{\citenamefont {Feeney}\ \emph
  {et~al.}(2011{\natexlab{c}})\citenamefont {Feeney}, \citenamefont {Peiris},\
  and\ \citenamefont {Pontzen}}]{Feeney_Pontzen_Peiris:2011}%
  \BibitemOpen
  \bibfield  {author} {\bibinfo {author} {\bibfnamefont {S.~M.}\ \bibnamefont
  {Feeney}}, \bibinfo {author} {\bibfnamefont {H.~V.}\ \bibnamefont {Peiris}},
  \ and\ \bibinfo {author} {\bibfnamefont {A.}~\bibnamefont {Pontzen}},\ }\href
  {\doibase 10.1103/PhysRevD.84.103002} {\bibfield  {journal} {\bibinfo
  {journal} {Phys. Rev.}\ }\textbf {\bibinfo {volume} {D84}},\ \bibinfo {pages}
  {103002} (\bibinfo {year} {2011}{\natexlab{c}})},\ \Eprint
  {http://arxiv.org/abs/1107.5466} {arXiv:1107.5466 [astro-ph.CO]} \BibitemShut
  {NoStop}%
\bibitem [{\citenamefont {de~Oliveira-Costa}\ and\ \citenamefont
  {Tegmark}(2006)}]{deOliveiraCosta:2006zj}%
  \BibitemOpen
  \bibfield  {author} {\bibinfo {author} {\bibfnamefont {A.}~\bibnamefont
  {de~Oliveira-Costa}}\ and\ \bibinfo {author} {\bibfnamefont {M.}~\bibnamefont
  {Tegmark}},\ }\href {\doibase 10.1103/PhysRevD.74.023005} {\bibfield
  {journal} {\bibinfo  {journal} {Phys. Rev.}\ }\textbf {\bibinfo {volume}
  {D74}},\ \bibinfo {pages} {023005} (\bibinfo {year} {2006})},\ \Eprint
  {http://arxiv.org/abs/astro-ph/0603369} {arXiv:astro-ph/0603369} \BibitemShut
  {NoStop}%
\bibitem [{\citenamefont {Anderson}\ \emph {et~al.}(1999)\citenamefont
  {Anderson}, \citenamefont {Bai}, \citenamefont {Bischof}, \citenamefont
  {Blackford}, \citenamefont {Demmel}, \citenamefont {Dongarra}, \citenamefont
  {Du~Croz}, \citenamefont {Greenbaum}, \citenamefont {Hammarling},
  \citenamefont {McKenney},\ and\ \citenamefont {Sorensen}}]{lapack}%
  \BibitemOpen
  \bibfield  {author} {\bibinfo {author} {\bibfnamefont {E.}~\bibnamefont
  {Anderson}}, \bibinfo {author} {\bibfnamefont {Z.}~\bibnamefont {Bai}},
  \bibinfo {author} {\bibfnamefont {C.}~\bibnamefont {Bischof}}, \bibinfo
  {author} {\bibfnamefont {S.}~\bibnamefont {Blackford}}, \bibinfo {author}
  {\bibfnamefont {J.}~\bibnamefont {Demmel}}, \bibinfo {author} {\bibfnamefont
  {J.}~\bibnamefont {Dongarra}}, \bibinfo {author} {\bibfnamefont
  {J.}~\bibnamefont {Du~Croz}}, \bibinfo {author} {\bibfnamefont
  {A.}~\bibnamefont {Greenbaum}}, \bibinfo {author} {\bibfnamefont
  {S.}~\bibnamefont {Hammarling}}, \bibinfo {author} {\bibfnamefont
  {A.}~\bibnamefont {McKenney}}, \ and\ \bibinfo {author} {\bibfnamefont
  {D.}~\bibnamefont {Sorensen}},\ }\href@noop {} {\emph {\bibinfo {title}
  {{LAPACK} Users' Guide}}},\ \bibinfo {edition} {3rd}\ ed.\ (\bibinfo
  {publisher} {Society for Industrial and Applied Mathematics},\ \bibinfo
  {address} {Philadelphia, PA},\ \bibinfo {year} {1999})\BibitemShut {NoStop}%
\bibitem [{\citenamefont {{Hinshaw}}\ \emph {et~al.}(2003)\citenamefont
  {{Hinshaw}}, \citenamefont {{Spergel}}, \citenamefont {{Verde}},
  \citenamefont {{Hill}}, \citenamefont {{Meyer}}, \citenamefont {{Barnes}},
  \citenamefont {{Bennett}}, \citenamefont {{Halpern}}, \citenamefont
  {{Jarosik}}, \citenamefont {{Kogut}}, \citenamefont {{Komatsu}},
  \citenamefont {{Limon}}, \citenamefont {{Page}}, \citenamefont {{Tucker}},
  \citenamefont {{Weiland}}, \citenamefont {{Wollack}},\ and\ \citenamefont
  {{Wright}}}]{Hinshaw_etal:2003}%
  \BibitemOpen
  \bibfield  {author} {\bibinfo {author} {\bibfnamefont {G.}~\bibnamefont
  {{Hinshaw}}}, \bibinfo {author} {\bibfnamefont {D.~N.}\ \bibnamefont
  {{Spergel}}}, \bibinfo {author} {\bibfnamefont {L.}~\bibnamefont {{Verde}}},
  \bibinfo {author} {\bibfnamefont {R.~S.}\ \bibnamefont {{Hill}}}, \bibinfo
  {author} {\bibfnamefont {S.~S.}\ \bibnamefont {{Meyer}}}, \bibinfo {author}
  {\bibfnamefont {C.}~\bibnamefont {{Barnes}}}, \bibinfo {author}
  {\bibfnamefont {C.~L.}\ \bibnamefont {{Bennett}}}, \bibinfo {author}
  {\bibfnamefont {M.}~\bibnamefont {{Halpern}}}, \bibinfo {author}
  {\bibfnamefont {N.}~\bibnamefont {{Jarosik}}}, \bibinfo {author}
  {\bibfnamefont {A.}~\bibnamefont {{Kogut}}}, \bibinfo {author} {\bibfnamefont
  {E.}~\bibnamefont {{Komatsu}}}, \bibinfo {author} {\bibfnamefont
  {M.}~\bibnamefont {{Limon}}}, \bibinfo {author} {\bibfnamefont
  {L.}~\bibnamefont {{Page}}}, \bibinfo {author} {\bibfnamefont {G.~S.}\
  \bibnamefont {{Tucker}}}, \bibinfo {author} {\bibfnamefont {J.~L.}\
  \bibnamefont {{Weiland}}}, \bibinfo {author} {\bibfnamefont {E.}~\bibnamefont
  {{Wollack}}}, \ and\ \bibinfo {author} {\bibfnamefont {E.~L.}\ \bibnamefont
  {{Wright}}},\ }\href {\doibase 10.1086/377225} {\bibfield  {journal}
  {\bibinfo  {journal} {Astrophys. J. Suppl. Ser.}\ }\textbf {\bibinfo {volume}
  {148}},\ \bibinfo {pages} {135} (\bibinfo {year} {2003})},\ \Eprint
  {http://arxiv.org/abs/arXiv:astro-ph/0302217} {arXiv:astro-ph/0302217}
  \BibitemShut {NoStop}%
\bibitem [{\citenamefont {{Skilling}}(2004)}]{Skilling:2004}%
  \BibitemOpen
  \bibfield  {author} {\bibinfo {author} {\bibfnamefont {J.}~\bibnamefont
  {{Skilling}}},\ }in\ \href {\doibase 10.1063/1.1835238} {\emph {\bibinfo
  {booktitle} {AIP Conference Proceedings of the 24th International Workshop on
  Bayesian Inference and Maximum Entropy Methods in Science and
  Engineering}}},\ \bibinfo {series} {American Institute of Physics Conference
  Series}, Vol.\ \bibinfo {volume} {735},\ \bibinfo {editor} {edited by\
  \bibinfo {editor} {\bibfnamefont {R.}~\bibnamefont {{Fischer}}}, \bibinfo
  {editor} {\bibfnamefont {R.}~\bibnamefont {{Preuss}}}, \ and\ \bibinfo
  {editor} {\bibfnamefont {U.~V.}\ \bibnamefont {{Toussaint}}}}\ (\bibinfo
  {year} {2004})\ pp.\ \bibinfo {pages} {395--405}\BibitemShut {NoStop}%
\bibitem [{\citenamefont {Bennett}\ \emph
  {et~al.}(2003{\natexlab{b}})\citenamefont {Bennett} \emph
  {et~al.}}]{Bennett:2003ca}%
  \BibitemOpen
  \bibfield  {author} {\bibinfo {author} {\bibfnamefont {C.}~\bibnamefont
  {Bennett}} \emph {et~al.} (\bibinfo {collaboration} {WMAP Collaboration}),\
  }\href {\doibase 10.1086/377252} {\bibfield  {journal} {\bibinfo  {journal}
  {Astrophys.J.Suppl.}\ }\textbf {\bibinfo {volume} {148}},\ \bibinfo {pages}
  {97} (\bibinfo {year} {2003}{\natexlab{b}})},\ \Eprint
  {http://arxiv.org/abs/astro-ph/0302208} {arXiv:astro-ph/0302208 [astro-ph]}
  \BibitemShut {NoStop}%
\bibitem [{\citenamefont {Wright}\ \emph {et~al.}(2009)\citenamefont {Wright}
  \emph {et~al.}}]{Wright:2008ib}%
  \BibitemOpen
  \bibfield  {author} {\bibinfo {author} {\bibfnamefont {E.}~\bibnamefont
  {Wright}} \emph {et~al.} (\bibinfo {collaboration} {WMAP Collaboration}),\
  }\href {\doibase 10.1088/0067-0049/180/2/283} {\bibfield  {journal} {\bibinfo
   {journal} {Astrophys.J.Suppl.}\ }\textbf {\bibinfo {volume} {180}},\
  \bibinfo {pages} {283} (\bibinfo {year} {2009})},\ \Eprint
  {http://arxiv.org/abs/0803.0577} {arXiv:0803.0577 [astro-ph]} \BibitemShut
  {NoStop}%
\bibitem [{\citenamefont {Pontzen}\ and\ \citenamefont
  {Peiris}(2010)}]{Pontzen_Peiris:2010}%
  \BibitemOpen
  \bibfield  {author} {\bibinfo {author} {\bibfnamefont {A.}~\bibnamefont
  {Pontzen}}\ and\ \bibinfo {author} {\bibfnamefont {H.~V.}\ \bibnamefont
  {Peiris}},\ }\href {\doibase 10.1103/PhysRevD.81.103008} {\bibfield
  {journal} {\bibinfo  {journal} {Phys. Rev.}\ }\textbf {\bibinfo {volume}
  {D81}},\ \bibinfo {pages} {103008} (\bibinfo {year} {2010})},\ \Eprint
  {http://arxiv.org/abs/1004.2706} {arXiv:1004.2706 [astro-ph.CO]} \BibitemShut
  {NoStop}%
\bibitem [{\citenamefont {Dickey}(1971)}]{Dickey:1971}%
  \BibitemOpen
  \bibfield  {author} {\bibinfo {author} {\bibfnamefont {J.}~\bibnamefont
  {Dickey}},\ }\href@noop {} {\bibfield  {journal} {\bibinfo  {journal} {Ann.
  Mathemat. Statist.}\ }\textbf {\bibinfo {volume} {42}},\ \bibinfo {pages}
  {204} (\bibinfo {year} {1971})}\BibitemShut {NoStop}%
\bibitem [{\citenamefont {{Tauber}}\ \emph {et~al.}(2010)\citenamefont
  {{Tauber}}, \citenamefont {{Mandolesi}}, \citenamefont {{Puget}} \emph
  {et~al.}}]{Tauber2010}%
  \BibitemOpen
  \bibfield  {author} {\bibinfo {author} {\bibfnamefont {J.~A.}\ \bibnamefont
  {{Tauber}}}, \bibinfo {author} {\bibfnamefont {N.}~\bibnamefont
  {{Mandolesi}}}, \bibinfo {author} {\bibfnamefont {J.}~\bibnamefont
  {{Puget}}},  \emph {et~al.},\ }\href {\doibase 10.1051/0004-6361/200912983}
  {\bibfield  {journal} {\bibinfo  {journal} {Astron. Astrophys.}\ }\textbf
  {\bibinfo {volume} {520}} (\bibinfo {year} {2010}),\
  10.1051/0004-6361/200912983}\BibitemShut {NoStop}%
\bibitem [{\citenamefont {{Niemack}}\ \emph {et~al.}(2010)\citenamefont
  {{Niemack}}, \citenamefont {{Ade}}, \citenamefont {{Aguirre}}, \citenamefont
  {{Barrientos}}, \citenamefont {{Beall}}, \citenamefont {{Bond}},
  \citenamefont {{Britton}}, \citenamefont {{Cho}}, \citenamefont {{Das}},
  \citenamefont {{Devlin}}, \citenamefont {{Dicker}}, \citenamefont
  {{Dunkley}}, \citenamefont {{D{\"u}nner}}, \citenamefont {{Fowler}},
  \citenamefont {{Hajian}}, \citenamefont {{Halpern}}, \citenamefont
  {{Hasselfield}}, \citenamefont {{Hilton}}, \citenamefont {{Hilton}},
  \citenamefont {{Hubmayr}}, \citenamefont {{Hughes}}, \citenamefont
  {{Infante}}, \citenamefont {{Irwin}}, \citenamefont {{Jarosik}},
  \citenamefont {{Klein}}, \citenamefont {{Kosowsky}}, \citenamefont
  {{Marriage}}, \citenamefont {{McMahon}}, \citenamefont {{Menanteau}},
  \citenamefont {{Moodley}}, \citenamefont {{Nibarger}}, \citenamefont
  {{Nolta}}, \citenamefont {{Page}}, \citenamefont {{Partridge}}, \citenamefont
  {{Reese}}, \citenamefont {{Sievers}}, \citenamefont {{Spergel}},
  \citenamefont {{Staggs}}, \citenamefont {{Thornton}}, \citenamefont
  {{Tucker}}, \citenamefont {{Wollack}},\ and\ \citenamefont
  {{Yoon}}}]{Niemack_etal:2010}%
  \BibitemOpen
  \bibfield  {author} {\bibinfo {author} {\bibfnamefont {M.~D.}\ \bibnamefont
  {{Niemack}}}, \bibinfo {author} {\bibfnamefont {P.~A.~R.}\ \bibnamefont
  {{Ade}}}, \bibinfo {author} {\bibfnamefont {J.}~\bibnamefont {{Aguirre}}},
  \bibinfo {author} {\bibfnamefont {F.}~\bibnamefont {{Barrientos}}}, \bibinfo
  {author} {\bibfnamefont {J.~A.}\ \bibnamefont {{Beall}}}, \bibinfo {author}
  {\bibfnamefont {J.~R.}\ \bibnamefont {{Bond}}}, \bibinfo {author}
  {\bibfnamefont {J.}~\bibnamefont {{Britton}}}, \bibinfo {author}
  {\bibfnamefont {H.~M.}\ \bibnamefont {{Cho}}}, \bibinfo {author}
  {\bibfnamefont {S.}~\bibnamefont {{Das}}}, \bibinfo {author} {\bibfnamefont
  {M.~J.}\ \bibnamefont {{Devlin}}}, \bibinfo {author} {\bibfnamefont
  {S.}~\bibnamefont {{Dicker}}}, \bibinfo {author} {\bibfnamefont
  {J.}~\bibnamefont {{Dunkley}}}, \bibinfo {author} {\bibfnamefont
  {R.}~\bibnamefont {{D{\"u}nner}}}, \bibinfo {author} {\bibfnamefont {J.~W.}\
  \bibnamefont {{Fowler}}}, \bibinfo {author} {\bibfnamefont {A.}~\bibnamefont
  {{Hajian}}}, \bibinfo {author} {\bibfnamefont {M.}~\bibnamefont {{Halpern}}},
  \bibinfo {author} {\bibfnamefont {M.}~\bibnamefont {{Hasselfield}}}, \bibinfo
  {author} {\bibfnamefont {G.~C.}\ \bibnamefont {{Hilton}}}, \bibinfo {author}
  {\bibfnamefont {M.}~\bibnamefont {{Hilton}}}, \bibinfo {author}
  {\bibfnamefont {J.}~\bibnamefont {{Hubmayr}}}, \bibinfo {author}
  {\bibfnamefont {J.~P.}\ \bibnamefont {{Hughes}}}, \bibinfo {author}
  {\bibfnamefont {L.}~\bibnamefont {{Infante}}}, \bibinfo {author}
  {\bibfnamefont {K.~D.}\ \bibnamefont {{Irwin}}}, \bibinfo {author}
  {\bibfnamefont {N.}~\bibnamefont {{Jarosik}}}, \bibinfo {author}
  {\bibfnamefont {J.}~\bibnamefont {{Klein}}}, \bibinfo {author} {\bibfnamefont
  {A.}~\bibnamefont {{Kosowsky}}}, \bibinfo {author} {\bibfnamefont {T.~A.}\
  \bibnamefont {{Marriage}}}, \bibinfo {author} {\bibfnamefont
  {J.}~\bibnamefont {{McMahon}}}, \bibinfo {author} {\bibfnamefont
  {F.}~\bibnamefont {{Menanteau}}}, \bibinfo {author} {\bibfnamefont
  {K.}~\bibnamefont {{Moodley}}}, \bibinfo {author} {\bibfnamefont {J.~P.}\
  \bibnamefont {{Nibarger}}}, \bibinfo {author} {\bibfnamefont {M.~R.}\
  \bibnamefont {{Nolta}}}, \bibinfo {author} {\bibfnamefont {L.~A.}\
  \bibnamefont {{Page}}}, \bibinfo {author} {\bibfnamefont {B.}~\bibnamefont
  {{Partridge}}}, \bibinfo {author} {\bibfnamefont {E.~D.}\ \bibnamefont
  {{Reese}}}, \bibinfo {author} {\bibfnamefont {J.}~\bibnamefont {{Sievers}}},
  \bibinfo {author} {\bibfnamefont {D.~N.}\ \bibnamefont {{Spergel}}}, \bibinfo
  {author} {\bibfnamefont {S.~T.}\ \bibnamefont {{Staggs}}}, \bibinfo {author}
  {\bibfnamefont {R.}~\bibnamefont {{Thornton}}}, \bibinfo {author}
  {\bibfnamefont {C.}~\bibnamefont {{Tucker}}}, \bibinfo {author}
  {\bibfnamefont {E.}~\bibnamefont {{Wollack}}}, \ and\ \bibinfo {author}
  {\bibfnamefont {K.~W.}\ \bibnamefont {{Yoon}}},\ }in\ \href {\doibase
  10.1117/12.857464} {\emph {\bibinfo {booktitle} {Society of Photo-Optical
  Instrumentation Engineers (SPIE) Conference Series}}},\ \bibinfo {series}
  {Society of Photo-Optical Instrumentation Engineers (SPIE) Conference
  Series}, Vol.\ \bibinfo {volume} {7741}\ (\bibinfo {year} {2010})\ \Eprint
  {http://arxiv.org/abs/1006.5049} {arXiv:1006.5049 [astro-ph.IM]} \BibitemShut
  {NoStop}%
\bibitem [{\citenamefont {{McMahon}}\ \emph {et~al.}(2009)\citenamefont
  {{McMahon}}, \citenamefont {{Aird}}, \citenamefont {{Benson}}, \citenamefont
  {{Bleem}}, \citenamefont {{Britton}}, \citenamefont {{Carlstrom}},
  \citenamefont {{Chang}}, \citenamefont {{Cho}}, \citenamefont {{de Haan}},
  \citenamefont {{Crawford}}, \citenamefont {{Crites}}, \citenamefont
  {{Datesman}}, \citenamefont {{Dobbs}}, \citenamefont {{Everett}},
  \citenamefont {{Halverson}}, \citenamefont {{Holder}}, \citenamefont
  {{Holzapfel}}, \citenamefont {{Hrubes}}, \citenamefont {{Irwin}},
  \citenamefont {{Joy}}, \citenamefont {{Keisler}}, \citenamefont {{Lanting}},
  \citenamefont {{Lee}}, \citenamefont {{Leitch}}, \citenamefont {{Loehr}},
  \citenamefont {{Lueker}}, \citenamefont {{Mehl}}, \citenamefont {{Meyer}},
  \citenamefont {{Mohr}}, \citenamefont {{Montroy}}, \citenamefont {{Niemack}},
  \citenamefont {{Ngeow}}, \citenamefont {{Novosad}}, \citenamefont {{Padin}},
  \citenamefont {{Plagge}}, \citenamefont {{Pryke}}, \citenamefont
  {{Reichardt}}, \citenamefont {{Ruhl}}, \citenamefont {{Schaffer}},
  \citenamefont {{Shaw}}, \citenamefont {{Shirokoff}}, \citenamefont
  {{Spieler}}, \citenamefont {{Stadler}}, \citenamefont {{Stark}},
  \citenamefont {{Staniszewski}}, \citenamefont {{Vanderlinde}}, \citenamefont
  {{Vieira}}, \citenamefont {{Wang}}, \citenamefont {{Williamson}},
  \citenamefont {{Yefremenko}}, \citenamefont {{Yoon}}, \citenamefont
  {{Zhan}},\ and\ \citenamefont {{Zenteno}}}]{McMahon_etal:2009}%
  \BibitemOpen
  \bibfield  {author} {\bibinfo {author} {\bibfnamefont {J.~J.}\ \bibnamefont
  {{McMahon}}}, \bibinfo {author} {\bibfnamefont {K.~A.}\ \bibnamefont
  {{Aird}}}, \bibinfo {author} {\bibfnamefont {B.~A.}\ \bibnamefont
  {{Benson}}}, \bibinfo {author} {\bibfnamefont {L.~E.}\ \bibnamefont
  {{Bleem}}}, \bibinfo {author} {\bibfnamefont {J.}~\bibnamefont {{Britton}}},
  \bibinfo {author} {\bibfnamefont {J.~E.}\ \bibnamefont {{Carlstrom}}},
  \bibinfo {author} {\bibfnamefont {C.~L.}\ \bibnamefont {{Chang}}}, \bibinfo
  {author} {\bibfnamefont {H.~S.}\ \bibnamefont {{Cho}}}, \bibinfo {author}
  {\bibfnamefont {T.}~\bibnamefont {{de Haan}}}, \bibinfo {author}
  {\bibfnamefont {T.~M.}\ \bibnamefont {{Crawford}}}, \bibinfo {author}
  {\bibfnamefont {A.~T.}\ \bibnamefont {{Crites}}}, \bibinfo {author}
  {\bibfnamefont {A.}~\bibnamefont {{Datesman}}}, \bibinfo {author}
  {\bibfnamefont {M.~A.}\ \bibnamefont {{Dobbs}}}, \bibinfo {author}
  {\bibfnamefont {W.}~\bibnamefont {{Everett}}}, \bibinfo {author}
  {\bibfnamefont {N.~W.}\ \bibnamefont {{Halverson}}}, \bibinfo {author}
  {\bibfnamefont {G.~P.}\ \bibnamefont {{Holder}}}, \bibinfo {author}
  {\bibfnamefont {W.~L.}\ \bibnamefont {{Holzapfel}}}, \bibinfo {author}
  {\bibfnamefont {D.}~\bibnamefont {{Hrubes}}}, \bibinfo {author}
  {\bibfnamefont {K.~D.}\ \bibnamefont {{Irwin}}}, \bibinfo {author}
  {\bibfnamefont {M.}~\bibnamefont {{Joy}}}, \bibinfo {author} {\bibfnamefont
  {R.}~\bibnamefont {{Keisler}}}, \bibinfo {author} {\bibfnamefont {T.~M.}\
  \bibnamefont {{Lanting}}}, \bibinfo {author} {\bibfnamefont {A.~T.}\
  \bibnamefont {{Lee}}}, \bibinfo {author} {\bibfnamefont {E.~M.}\ \bibnamefont
  {{Leitch}}}, \bibinfo {author} {\bibfnamefont {A.}~\bibnamefont {{Loehr}}},
  \bibinfo {author} {\bibfnamefont {M.}~\bibnamefont {{Lueker}}}, \bibinfo
  {author} {\bibfnamefont {J.}~\bibnamefont {{Mehl}}}, \bibinfo {author}
  {\bibfnamefont {S.~S.}\ \bibnamefont {{Meyer}}}, \bibinfo {author}
  {\bibfnamefont {J.~J.}\ \bibnamefont {{Mohr}}}, \bibinfo {author}
  {\bibfnamefont {T.~E.}\ \bibnamefont {{Montroy}}}, \bibinfo {author}
  {\bibfnamefont {M.~D.}\ \bibnamefont {{Niemack}}}, \bibinfo {author}
  {\bibfnamefont {C.~C.}\ \bibnamefont {{Ngeow}}}, \bibinfo {author}
  {\bibfnamefont {V.}~\bibnamefont {{Novosad}}}, \bibinfo {author}
  {\bibfnamefont {S.}~\bibnamefont {{Padin}}}, \bibinfo {author} {\bibfnamefont
  {T.}~\bibnamefont {{Plagge}}}, \bibinfo {author} {\bibfnamefont
  {C.}~\bibnamefont {{Pryke}}}, \bibinfo {author} {\bibfnamefont
  {C.}~\bibnamefont {{Reichardt}}}, \bibinfo {author} {\bibfnamefont {J.~E.}\
  \bibnamefont {{Ruhl}}}, \bibinfo {author} {\bibfnamefont {K.~K.}\
  \bibnamefont {{Schaffer}}}, \bibinfo {author} {\bibfnamefont
  {L.}~\bibnamefont {{Shaw}}}, \bibinfo {author} {\bibfnamefont
  {E.}~\bibnamefont {{Shirokoff}}}, \bibinfo {author} {\bibfnamefont {H.~G.}\
  \bibnamefont {{Spieler}}}, \bibinfo {author} {\bibfnamefont {B.}~\bibnamefont
  {{Stadler}}}, \bibinfo {author} {\bibfnamefont {A.~A.}\ \bibnamefont
  {{Stark}}}, \bibinfo {author} {\bibfnamefont {Z.}~\bibnamefont
  {{Staniszewski}}}, \bibinfo {author} {\bibfnamefont {K.}~\bibnamefont
  {{Vanderlinde}}}, \bibinfo {author} {\bibfnamefont {J.~D.}\ \bibnamefont
  {{Vieira}}}, \bibinfo {author} {\bibfnamefont {G.}~\bibnamefont {{Wang}}},
  \bibinfo {author} {\bibfnamefont {R.}~\bibnamefont {{Williamson}}}, \bibinfo
  {author} {\bibfnamefont {V.}~\bibnamefont {{Yefremenko}}}, \bibinfo {author}
  {\bibfnamefont {K.~W.}\ \bibnamefont {{Yoon}}}, \bibinfo {author}
  {\bibfnamefont {O.}~\bibnamefont {{Zhan}}}, \ and\ \bibinfo {author}
  {\bibfnamefont {A.}~\bibnamefont {{Zenteno}}},\ }in\ \href {\doibase
  10.1063/1.3292391} {\emph {\bibinfo {booktitle} {American Institute of
  Physics Conference Series}}},\ \bibinfo {series} {American Institute of
  Physics Conference Series}, Vol.\ \bibinfo {volume} {1185},\ \bibinfo
  {editor} {edited by\ \bibinfo {editor} {\bibfnamefont {B.}~\bibnamefont
  {{Young}}}, \bibinfo {editor} {\bibfnamefont {B.}~\bibnamefont {{Cabrera}}},
  \ and\ \bibinfo {editor} {\bibfnamefont {A.}~\bibnamefont {{Miller}}}}\
  (\bibinfo {year} {2009})\ pp.\ \bibinfo {pages} {511--514}\BibitemShut
  {NoStop}%
\bibitem [{\citenamefont {{Crill}}\ \emph {et~al.}(2008)\citenamefont
  {{Crill}}, \citenamefont {{Ade}}, \citenamefont {{Battistelli}},
  \citenamefont {{Benton}}, \citenamefont {{Bihary}}, \citenamefont {{Bock}},
  \citenamefont {{Bond}}, \citenamefont {{Brevik}}, \citenamefont {{Bryan}},
  \citenamefont {{Contaldi}}, \citenamefont {{Dor{\'e}}}, \citenamefont
  {{Farhang}}, \citenamefont {{Fissel}}, \citenamefont {{Golwala}},
  \citenamefont {{Halpern}}, \citenamefont {{Hilton}}, \citenamefont
  {{Holmes}}, \citenamefont {{Hristov}}, \citenamefont {{Irwin}}, \citenamefont
  {{Jones}}, \citenamefont {{Kuo}}, \citenamefont {{Lange}}, \citenamefont
  {{Lawrie}}, \citenamefont {{MacTavish}}, \citenamefont {{Martin}},
  \citenamefont {{Mason}}, \citenamefont {{Montroy}}, \citenamefont
  {{Netterfield}}, \citenamefont {{Pascale}}, \citenamefont {{Riley}},
  \citenamefont {{Ruhl}}, \citenamefont {{Runyan}}, \citenamefont
  {{Trangsrud}}, \citenamefont {{Tucker}}, \citenamefont {{Turner}},
  \citenamefont {{Viero}},\ and\ \citenamefont {{Wiebe}}}]{Crill_etal:2008}%
  \BibitemOpen
  \bibfield  {author} {\bibinfo {author} {\bibfnamefont {B.~P.}\ \bibnamefont
  {{Crill}}}, \bibinfo {author} {\bibfnamefont {P.~A.~R.}\ \bibnamefont
  {{Ade}}}, \bibinfo {author} {\bibfnamefont {E.~S.}\ \bibnamefont
  {{Battistelli}}}, \bibinfo {author} {\bibfnamefont {S.}~\bibnamefont
  {{Benton}}}, \bibinfo {author} {\bibfnamefont {R.}~\bibnamefont {{Bihary}}},
  \bibinfo {author} {\bibfnamefont {J.~J.}\ \bibnamefont {{Bock}}}, \bibinfo
  {author} {\bibfnamefont {J.~R.}\ \bibnamefont {{Bond}}}, \bibinfo {author}
  {\bibfnamefont {J.}~\bibnamefont {{Brevik}}}, \bibinfo {author}
  {\bibfnamefont {S.}~\bibnamefont {{Bryan}}}, \bibinfo {author} {\bibfnamefont
  {C.~R.}\ \bibnamefont {{Contaldi}}}, \bibinfo {author} {\bibfnamefont
  {O.}~\bibnamefont {{Dor{\'e}}}}, \bibinfo {author} {\bibfnamefont
  {M.}~\bibnamefont {{Farhang}}}, \bibinfo {author} {\bibfnamefont
  {L.}~\bibnamefont {{Fissel}}}, \bibinfo {author} {\bibfnamefont {S.~R.}\
  \bibnamefont {{Golwala}}}, \bibinfo {author} {\bibfnamefont {M.}~\bibnamefont
  {{Halpern}}}, \bibinfo {author} {\bibfnamefont {G.}~\bibnamefont {{Hilton}}},
  \bibinfo {author} {\bibfnamefont {W.}~\bibnamefont {{Holmes}}}, \bibinfo
  {author} {\bibfnamefont {V.~V.}\ \bibnamefont {{Hristov}}}, \bibinfo {author}
  {\bibfnamefont {K.}~\bibnamefont {{Irwin}}}, \bibinfo {author} {\bibfnamefont
  {W.~C.}\ \bibnamefont {{Jones}}}, \bibinfo {author} {\bibfnamefont {C.~L.}\
  \bibnamefont {{Kuo}}}, \bibinfo {author} {\bibfnamefont {A.~E.}\ \bibnamefont
  {{Lange}}}, \bibinfo {author} {\bibfnamefont {C.}~\bibnamefont {{Lawrie}}},
  \bibinfo {author} {\bibfnamefont {C.~J.}\ \bibnamefont {{MacTavish}}},
  \bibinfo {author} {\bibfnamefont {T.~G.}\ \bibnamefont {{Martin}}}, \bibinfo
  {author} {\bibfnamefont {P.}~\bibnamefont {{Mason}}}, \bibinfo {author}
  {\bibfnamefont {T.~E.}\ \bibnamefont {{Montroy}}}, \bibinfo {author}
  {\bibfnamefont {C.~B.}\ \bibnamefont {{Netterfield}}}, \bibinfo {author}
  {\bibfnamefont {E.}~\bibnamefont {{Pascale}}}, \bibinfo {author}
  {\bibfnamefont {D.}~\bibnamefont {{Riley}}}, \bibinfo {author} {\bibfnamefont
  {J.~E.}\ \bibnamefont {{Ruhl}}}, \bibinfo {author} {\bibfnamefont {M.~C.}\
  \bibnamefont {{Runyan}}}, \bibinfo {author} {\bibfnamefont {A.}~\bibnamefont
  {{Trangsrud}}}, \bibinfo {author} {\bibfnamefont {C.}~\bibnamefont
  {{Tucker}}}, \bibinfo {author} {\bibfnamefont {A.}~\bibnamefont {{Turner}}},
  \bibinfo {author} {\bibfnamefont {M.}~\bibnamefont {{Viero}}}, \ and\
  \bibinfo {author} {\bibfnamefont {D.}~\bibnamefont {{Wiebe}}},\ }in\ \href
  {\doibase 10.1117/12.787446} {\emph {\bibinfo {booktitle} {Society of
  Photo-Optical Instrumentation Engineers (SPIE) Conference Series}}},\
  \bibinfo {series} {Society of Photo-Optical Instrumentation Engineers (SPIE)
  Conference Series}, Vol.\ \bibinfo {volume} {7010}\ (\bibinfo {year} {2008})\
  \Eprint {http://arxiv.org/abs/0807.1548} {arXiv:0807.1548} \BibitemShut
  {NoStop}%
\end{thebibliography}%


\end{document}